\documentclass[10pt,journal,compsoc]{IEEEtran}
\ifCLASSOPTIONcompsoc
  \usepackage[nocompress]{cite}
\else
  \usepackage{cite}
\fi
\ifCLASSINFOpdf
\else
\fi
\usepackage{amsmath,amssymb,amsfonts}
\usepackage{balance}

\makeatletter
\def\bbordermatrix#1{\begingroup \m@th
  \@tempdima 8.75\p@
  \setbox\z@\vbox{%
    \def\cr{\crcr\noalign{\kern2\p@\global\let\cr\endline}}%
    \ialign{$##$\hfil\kern2\p@\kern\@tempdima&\thinspace\hfil$##$\hfil
      &&\quad\hfil$##$\hfil\crcr
      \omit\strut\hfil\crcr\noalign{\kern-\baselineskip}%
      #1\crcr\omit\strut\cr}}%
  \setbox\tw@\vbox{\unvcopy\z@\global\setbox\@ne\lastbox}%
  \setbox\tw@\hbox{\unhbox\@ne\unskip\global\setbox\@ne\lastbox}%
  \setbox\tw@\hbox{$\kern\wd\@ne\kern-\@tempdima\left(\kern-\wd\@ne
    \global\setbox\@ne\vbox{\box\@ne\kern2\p@}%
    \vcenter{\kern-\ht\@ne\unvbox\z@\kern-\baselineskip}\,\right)$}%
  \null\;\vbox{\kern\ht\@ne\box\tw@}\endgroup}
\makeatother

\usepackage{enumitem}  
\usepackage{subfigure}
\usepackage{graphicx}
\usepackage{color}
\usepackage{algorithm2e}
\usepackage{hyperref}
\usepackage{xurl}  
\usepackage[x11names]{xcolor}
\usepackage{ifthen}
\hypersetup{colorlinks, linkcolor=purple, urlcolor=blue, citecolor=Green4}

\usepackage{tikz}            
\usepackage{fp}            
\usetikzlibrary{calc}          
\usepackage{verbatim}
\usepackage{tkz-euclide}   
\usetikzlibrary{math}

\usepackage{amsthm}   
\usepackage{mathtools} 
\usepackage{thmtools}  

\usepackage{booktabs}    
\usepackage{multirow}

\usepackage[nice]{nicefrac}

\newtheorem{definition}{Definition}
\newtheorem{lemma}{Lemma}

\graphicspath{{./figures/}}

\newcommand{\E}{\mathbf{E}}

\newcommand{\vectr}[1]{\boldsymbol{#1}}

\newcommand{\calx}{\mathcal{X}}   
\newcommand{\caly}{\mathcal{Y}}   
\newcommand{\calu}{\mathcal{U}}   
\newcommand{\calv}{\mathcal{V}}   

\newcommand{\calz}{\mathcal{Z}}   

\newcommand{\calc}{\mathcal{C}}   

\newcommand{\dist}{\mathbf{d}} 

\newcommand{\reals}{\mathbb{R}}    
\newcommand{\binfield}{\mathbb{F}_2}
\newcommand{\calb}{\mathcal{B}}     
\newcommand{\integers}{\mathbb{Z}} 

\newcommand{\vtheta}{\vectr{\theta}}     
\newcommand{\vphi}{\vectr{\phi}}           
\newcommand{\vq}{\vectr{q}}           

\newcommand{\mech}{M}  
\newcommand{\vw}{\vectr{w}}        
\newcommand{\vzeros}{\vectr{0}}        
\newcommand{\vones}{\vectr{1}}        
\newcommand{\mg}{G}    
\newcommand{\dmax}{d_{\textit{max}}}     
\newcommand{\inv} {INV}                                          
\newcommand{\invn} {INV-N}                                          
\newcommand{\invp} {INV-P}                                          
\newcommand{\rapn} {\textsc{Rap}-N}                                          
\newcommand{\rapp} {\textsc{Rap}-P}                                          
\newcommand{\ibu}{IBU}                                      
\newcommand{\gibu}{GIBU}                                         

\newcommand{\leps}{\varepsilon}    
\newcommand{\geps}{\varepsilon_g}  

\newboolean{showrevisions}
\setboolean{showrevisions}{true}  
\ifthenelse{\boolean{showrevisions}}{%

}{%
}

\makeatletter
\newcounter{subfig}[figure] 
\renewcommand{\thesubfig}{\alph{subfig}}
\newcommand{\mysubfig}[4][\thesubfig]{
  \begin{minipage}[t]{#3}%
    \centering
    \includegraphics[width=\linewidth]{#4}\\[-2pt]
    \refstepcounter{subfig}%
    \if\relax\detokenize{#1}\relax
    \else
      \footnotesize(#1)
    \fi
    \label{#2}%
  \end{minipage}%
}
\makeatother

\begin{document}
%

\title{On the Consistency and Performance of the Iterative Bayesian Update}
%
%
%
%

\author{Ehab~ElSalamouny
        and~Catuscia~Palamidessi 
\IEEEcompsocitemizethanks{\IEEEcompsocthanksitem E. ElSalamouny is with Inria, France, and Suez Canal University, Egypt. C. Palamidessi is with Inria and LIX, \'{E}cole Polytechnique, France.\protect\\
 }
\thanks{Published in \emph{IEEE Transactions on Dependable and Secure Computing}, 2026.\\
Copyright \copyright\ 2026 IEEE. Personal use of this material is permitted. However,
permission to use this material for any other purposes must be obtained
from the IEEE.\\
DOI \href{https://doi.org/10.1109/TDSC.2026.3718449}{10.1109/TDSC.2026.3718449}
} }

\IEEEtitleabstractindextext{%
\begin{abstract}

In many situations, estimating the distribution of users' data concerning certain attributes is important. To facilitate this estimation while safeguarding users' privacy, the local privacy model 
is commonly employed, in which each user applies a local protection mechanism to release a noisy version of their original data to the data collector. 
The original distribution is then estimated using methods such as Matrix Inversion (\inv{}), \textsc{Rappor}'s estimator, and iterative Bayesian update (\ibu{}). 
In this article, we experimentally demonstrate that \ibu{} significantly outperforms the other methods when user data is protected through metric privacy mechanisms. 
We also explain the mathematical reason for the suboptimal performance of \inv{} under those metric privacy mechanisms. Conversely, \ibu{} exhibits performance 
similar to \inv{} under typical mechanisms of local differential privacy, specifically the $k$-RR and \textsc{Rappor}. 
In addition, we investigate \ibu{}'s consistency, which is a crucial property as it means that the estimate converges to the true distribution as the number of data points increases. 
In the literature, \ibu{} is claimed to be consistent, but there is no proof for this claim. We provide a formal proof of consistency leveraging the fact that \ibu{} is a maximum likelihood estimator.
Finally, we examine scenarios involving an infinite alphabet for sensitive data and propose a method allowing \ibu{} to operate effectively in these situations.
\end{abstract}

%

}

\maketitle

\IEEEdisplaynontitleabstractindextext

%
\IEEEpeerreviewmaketitle

\IEEEraisesectionheading{\section{Introduction}\label{sec:introduction}}

\IEEEPARstart{W}ith the growing use of internet-connected devices, personal data is collected in ever larger volumes and for diverse purposes. Big-data technologies offer substantial benefits for industry and society, but the collection and processing of personal data raise serious privacy issues. This has led researchers to seek methods that protect sensitive information while preserving data utility. In this context, \emph{differential privacy}~\cite{Dwork:14:Algorithmic} has become the most widely adopted framework over the past two decades.

In this paper, we study the \emph{local} privacy model, where each user applies a privacy mechanism to their data to produce and send a noisy version to a server. Because the mechanism runs on the user’s device, the model is called local.
A well-known such model is \emph{local differential privacy} (LDP)~\cite{Duchi:13:FOCS,Kairouz:16:JMLR}. Another is \emph{metric privacy}~\cite{Chatzikokolakis:13:PETS,Andres:13:CCS}, or $d$-privacy, a generalization of differential privacy tailored to data with a natural distance, such as ages, meter readings, and locations. Unlike LDP, which can map a data point to any value in the domain, metric privacy preserves proximity, yielding a better privacy-utility trade-off when utility depends on the distance between true values and their estimates.

Concerning utility, we focus on estimating the distribution of the original data from the collected noisy releases. This constitutes the most general kind of aggregate information that can be extracted from the data, from which a wide range of further analyses can be derived. Estimating the distribution has also many applications on its own. For instance, in machine learning, characterizing the distribution of training data allows developers to identify systematic biases or ``distribution shifts'' that could lead to unfair or inaccurate outcomes.

While much prior work studies histogram estimation, we focus on probability distributions, i.e., estimates on the simplex that are non-negative and sum to one. These constraints ensure compliance with the axioms of probability, which is essential for tasks such as sampling, likelihood computation, Bayesian inference, and information-theoretic analysis. They also enable meaningful comparison via divergences (KL, TV, Wasserstein, etc.), whereas unconstrained histograms can produce misleading or distorted values for these metrics.

One proposed LDP protocol for privately collecting data and performing this estimation is \textsc{Rappor}~\cite{Erlingsson:14:CCS}. It has two components: a mechanism for obfuscating users' data and an estimator that recovers the original distribution from the obfuscated outputs. However, depending on the sensitive attribute, users may need a different sanitization mechanism than \textsc{Rappor}. Thus, we need a generic estimator that can handle noisy data from arbitrary mechanisms. Two kinds of generic estimators have been proposed in the 
literature: the family of estimators based on the matrix inversion (\inv{})~\cite{Agrawal:05:ICMD,Kairouz:16:ICML}
and the iterative Bayesian update (\ibu{})~\cite{Agrawal:01:PDS,Agrawal:05:ICMD}\footnote{\ibu{} was originally proposed in \cite{Agrawal:01:PDS}; however, 
the name “iterative Bayesian update” (\ibu{}) was introduced later in \cite{Agrawal:05:ICMD}.}.
The \inv{} estimators have the advantage of relying on simple linear algebra, followed by 
a post-processing step to obtain a valid distribution. 
This post-processing can be either a normalization or a projection onto the space 
of probability distributions; 
we refer to the corresponding variants as \invn{} and \invp{}, respectively.
In the rest of this paper, we will use \inv{}  to refer to both 
\invn{} and \invp{}. 
In contrast, \ibu{} is an iterative algorithm based on the Expectation–Maximization method from statistics, 
and it produces a maximum likelihood estimate (MLE) of the original distribution from the noisy data, 
as proved in \cite{ehab:eurosp:2020}.

\begin{figure}
\centering 
\subfigure[Using \invn{}]{
      \label{fig:geom_ages_invn}
      \includegraphics[width=0.22\textwidth]{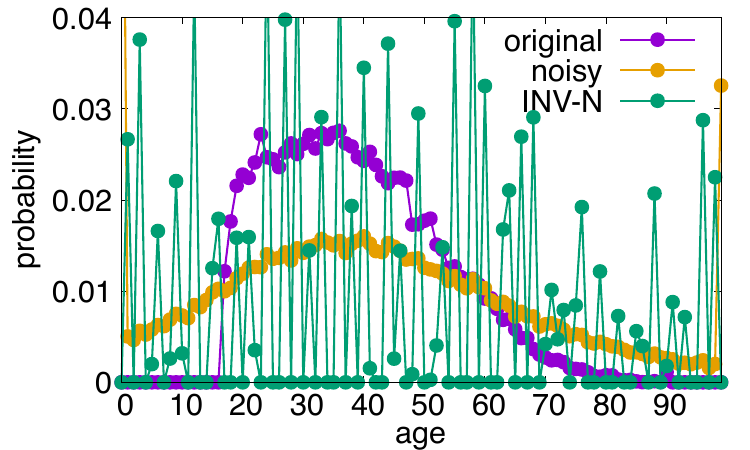}
      }
\vspace{-10pt}
\subfigure[Using \invp{}]{
      \label{fig:geom_ages_invp}
      \includegraphics[width=0.22\textwidth]{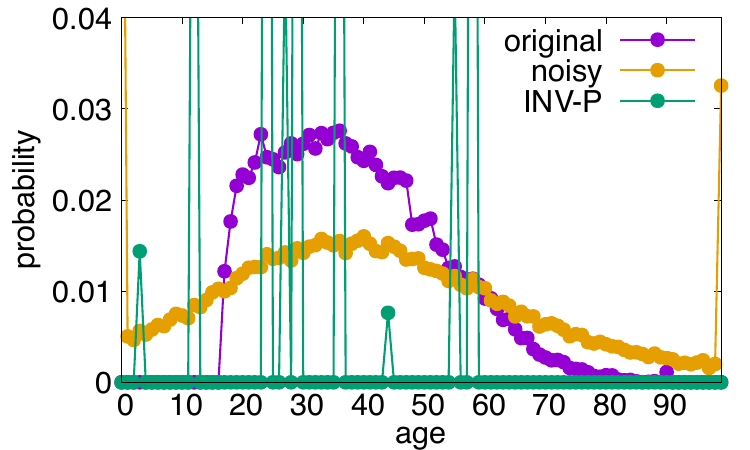}
      }
\vspace{-10pt}
\subfigure[Using \ibu{}]{
      \label{fig:geom_ages_ibu}
      \includegraphics[width=0.22\textwidth]{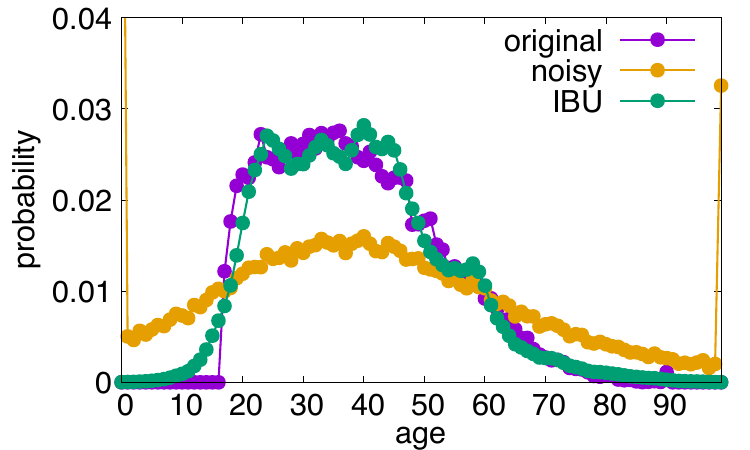}
      }
\caption{Estimating the real distribution of persons' ages registered in the Adults dataset~\cite{adult:1996}. This estimation  
uses sanitized data produced by a truncated linear geometric mechanism with $\lambda=0.05$.}
\label{fig:geometric_on_ages}
\end{figure}

In this paper, we focus on \ibu{}. Our interest in this estimator is motivated by our experimental results showing that \ibu{} has a much wider range of applicability than other methods. 
When the data are obfuscated using the $k$-RR, a typical mechanism satisfying LDP, both \ibu{} and \inv{} produce comparably accurate estimates. Likewise, when the basic \textsc{Rappor} mechanism, which also satisfies LDP, obfuscates the data, both the \textsc{Rappor} estimator and \ibu{} yield similarly good estimates.
However, when the data are obfuscated using typical mechanisms of metric privacy, we find that \ibu{} 
decisively outperforms \inv{}.
For instance, Figure~\ref{fig:geometric_on_ages} shows the results of estimating the distribution of the users' ages after sanitizing the original values by a linear geometric mechanism. As we can see, the \inv{} methods perform poorly, while \ibu{}'s estimate is quite precise. 
Full details of our experiments are given in Section~\ref{sec:experiments}.
We demonstrate the superiority of \ibu{} over \inv{} for two classes of metric privacy mechanisms. 
The first class includes mechanisms applicable to one-dimensional data, such as the linear geometric, 
Laplace, and exponential mechanisms. The second class targets two-dimensional (location) data, where metric privacy is also called geo-indistinguishability~\cite{Andres:13:CCS}. Mechanisms in this class include 
the planar Laplace~\cite{Andres:13:CCS}, planar geometric~\cite{Chatzikokolakis:17:POPETS}, and the planar version of the exponential mechanism~\cite{Chatzikokolakis:15:POPETS}, in addition to the \emph{taxicab mechanism} introduced in this paper.
Thus, relative to \ibu{}, we can claim that while \inv{} performs well under typical local differential privacy mechanisms, it performs poorly under metric privacy mechanisms. We support this conclusion by deriving an upper bound on the error of \inv{} (before post-processing) under $k$-RR (\autoref{prop:invKrrBound}) and lower bounds under the linear geometric and taxicab mechanisms (Propositions \ref{prop:invTLinGeomBound} and \ref{prop:invTTaxicabBound}, respectively).

A practical estimator needs to be \emph{consistent}, meaning that its estimate converges to the 
true distribution as the number of data points increases. In general, consistency depends on 
the privacy mechanism used to sanitize the data. For instance, \inv{} is consistent when used 
with the $k$-RR mechanism~\cite{Kairouz:16:JMLR}, as in this case the estimation error decreases 
with the number of noisy releases~\cite{Kairouz:16:ICML}. 
Similarly, the authors of \cite{Kairouz:16:ICML} derived an expression for the estimation error 
of the \textsc{Rappor} protocol, showing that its estimate also converges to the true distribution.

Regarding \ibu{}, its consistency was claimed in \cite{Agrawal:01:PDS} (Observation 4.1); however, the justification was incorrect, and the claim itself is wrong: consistency was claimed in general, whereas, as we will show, there exist mechanisms for which \ibu{} is not consistent. 

In this paper, we describe precisely the conditions that guarantee the consistency of \ibu{}.
Unlike the other estimators, it is hard to justify \ibu{}'s consistency based on an analytic expression for the estimation error because of two main reasons:
firstly, \ibu{} is an iterative procedure; secondly, we want to study \ibu{}'s consistency with any privacy mechanism instead of 
restricting to specific ones as it was the case for \inv{} and \textsc{Rappor} in \cite{Kairouz:16:ICML}. 
Therefore, we take a different approach: we use the fact that \ibu{} yields an MLE~\cite{ehab:eurosp:2020} and we apply the classical
theory of the maximum likelihood estimation (e.g.~\cite{Whitney:94:BOOK}) to the local privacy model to identify the required conditions of \ibu{}'s consistency. 

The consistency depends on whether the used privacy mechanism assures 
the \emph{identification} of the original distribution. This property 
holds if every two different distributions of the original 
data induce different distributions on the mechanism's outputs. We characterize the 
identification property, and show that it is equivalent to the \emph{strict concavity} of 
the log-likelihood function for \emph{some finite set of observations} (\autoref{thm:identification}). 
Additional conditions on the underlying privacy mechanism are also needed to ensure \ibu{}'s consistency. 
We describe these conditions and show that they are satisfied by state-of-the-art mechanisms. 

Finally, we consider the situation when the alphabet of the secrets (original users' data) is infinite, for instance, when the bounds of the sensitive attribute are unknown. Applying \ibu{} in this case is problematic because \ibu{} was designed for only finite alphabets. 
To address this problem, we present a method that identifies a finite subset of the alphabet, which we call ``likely subset'', such that the MLE computed on this subset is also an MLE for the original alphabet.

\subsection{Contributions}
\begin{itemize}

\item We identify sufficient conditions for \ibu{}'s consistency.
In particular, we show that the log-likelihood function must be strictly concave for some set of observations.

\item We show an example in which the above conditions are not satisfied, and 
the consistency does not hold, thus proving that the general claim of \cite{Agrawal:01:PDS} is wrong. 

\item We prove that \ibu{} is consistent when used with the $k$-RR, \textsc{Rappor}, and linear geometric mechanisms (both truncated and non-truncated). We also introduce the taxicab mechanism for planar (location) data and show that \ibu{} remains consistent under this mechanism.

\item We compare \ibu{} and \inv{} on real distributions and privacy mechanisms, experimentally showing that the estimation performance of \inv{} is poor under metric privacy mechanisms. 
We formally justify this observation by deriving lower bounds for the \inv{} error 
under two mechanisms of this kind, namely the linear geometric and the taxicab mechanism,  
and we show that these bounds are significantly large.

\item For an infinite alphabet $\calx$ and particular mechanisms, we characterize the ``likely'' finite subset, such that running \ibu{} on this subset computes the MLE on the original $\calx$. 

\end{itemize}

\subsection{Structure of the paper}
Section~\ref{sec:related_work} discusses related work, followed by preliminaries in Section~\ref{sec:preliminaries}. 
Section~\ref{sec:properties} summarizes the relationships established in this paper among the properties of the local model.
Section~\ref{sec:strictconcavity} characterizes the log-likelihood strict concavity, followed by 
discussing the MLE uniqueness in Section~\ref{sec:uniqueness} and the consistency of \ibu{} in Section~\ref{sec:conv_real}. 
Section~\ref{sec:experiments} experimentally evaluates \ibu{} against existing estimators.
Section~\ref{sec:infinite_alphabet} extends the applicability of \ibu{} to infinite alphabets.
Finally, Section~\ref{sec:conclusion} concludes the paper and outlines directions for future work.
Proofs are provided in the 
appendices,
and our implementation used in the experiments 
is available at \url{https://gitlab.com/eelsalam/cpriv}.

\section{Related work}\label{sec:related_work}

Most of the literature on estimating the distribution of private data focused on LDP protocols~\cite{Duchi:13:FOCS,Dwork:14:Algorithmic,Kairouz:16:JMLR}. We refer to 
\cite{Raab:25:PoPETS} for a detailed survey.
The authors of \cite{Duchi:13:NIPS,Duchi:13:FOCS} identified a lower bound for the minimax error of the estimated distributions under LDP, and they proposed a protocol that attains this lower bound. 
In their protocol, every user reports to the server a noisy-bit vector in the same manner as the basic \textsc{Rappor} mechanism~\cite{Erlingsson:14:CCS}, which we recall 
in Section \ref{sec:rappor}. The authors of \cite{Bassily:15:STOC} could reduce the communication cost 
(the length of the reported vector) using the idea of \emph{Random Projection Matrix} which was also used in \cite{Hsu:2012:ICALP} for efficiently 
identifying heavy hitters. This report-length issue was mitigated also for \textsc{Rappor}~\cite{Erlingsson:14:CCS} 
by mapping the \emph{lengthy} bit-vector, encoding the user's original datum, into a (smaller) Bloom filter which is obfuscated and then 
reported to the server. Wang et al.~\cite{Wang:17:USENIX} generalized both \textsc{Rappor}
and the idea of random projection matrix;
in one generic framework for LDP protocols. This framework enabled the authors 
to introduce Optimized Local Hashing (OLH) which has low communication cost and better accuracy than earlier protocols. 
Kairouz et al. provided in \cite{Kairouz:16:ICML} an analysis for the estimation error of the basic \textsc{Rappor} protocol, and also for the \inv{} 
estimators after sanitizing users' data using the $k$-RR mechanism. All the above proposals used the idea of randomized response 
\cite{Warner:65:jastat} to satisfy LDP, and their authors 
derived error bounds that show their consistency. 
A comprehensive overview of the distribution estimation methods for LDP can be found in \cite{wang2020locally}. We point out that some of the methods in that paper correspond to those we consider in our paper. In particular, Norm-Mul corresponds to \invn{}, and Norm-Sub/CLS correspond to \invp{}, while MLE-Apx corresponds to IBU, but only for the $k$-RR mechanism.  Other methods for improving the estimate, such as the Bayes shrinkage and Power, use prior information or assumptions on the true distribution, and as such are outside the scope of our paper, since we focus on estimation methods that do not need any auxiliary information. Likewise, methods such as the Hadamard response are designed for specific LDP mechanisms and thus cannot be applied to the metric privacy mechanisms considered in our paper. Instead, we use IBU and INV, both of which are applicable in this setting.

As already explained, \ibu{} can be used in conjunction with \emph{any} mechanism, 
in particular those satisfying metric privacy~\cite{Chatzikokolakis:13:PETS,Chatzikokolakis:17:POPETS,Andres:13:CCS,Oya:17:CCS,imola2024metric_ccs}. 
These mechanisms have been applied to various settings, such as 
geolocation~\cite{Andres:13:CCS,Biswas:24:POPETS,Atmaca:24:OJVT,athanasiou:25:POPETS}, smart meters~\cite{Chatzikokolakis:13:PETS}, natural language processing~\cite{meisenbacher:24:LREC},
deep learning \cite{SHEN:2024} and federated learning~\cite{DBLP:journals/sncs/GalliJBPC23}. Their flexibility has also been shown by applications on special metric domain~\cite{hildebrandt2025so3,faustini2025empirical}.
See \cite{xie2025decade} for a comprehensive overview of the metric privacy literature. 
\ibu{} has also been employed to eliminate bias from training data in the context of fairness \cite{Binkyte:24:FAccT}.
Under this generality, we describe the consistency of \ibu{}. 
Concerning the estimation performance of \ibu{}, the authors of \cite{Arcolezi:2023:DBSEC} have recently provided an experimental assessment of \ibu{}'s performance 
on several LDP protocols,  including 
Optimal Unary Incoding (OUE)\cite{Wang:17:USENIX}, 
Binary Local Hashing (BLH) \cite{Bassily:15:STOC},
and OLH \cite{Wang:17:USENIX},
and have compared it to the estimation procedure proposed in the literature for these protocols, obtaining results that show the superiority of \ibu{}. Our paper complements these results, showing that the superiority of \ibu{} is even 
more prominent when the data are obfuscated by the ``classical'' metric privacy mechanisms Laplace and Geometric~\cite{Chatzikokolakis:13:PETS,Andres:13:CCS,Chatzikokolakis:17:POPETS}.
Furthermore, 
\ibu{} has been shown to perform well also on other mechanisms: \cite{Biswas:24:POPETS} has reported experiments on the application of \ibu{} on 
data obfuscated by the Blahut-Arimoto mechanism~\cite{Oya:17:CCS}, presenting high precision estimates. 
In \cite{Atmaca:24:OJVT}, the authors have used \ibu{} to improve the estimation of vehicular data 
obfuscated by a privacy mechanism based on a metric induced by a graph representing road distances. 

\section{Preliminaries}\label{sec:preliminaries}

\subsection{The local privacy model}\label{sec:local_privacy_model}

\begin{figure}
\center
\includegraphics{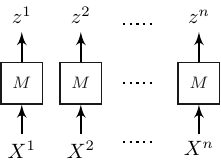}
\caption{The local privacy model. 
Every user's datum $X^i$ is obfuscated by the mechanism $\mech$
to produce a noisy datum $z^i$.} 
\label{fig:local_priv_model_ibu}
\end{figure}

We consider $n$ users and a sensitive attribute that takes one value for each user. 
Let $\calx$ be the \emph{alphabet}, i.e., the set of possible values, of this attribute. 
We assume that the users' values are independently drawn from some probability distribution 
$\vtheta = (\theta_x: \forall x \in \calx)$, hence represented by  
i.i.d. 
random variables $X^i \sim \vtheta$ where $i\in [n] = \{1,2,\dots, n\}$. 
Suppose that a data curator wants to estimate the distribution $\vtheta$ while satisfying a certain level of privacy for users. 
For this purpose, every user $i$ obfuscates their original datum $X^i$ 
using a privacy mechanism $\mech:\calx \to \calz$ to produce a noisy 
observation $z^i\in\calz$ as in Figure~\ref{fig:local_priv_model_ibu}. 
The noisy observations $(z^i)_{i\in[n]}$ are then sent to the curator, who estimates the original distribution $\vtheta$ from the collected data. 
We denote by $\vq=(q_z: z\in \calz)$ the empirical \emph{noisy} distribution on $\calz$, 
where $q_z$ is the fraction of users whose reported value is $z$. 
This model is called `local' because the privacy mechanism is applied \emph{locally} by each user to 
to their own data point. 

In this paper, we assume that the same mechanism $\mech$ is used by all users. 
This mechanism is represented as a matrix whose rows are indexed by the elements of $\calx$ 
and whose columns are indexed by the elements of $\calz$. 
Each entry $\mech_{xz}$ denotes the probability that the mechanism outputs $z$ 
when the original value is $x$. 
Note that $\mech$ is \emph{stochastic} since $\sum_z \mech_{xz} =1$. 
We write $\mech_{x *}$ for the $x$-th row of $\mech$, and $\mech_{* z}$ for the $z$-th column. 
Table \ref{tab:emnotations} summarizes these notations.

The assumptions that the original data $(X^i)_{i\in[n]}$ are i.i.d. random variables, and that all users employ the same mechanism $\mech$ define the statistical model considered throughout this paper. In particular, they define a setting in which the \ibu{} algorithm is guaranteed to compute the MLE \cite{ehab:eurosp:2020}. Moreover, because each user independently applies the same randomized mechanism to their datum, the noisy observations $(Z^i)_{i\in[n]}$ are also i.i.d. random variables with distribution $\vtheta\mech$ over $\mathcal{Z}$. We rely on this induced i.i.d. property in establishing the consistency results of Section~\ref{sec:conv_real}.

Finally, we present specific properties of the local model that are important for our consistency results. 
\begin{definition}[\!\!\cite{Whitney:94:BOOK}]\label{def:id_cont_dom} 
For a distribution $\vphi\in\calc$, let $\Pr[z\mid\vphi]$ denote the probability
that the mechanism outputs $z\in\calz$ when its input is drawn from $\calx$ according to
$\vphi$. Let $\Pr[\cdot\mid\vphi]$ denote the induced distribution on
$\calz$. For a true distribution $\vtheta\in\calc$, let
$Z\sim\Pr[\cdot\mid\vtheta]$. The mechanism is said to satisfy:
\begin{enumerate}[label=(\roman*)]
\item\label{def:id_cont_dom:id} \textbf{Identification}
 if the mapping $\vphi\mapsto\Pr[\cdot\mid\vphi]$ is injective; that is,
$\Pr[\cdot\mid\vphi]\neq\Pr[\cdot\mid\vtheta]$ whenever
$\vphi\neq\vtheta$.

\item\label{def:id_cont_dom:cont} \textbf{Continuity}
if for every $\vtheta\in\calc$, the function
$\vphi\mapsto\log\Pr[Z\mid\vphi]$ is continuous at every
$\vphi\in\calc$ with probability one.

\item\label{def:id_cont_dom:dom} \textbf{Dominance:}
if for every $\vtheta\in\calc$
\[
\E_{\vtheta}\!\left[
\sup_{\vphi\in\calc}
\left|
\log\Pr[Z\mid\vphi]
\right|
\right]
<\infty.
\]
\end{enumerate}
\end{definition}

\begin{table}[t]\caption{Basic notations} 
\begin{tabular}{r p{0.35 \textwidth}}
\toprule
$\calx$ & the alphabet (possible values) of an attribute. \\[0.3ex]
$\calc$ & the set of all probability distributions over $\calx$. \\[0.3ex]
$\vtheta, \vphi$ & probability distributions over $\calx$. \\[0.3ex]
$\vtheta^t$ & the distribution estimated by \ibu{} at iteration $t$.\\[0.3ex]
$\theta_x$ &the $x$-th component of the distribution $\vtheta$. \\[0.3ex]
$X^i$ & the original datum (ranging on $\calx$) of user $i$.\\[0.3ex]
$\calz$ &  alphabet of the noisy data points.\\[0.3ex] 
$Z^i$ & the noisy data point (ranging on $\calz$) from user $i$ \\[0.3ex]
$z^i$ & the observed value of $Z^i$. \\[0.3ex] 
$\mech: \calx \to \calz$ & an obfuscation mechanism.   \\[0.3ex]
$\mech_{x z}$ & probability that $\mech$ yields  $z\in\calz$ from  $x\in\calx$. \\[0.3ex]
$\mech_{x *}$ & the row of $\mech$ corresponding to $x\in\calx$.  \\[0.3ex]
$\mech_{* z}$ & the column of $\mech$ corresponding to $z\in\calz$. \\[0.3ex]
$\vq$ & a noisy distribution on $\calz$. \\[0.3ex]
$q_z$ & the $z$-th component of the distribution $\vq$. \\[0.3ex]
$L(\vphi)$ & the log-likelihood of  $\vphi$ w.r.t. the observed 
                        data.\\[0.3ex]
$\mg$     & the observations probability matrix \\[0.3ex]
$G_{xi}$ & the probability of reporting $z^i$ when the original datum of user $i$ is $x$.\\ 
\bottomrule
\end{tabular}
\label{tab:emnotations}
\end{table}

\subsection{The log-likelihood in the local model}
In the local model, the log-likelihood function is defined over $\calc$, 
the set of all probability distributions on $\calx$. Let $(Z^i)_{i\in[n]}$ be random variables representing 
the noisy data reported from the $n$ users, and $(z^i)_{i\in[n]}$ be the observed values of these variables. 
Given the observed data, the log-likelihood of a distribution $\vphi\in\calc$ is 
\begin{equation}\label{eq:L}
L(\vphi) = \log \Pr[\land_{i\in[n]} Z^i = z^i | \vphi] \quad\quad \forall \vphi\in\calc. 
\end{equation}
For our analysis, we will express $L(\cdot)$ using the observation probability matrix $\mg$, as follows.
\begin{definition}[Observation probability matrix, \cite{ehab:eurosp:2020}]\label{def:probability_matrix}
Let $(z^i)_{i \in [n]}$ be the observations produced by a mechanism $\mech$. 
The observation probability matrix $\mg$ assigns to each 
$x \in \calx$ and $i \in [n]$ the conditional probability $G_{x i}$ of observing 
$z^i$ when the original value of user $i$ is $x$, that is,  
$
G_{x i} = \Pr[Z^i = z^i | X^i = x] = \mech_{x z^i}.   
$
\end{definition}
As decribed in \cite{ehab:eurosp:2020}, by \eqref{eq:L} and the independence of $(Z^i)_{i \in [n]}$, we can write $L(\vphi)$  
in terms of $\mg$ as
\begin{equation}\label{eq:L_ibu}
L(\vphi) = \sum_{i=1}^n L_i(\vphi), \quad\text{where}\quad L_i(\vphi) = \log \sum_{x \in \calx} \phi_x \, \mg_{x i}.
\end{equation}
\subsection{The Iterative Bayesian Update }\label{sec:ibu}
The iterative Bayesian update (\ibu{}) 
is an iterative method proposed in \cite{Agrawal:01:PDS} to solve the problem presented in Section \ref{sec:local_privacy_model}, namely, to estimate the original distribution 
$\vtheta$ over the alphabet $\calx$ from the reported noisy data. 

Let $\mech$ be the mechanism that generated the noisy data. The method starts from a full-support 
distribution $\vtheta^0$ on $\calx$ (e.g., the uniform distribution), 
and iteratively updates it. At iteration $t+1$, 
the estimate $\vtheta^{t+1}$ is computed from the preceding one $\vtheta^t$, 
$\vq$, and the mechanism matrix $\mech$.  
Throughout the iterations, the log-likelihood $L(\vtheta^t)$ increases 
monotonically and converges to an MLE in the limit, as proved in \cite{ehab:eurosp:2020}.
Starting with a full support distribution is important for this convergence:
if $\theta^0_x=0$, then $\theta^t_x$ remains $0$ in all subsequent iterations, 
regardless of the correct $x$-th component of the MLE.

Algorithm \ref{alg:ibu}  presents a practical implementation of \ibu{}, 
which terminates when the increase in the log-likelihood between 
successive iterations falls below a threshold $\delta$, and returns the $\theta^t$ computed at that point.
Throughout the paper,  we distinguish between the ideal IBU (infinite) iterative process and its practical implementation. Our consistency results in Section \ref{sec:conv_real}  concern the ideal 
\ibu{}. On the other hand, \autoref{alg:ibu} terminates after finitely many iterations according to the threshold $\delta$, and therefore returns an approximation of the MLE, that converges to the exact estimate as $\delta$ decreases. In the rest of the paper, we write simply \ibu{} to refer to the ideal process, and ``the \ibu{} algorithm'' when we refer to its practical implementation. 

\begin{algorithm}[t]
\KwData{ 
$\mech$, $\vq=(q_z: z\in \calz)$, and $\delta>0$ \;
}
\KwResult{An MLE for the real distribution $\vtheta$ .}
Set $t\gets0$ and $\vtheta^0\gets$ any distribution over $\calx$ with $\theta^0_x>0$ for every $x \in \calx$\; 
\Repeat{ $| L(\vtheta^t)  - L(\vtheta^{t-1}) | < \delta$ }{
$\theta^{t+1}_x \gets \sum_{z\in \calz} q_z\, \frac {\theta^t_x \mech_{x z}}{\sum_{u \in \calx} \theta^t_u \mech_{u z}} \quad \forall x\in \calx$\;
$t \gets t+1$ \;
}
\Return $\vtheta^t$
\vspace{3mm}
\caption{The \ibu{} algorithm 
}
\label{alg:ibu}
\end{algorithm}
\subsection{Matrix Inversion estimators}  
The matrix inversion (\inv), proposed in \cite{Kairouz:16:ICML}, refers to a family of generic estimators that work on noisy data from any mechanism having an invertible matrix. 
Suppose that a mechanism $\mech:\calx \to \calz$ is applied to the original data 
of all users to produce the noisy distribution $\vq$ over $\calz$. 
Then from $\vq$ we can obtain the following real-valued vector $\vw \in \reals^{|\calx|}$.
\[
\vw=\vq \, \mech^{-1}
\]
where $\mech^{-1}$ is the inverse of $\mech$. 
Note that in general $\vw$ may not be a distribution when some of its components are negative. 
To obtain a valid distribution on $\calx$ from $\vw$, two methods have been proposed in \cite{Kairouz:16:ICML}, yielding 
two variants of \inv{}:  \\[-1ex]

\begin{quote}
{\bf Normalization (\invn{}):} setting the negative components of $\vw$ to $0$ and then normalizing so that the sum of the components is $1$.
\end{quote}

\begin{quote}
{\bf Projection (\invp{}):} projecting $\vw$ onto the set of distributions in $\reals^{|\calx|}$, 
i.e., by selecting from this set the distribution that is nearest to $\vw$ with respect to the Euclidean distance, for example, using the algorithm in 
\cite{Wang2013ProjectionOT}. 
\end{quote}

\subsection{The \textsc{Rappor} protocol}\label{sec:rappor}
{\bf \textsc{Rappor} mechanism} 
is built on the idea of randomized response~\cite{Warner:65:jastat} 
to enable data collection from end-users with LDP guarantees~\cite{Erlingsson:14:CCS}. 
In its basic form, known as Basic (One-Time) \textsc{Rappor}, the mechanism 
encodes the user's private value $x \in \calx$ into a bit vector 
$b \in \{0,1\}^{|\calx|}$, where only one bit $b_x$ corresponding to $x$ 
is set to $1$, and all other bits are set to $0$. 
Each bit $b_y$ is then independently obfuscated, retaining its value with probability 
$p = e^{\leps/2}/(1+e^{\leps/2})$ and flipping to $1-b_y$ otherwise. 
The resulting noisy bit vector, $\beta$, is finally reported to the server.

{\bf \textsc{Rappor}'s estimator} 
is a specialized estimator that works on the noisy bit vectors produced by the \textsc{Rappor} mechanism. 
Recall that the $x$-th bit of each reported vector privately indicates whether or not the corresponding user's value is $x$. Accordingly, the estimator first constructs the histogram of the $x$-th bit over $\{0,1\}$ across all users, and then applies one of the \inv{} variants to estimate the probability of $x$. Repeating this process for every $x\in\calx$ yields an estimate of the full distribution over $\calx$.
Further details on \textsc{Rappor}'s estimator can be found in \cite{Erlingsson:14:CCS,Kairouz:16:ICML}. 
\section{Roadmap of the main theoretical results}\label{sec:properties} 
The main theoretical properties analyzed in our paper fall into two categories: \emph{empirical}, 
meaning that they depend on the noisy data $(z^i)_{i\in[n]}$ \emph{and} on the mechanism $\mech$, 
and \emph{mechanism-level}, meaning that they depend solely on the mechanism. 
As an example, because the log-likelihood function $L(\cdot)$ is determined by both the noisy data 
and the mechanism, the strict concavity of $L(\cdot)$ is an empirical property. On the other hand, 
as we will see, the consistency of \ibu{} is a mechanism-level property.

\autoref{thm:strictconcavity} characterizes the strict concavity of $L(\cdot)$, and  
\autoref{cor:strictlyConcave_one_mech} provides a sufficient condition for $L(\cdot)$ to satisfy it. Corollaries 
\ref{cor:strictlyConcave-krr}, \ref{cor:strictlyConcaveLinearGeometric}, \ref{cor:strictlyConcaveTLinearGeometric},
\ref{cor:strictlyConcaveTaxicab},
and \ref{cor:strictlyConcaveTTaxicab} 
apply \autoref{cor:strictlyConcave_one_mech} to a variety of mechanisms. 
\autoref{prop:uniqueness} provides sufficient conditions for the uniqueness of the MLE.
\autoref{thm:mlconsistencygen} (from \cite{Whitney:94:BOOK}) provides sufficient conditions for the consistency of \ibu{}, 
and \autoref{thm:identification} characterises one of these conditions (identification). 
Finally, Corollaries \ref{cor:consistency1}, \ref{cor:consistency2} and \ref{cor:consistency3} apply \autoref{thm:mlconsistencygen} to a variety of mechanisms, thereby establishing the consistency of \ibu{} for each of them.

Figure~\ref{fig:summary_properties} summarizes the logical relationships between the 
various properties, thereby providing a roadmap for the proof of \ibu{}'s consistency.
In this figure, the empirical properties are represented by ovals (rounded rectangles), 
and the mechanism-level ones by rectangles.

\begin{figure}[!ht]
\centering
\begin{tikzpicture}[auto, node distance=0.4cm, >=latex', font=\footnotesize]
\tikzset{
  block/.style = {draw, fill=white, rectangle, minimum height=2.5em, text width=3.3cm, align=center, inner sep=4pt},
  longblock/.style = {draw, fill=white, rectangle, minimum height=2.0em, text width=5cm, align=center, inner sep=4pt}
}
\node[block, thick, rounded corners=10pt] (properness) {
For $\mech$ and noisy data $(z^i)_{i\in[n]}$, $L(\vphi)>-\infty$ for some $\vphi\in\calc$
}; 
\node[block, thick, rounded corners=10pt,
below=0.3cm of properness
] (strictconcavity1) {
$L(\cdot)$ is strictly concave for $\mech{}$ and noisy data $(z^i)_{i\in[n]}$
};
\node[block, thick, below=0.3cm of strictconcavity1] (strictconcavity2) {
There exist noisy data $(z^i)_{i\in[n]}$ such that $L(\cdot)$ is strictly concave
}; 
\node[block, thick, below=0.8cm of strictconcavity2] (identification) {
$\mech$ satisfies identification
};
\node[block, thick, 
right=0.8cm of strictconcavity2, yshift=-0.6cm] (linearindependence){
$\mech$ has $|\calx|$ linearly independent columns
};
\node[block, thick, right=0.8cm of identification, yshift=0.0cm] (dominance) {
$\mech$ satisfies  continuity and dominance
};
\draw [decorate, decoration={brace, amplitude=6pt, raise=1pt}, line width=1.0pt]
     (properness.north east) -- (strictconcavity1.south east) coordinate[midway, xshift=0.0cm] (y);

\draw [decorate, decoration={brace, amplitude=6pt, raise=4pt}, line width=1.0pt]
     (dominance.south east) -- (identification.south west) coordinate[midway, xshift=0.2cm] (z);
\node[block, thick, rounded corners=10pt, 
right=0.8cm of y] (uniqueness) {
The MLE is unique for $\mech$ and the noisy data $(z^i)_{i\in[n]}$
}; 
\node[longblock, thick, below=0.8cm of z] (consistency) {
\ibu{} is consistent under $\mech$
};

\draw[-implies, double equal sign distance] ($(y)+(0.3cm,0cm)$)
  -- node[pos=0.8,above,yshift=0.9cm,rotate=90]
{\scriptsize\autoref{prop:uniqueness}} 
++(0.5cm,0cm);
%
\draw[-implies, double equal sign distance] (strictconcavity1) 
-- node[right] { } (strictconcavity2);
%
\draw[implies-implies, double equal sign distance] 
(strictconcavity2.east)++(0,-0.2cm) 
-- node[pos=0, above right, yshift=0.1cm]
{\scriptsize\autoref{thm:identification} (ii-iii)} 
++(0.7cm, 0cm);
%
\draw[implies-implies, double equal sign distance] (strictconcavity2) 
-- node[left] {\scriptsize\autoref{thm:identification} (i-ii)} (identification);

\draw[-implies, double equal sign distance] ($(z)+(-0.2cm,-0.4cm)$)
  -- node[pos=0.5,right]
{\scriptsize\autoref{thm:mlconsistencygen}~\cite{Whitney:94:BOOK}} 
++(0,-0.35cm);

\draw[implies-implies, double equal sign distance] 
(linearindependence.west)++(0,-0.2cm) 
-- 
++(-0.8cm,-0.6cm);

\end{tikzpicture}
\caption{Relations between empirical (ovals) and mechanism-level (rectangles) properties.}
\label{fig:summary_properties}
\end{figure}

\section{Strict concavity of the log-likelihood}
\label{sec:strictconcavity}
\begin{figure}
\centering 
\subfigure[$L(\cdot)$ with $\{\it{2}\}$ ]{
      \label{fig:s_concavity_depends_on_observations1}
      \includegraphics[width=0.24\textwidth]{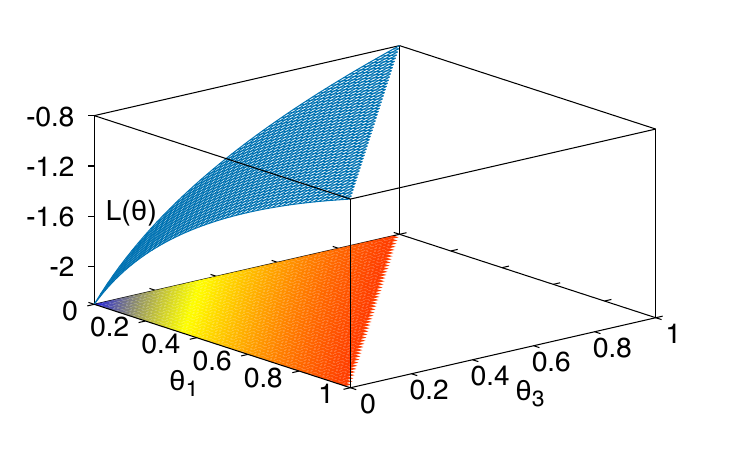}
      }
\hspace{-20pt}
\subfigure[$L(\cdot)$ with $\{\it{2},\it{1}\}$]{
      \label{fig:s_concavity_depends_on_observations2}
      \includegraphics[width=0.24\textwidth]{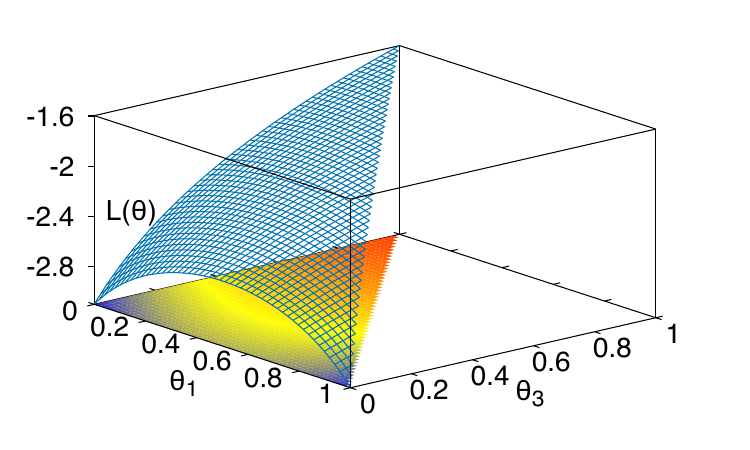}
      }
\caption{The log-likelihood $L(\cdot)$ with the mechanism~\eqref{eq:sconcavity_depends_on_observations} and two sets of observations shown in (a) and (b). $L(\cdot)$ is plotted for every distribution $\vtheta$ on the alphabet $\{\it{1},\it{2},\it{3}\}$ where $\vtheta$ is fully described by its two components $\theta_1, \theta_3$ since the remaining one $\theta_2=1- \theta_1-\theta_3$. }
\label{fig:sconcavity_depends_on_observations}
\end{figure}

While the log-likelihood function $L(\cdot)$ is concave by \eqref{eq:L_ibu}, it is not 
strictly concave in general. In the following, we investigate the strict 
concavity of $L(\cdot)$ as it is necessary for the consistency 
of \ibu{}, as shown in Section~\ref{sec:conv_real}. Moreover, strict concavity 
is closely related to the uniqueness of the MLE. This connection allows us to study the 
relationship between the MLE uniqueness and \ibu{}’s consistency, leading to the conclusions 
presented in Section~\ref{sec:uniqueness}.
To characterize the strict concavity of $L(\cdot)$, we
start by analyzing its value at \emph{convex} combination of two distributions 
$\vtheta, \vphi$.
 \begin{restatable}{lemma}{concavity}
\label{lemma:concavity}
For a given observation probability matrix $\mg$ and two distributions $\vtheta, \vphi \in \calc$, 
it holds for every $0<\gamma<1$ that 
\begin{equation}
\label{eq:concavity}
L(\gamma \vtheta + (1-\gamma) \vphi) \geq \gamma L(\vtheta) +(1-\gamma) \, L(\vphi),
\end{equation}
where the equality holds if and only if $(\vtheta-\vphi)\, \mg = \vzeros$. 
\end{restatable}
By definition (see e.g.~\cite{bertsekas:2003:BOOK}), 
$L(\cdot)$ is strictly concave if the inequality~\eqref{eq:concavity} is \emph{strict} 
for every pair of ditinct distributions $\vtheta, \vphi$. 
Since \autoref{lemma:concavity} characterizes precisely when this condition holds, 
we obtain two equivalent characterizations for the strict concavity of $L(\cdot)$ 
via the following theorem.  
\begin{restatable}[strict concavity]{theorem}{strictconcavity}
\label{thm:strictconcavity}
For a given mechanism and noisy data, let $\mg$ be the corresponding
observation probability matrix. Then the following conditions are equivalent.
\begin{enumerate}[label=(\roman*)]
\item\label{thm_item_1} $L(\cdot)$ is strictly concave for $\mg$ on $\calc$.
\item\label{thm_item_2} All distributions $\vtheta, \vphi$ with $\vtheta\neq\vphi$ satisfies $\vtheta \,\mg \neq \vphi \,\mg$.   
\item\label{thm_item_3}  Every nonzero vector $\vectr{w}\in \reals^{|\calx|}$ with $\sum_{x\in\calx} w_x=0$
satisfies $\vectr{w} \,\mg \neq \vzeros$. 
\end{enumerate}
\end{restatable}
Note that Condition \ref{thm_item_3} is satisfied if $\mg$ has $|\calx|$ linearly independent columns. 
Moreover, by Definition~\ref{def:probability_matrix}, each column of $\mg$ coincides with a column 
of the mechanism $\mech$. 
Therefore, we can relate the strict concavity of $L(\cdot)$ 
to the linear independence of $\mech$'s columns.  
Let $\tilde\calz$ be any set of distinct observations 
contained in the noisy data, and recall that $\mech_{*z}$ denotes 
the column of $\mech$ corresponding to an observable $z\in\calz$. Then we have
\begin{restatable}{corollary}{strictlyConcaveOneMech}
\label{cor:strictlyConcave_one_mech}
Let $\mech$ be a mechanism that works on an alphabet $\calx$.
Suppose that the noisy data contains a set $\tilde\calz$ of 
$|\calx|$ distinct observables such that the mechanism columns $\{\mech_{*z} : z\in \tilde\calz\}$ are linearly independent. 
Then $L(\cdot)$ is strictly concave on $\calc$.  
\end{restatable}
We remark that, in contrast to the claim in \cite{Agrawal:01:PDS}, the log-likelihood function 
is \emph{not} always strictly concave. Whether this property holds depends indeed on both 
the mechanism and the noisy data. As an  example, consider the alphabet 
$\calx = \{\it{1},\it{2},\it{3}\}$ and assume that the data are perturbed by the mechanism 
$\mech:\calx\to\calx$ whose matrix is specified in~\eqref{eq:sconcavity_depends_on_observations}.
Recall that every $\mech_{xz}$ (for $x,z\in\calx$) is the probability of reporting $z$ when the 
real datum is $x$ (See Section \ref{sec:local_privacy_model}).
\begin{equation}\label{eq:sconcavity_depends_on_observations}
    \bbordermatrix{   & \it{1}    & \it{2}    & \it{3}    \cr 
                    \it{1} & 0.10 & 0.45 & 0.45 \cr  
                    \it{2} & 0.45 & 0.10 & 0.45 \cr
                    \it{3} & 0.45 & 0.45 & 0.10 \cr}
\end{equation}
If the set of reported observations is $\{\it{2}\}$, the log-likelihood $L(\cdot)$ 
is not strictly concave as illustrated in Figure~\ref{fig:s_concavity_depends_on_observations1}. 
In particular, all convex combinations of the distributions $(1,0,0), (0,0,1)$ on $\calx$,
have the same value of $L(\cdot)$. 
However, if the reported observations are $\{\it{2},\it{1}\}$, $L(\cdot)$ is strictly concave 
as shown in Figure~\ref{fig:s_concavity_depends_on_observations2}. 
The latter case of $\{\it{2},\it{1}\}$  also shows that the condition in 
\autoref{cor:strictlyConcave_one_mech} is not necessary 
for the strict concavity: $L(\cdot)$ is strictly concave in this case although the 
condition of \autoref{cor:strictlyConcave_one_mech} 
is not satisfied (the distinct observations are only two, i.e., fewer than $|\calx|$). 

\autoref{thm:strictconcavity} and \autoref{cor:strictlyConcave_one_mech} are useful 
to show the strict concavity of $L(\cdot)$ in a variety of settings. This includes 
mechanisms with square matrices, such as $k$-RR~\cite{Kairouz:16:JMLR}, and the 
truncated versions of the geometric~\cite{Ghosh:09:STOC} and taxicab 
(Section~\ref{sec:taxicab}) mechanisms; mechanisms with many more columns than rows, 
such as \textsc{Rappor}~\cite{Erlingsson:14:CCS}; 
and even mechanisms with infinitely many columns, such as the (non-truncated) 
geometric~\cite{Ghosh:09:STOC} and taxicab (Section~\ref{sec:taxicab}) ones.
In the following subsections, we study the strict concavity of $L(\cdot)$ under these mechanisms. 
\subsection{k-RR mechanism}\label{sec:strictlyConcave:krr}
The $k$-RR mechanism was proposed in \cite{Kairouz:16:JMLR} as an extension of the 
randomized response ~\cite{Warner:65:jastat}, 
designed for binary alphabets ($k=2$), 
to arbitrary $k$-size alphabets.  
The definition of the probability that $k$-RR obfuscates $x\in\calx$ into $z\in\calx$, 
as specified in \cite{Kairouz:16:JMLR}, is: 

\begin{equation}\label{eq:krr}
\Pr[z|x] = \frac{1}{k-1+ e^\leps} 
\left\{ 
\begin{array}{l l}
e^\leps & \mbox{if $z=x$}\\
1                & \mbox{if $z\neq x$} 
\end{array} 
\right.
\end{equation} 
where the parameter $\leps>0$ determines the privacy level guaranteed by the mechanism. 
The $k$-RR satisfies $\leps$-LDP~\cite{Duchi:13:FOCS,Kairouz:16:JMLR}.  
That is, for any observation $z$, the ratio $\Pr[z |x]/\Pr[z |x']$, describing the distinguishability 
between every two values $x,x' \in \calx$, is at most $e^{\leps}$.

The matrix of the $k$-RR mechanism is square and has linearly independent columns 
(see \autoref{lem:krrLinIndependence} in Appendix \ref{sec:proofs_sconcavity}). 
So, if the noisy data include all elements of the alphabet, the condition of~\autoref{cor:strictlyConcave_one_mech} 
is satisfied, causing the log-likelihood $L(\cdot)$ to be strictly concave in this case. 
\begin{restatable}{corollary}{strictlyConcavekrr}
\label{cor:strictlyConcave-krr}
Consider a finite alphabet $\calx$, and a $k$-RR mechanism $\mech:\calx \to \calx$. 
Then $L(\cdot)$ is strictly concave for any noisy data that contains all the elements of $\calx$.
\end{restatable}
\subsection{linear geometric mechanisms}\label{sec:geometric}
In many applications, the users' data are discrete points in one-dimensional space, 
e.g., ages, counts, meter readings, etc. This space may be viewed as the set of 
integers $\integers$ equipped with a \emph{spacing} $s>0$ that defines the distance 
between every two successive points. The value and measurement unit of $s$ depend on the application.
In this situation, the alphabet $\calx$ of users' data is a subset of $\integers$, and the distance 
between any $x,x' \in \calx$ is $\dist(x,x') = s |x-x'|$.
As formulated in \cite{Ghosh:09:STOC}, the linear geometric mechanism with 
a privacy parameter $\lambda >0$, when applied to 
an element $x\in\calx$, produces an integer $z\in\integers$ with probability 
\begin{equation}\label{eq:linearGeometric}
\Pr[z |x] = a \, e^{- \lambda \, s\, |z-x|} \quad \mbox{where}\; \;   
a=\frac{1-e^{-\lambda \, s}}{1+e^{-\lambda \, s}}.
\end{equation}
This mechanism satisfies $(\lambda\dist)$-privacy on $\calx$, 
i.e., \emph{metric privacy} \cite{Chatzikokolakis:13:PETS} 
with the metric ($\lambda\dist$).
That is, for any observable $z$, the ratio $\Pr[z |x]/\Pr[z |x']$, 
describing the distinguishability between $x,x' \in \calx$,  
is at most $e^{ \lambda\dist(x,x') }$. 

Now, for a finite $\calx$, we can show that the mechanism columns corresponding 
to the elements of $\calx$ are linearly independent 
(see \autoref{lem:linearGeometricLinIndependence} in Appendix~\ref{sec:proofs_sconcavity}). 
From this and \autoref{cor:strictlyConcave_one_mech}, we obtain 

\begin{restatable}{corollary}{strictlyConcaveLinearGeometric}
\label{cor:strictlyConcaveLinearGeometric}
Consider a finite alphabet $\calx\subset\integers$, and a linear geometric mechanism $\mech:\calx \to \integers$.  
Then $L(\cdot)$ is strictly concave for any noisy data that contains all the elements of $\calx$. 
\end{restatable} 

A variant of the linear geometric mechanism is its \emph{truncated} version, 
which restricts the outputs to a finite range of integers. 
Let $z_1, z_2$ be two integers with $z_1< z_2$. 
As formulated in ~\cite{Ghosh:09:STOC}, the mechanism maps every $x\in \calx$ 
to an integer $z$ in the range from $z_1$ to $z_2$, with the probability
\begin{align}
P(z |x) &= a(z) \, e^{-\lambda \, s\,  |z-x|}, \label{eq:TLinearGeometric}\\
a(z) &= \frac{1}{1+e^{-\lambda\, s}} 
\left\{ 
 \begin{array}{l l}
1,            &z \in \{z_1, z_2\}, \\
1-e^{-\lambda\, s}, &z_1< z< z_2, \\
0,            & \text{otherwise.}
\end{array}\nonumber
\right.
\end{align}
Also in this case, the mechanism satisfies $(\lambda\dist)$-privacy on $\calx$. 
Moreover, if $\calx$ is contained in the mechanism's output range (i.e., from $z_1$ to $z_2$), 
the mechanism columns corresponding to the elements of $\calx$ are linearly independent 
(see \autoref{lem:TLinearGeometricLinIndependence} in Appendix \ref{sec:proofs_sconcavity}).  
From this fact and \autoref{cor:strictlyConcave_one_mech}, we obtain the following corollary.  
\begin{restatable}{corollary}{strictlyConcaveTLinearGeometric}
\label{cor:strictlyConcaveTLinearGeometric}
Let $\calz$ be a finite range of integers. Consider an alphabet $\calx\subseteq\calz$,  
and a truncated linear geometric mechanism $\mech:\calx \to \calz$.  
Then $L(\cdot)$ is strictly concave for any noisy data that contains all the elements of $\calx$. 
\end{restatable} 
\subsection{The taxicab mechanism}\label{sec:taxicab}
Similar to the linear geometric mechanism defined on a one-dimensional alphabet represented 
as a subset of $\integers$, we introduce a mechanism operating on a two-dimensional (planar) alphabet 
$\calx \subseteq \integers^2$. Each element of this alphabet is a pair of integers $(u,v)$ 
representing its horizontal and vertical coordinates. Let $s$ denote the \emph{spacing} between 
successive horizontal or vertical coordinates. The taxicab mechanism with parameter $\lambda>0$ 
maps each $(u,v)\in\calx$ to an output $(\mu,\nu)\in\integers^2$ with probability
\begin{align}
\Pr[ (\mu, \nu) | (u , v) ] &= a^2 \, e^{-\lambda \, s\, ( | \mu- u | + | \nu - v | )}, 
\label{eq:taxicab} \\
\mbox{where} \quad  &a=\frac{1-e^{-\lambda \, s}}{1+e^{-\lambda \, s}} \nonumber.
\end{align}
Note that this planar mechanism can be viewed as the composition of two linear geometric mechanisms~\eqref{eq:linearGeometric} independently obfuscating the two coordinates $u,v$. 
We also introduce the \emph{truncated taxicab mechanism}, 
whose output $(\mu,\nu)$ satisfies $\mu_1 \leq \mu \leq \mu_2$ and $\nu_1 \leq \nu \leq \nu_2$ 
for fixed integers $\mu_1, \mu_2, \nu_1, \nu_2$. 
Precisely, this mechanism maps every $(u , v) \in \calx$ to an output $(\mu , \nu)\in \integers^2$ 
with probability
\begin{align}
&\Pr[(\mu,\nu)\mid(u,v)] = a(\mu)\,b(\nu)\,e^{-\lambda s\,(|\mu-u| + |\nu-v|)}, 
\label{eq:truncatedTaxicab}\\[4pt]
&\quad a(\mu) = g(\mu; \mu_1,\mu_2), \qquad b(\nu) = g(\nu; \nu_1,\nu_2), \nonumber\\[4pt]
&g(z; z_1, z_2) = \frac{1}{1+e^{-\lambda s}}
\begin{cases}
1, & z \in \{z_1, z_2\},\\
1-e^{-\lambda s}, & z_1 < z < z_2,\\
0, & \text{otherwise}.
\end{cases} \nonumber
\end{align}
The mechanism~\eqref{eq:truncatedTaxicab} is a composition of 
two truncated linear geometric mechanisms~\eqref{eq:TLinearGeometric}
obfuscating the integer coordinates $u,v$ independently. 
It can be shown that both the mechanisms \eqref{eq:taxicab} and \eqref{eq:truncatedTaxicab} satisfy $(\sqrt{2} \lambda)$-geo-indistinguishability~\cite{Andres:13:CCS}, namely, 
$(\sqrt{2}\, \lambda \dist)$-privacy, where $\dist$ is the Euclidean distance   on $\calx$, defined as 
$\dist((u,v), (u',v')) = s \sqrt{(u'-u)^2 + (v'-v)^2}$. 

\begin{restatable}{proposition}{geoindTaxicab}\label{prop:geoindTaxicab}
The truncated and non-truncated taxicab mechanisms with parameter $\lambda$   
satisfy $(\sqrt{2}\, \lambda)$-geo-indistinguishability. 
\end{restatable} 

For the (non-truncated) taxicab mechanism, the columns corresponding to the 
elements of $\calx$ are linearly independent 
(see~\autoref{lem:taxicabLinIndependence} in Appendix \ref{sec:proofs_sconcavity}). 
This fact, together with \autoref{cor:strictlyConcave_one_mech}, 
implies the following result. 
\begin{restatable}{corollary}{strictlyConcaveTaxicab}
\label{cor:strictlyConcaveTaxicab}
Consider a finite alphabet $\calx\subset\integers^2$, and a taxicab mechanism $\mech:\calx \to \integers^2$.  
Then $L(\cdot)$ is strictly concave for any noisy data that contains all the elements of $\calx$. 
\end{restatable} 
For the truncated taxicab mechanism, outputs are confined to a finite region in $\integers^2$, 
defined by two finite ranges $\calz_1, \calz_2$, that bound the horizontal and vertical coordinates, 
respectively. That is, every output $(\mu,\nu)$ of the mechanism lies in $ \calz_1\times\calz_2$. If this set contains the alphabet $\calx$, then the mechanism columns corresponding to the elements of $\calx$ are linearly independent 
(see~\autoref{lem:TTaxicabLinIndependence} in Appendix \ref{sec:proofs_sconcavity}). 
Combined with \autoref{cor:strictlyConcave_one_mech}, this yields the following.
\begin{restatable}{corollary}{strictlyConcaveTTaxicab}
\label{cor:strictlyConcaveTTaxicab}
Let $\calz_1, \calz_2$ be two finite ranges of integers. Consider an alphabet $\calx\subseteq(\calz_1\times\calz_2)$,  
and a truncated taxicab mechanism $\mech:\calx \to (\calz_1\times\calz_2)$.  
Then $L(\cdot)$ is strictly concave for any noisy data that contains all the elements of $\calx$. 
\end{restatable} 

We call the mechanisms~\eqref{eq:taxicab} and \eqref{eq:truncatedTaxicab} `taxicab' because the obfuscation 
probability is determined by the quantity $s (| \mu- u | + | \nu - v |)$, which is the \emph{taxicab} (also called \emph{Manhattan})
distance between $( u, v)$ and $(\mu , \nu)$. These mechanisms are therefore different from the planar geometric 
mechanism \cite{Chatzikokolakis:17:POPETS} and its truncated version, where the obfuscation probability is determined instead 
by the \emph{Euclidean} distance between $( u, v)$ and $(\mu , \nu)$ 
(See also Appendix \ref{sec:plan_geom}).
%
\subsection{Basic \textsc{Rappor} mechanism}\label{sec:strict_concavity:rappor}
\begin{figure}
\centering
\includegraphics[width=0.3\textwidth]{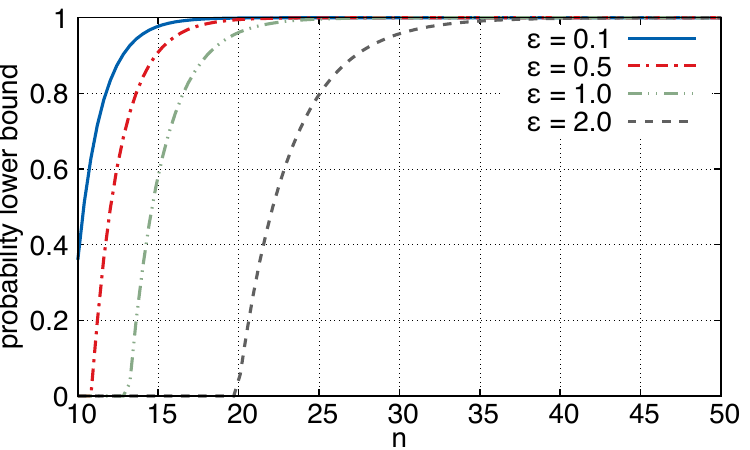}
\caption{A lower bound for the probability that the log-likelihood function of \textsc{Rappor} is strictly concave
given any set of $n$ original data drawn from the alphabet $\calx$ where $|\calx|=10$.} 
\label{fig:probability_of_s_concavity_rappor}
\end{figure}

To describe the strict concavity of the log-likelihood $L(\cdot)$ under the basic \textsc{Rappor} (see Section~\ref{sec:rappor}),
we first evaluate the probability $\Pr[\beta | X=x]$ of producing a noisy bit vector $\beta$ when the original user's datum is $x$. 
\begin{restatable}{lemma}{probrappor} 
\label{lemma:probrappor}
Consider a basic \textsc{Rappor} mechanism that works on alphabet $\calx$ and has parameter $\leps>0$.
Let $p = e^{\leps/2}/(1+e^{\leps/2})$. For every bit vector $\beta\in\{0,1\}^{|\calx|}$ and $y\in\calx$, 
let $\beta_y$ be the value of the $y$-th bit and $S(\beta)= \sum_{y\in\calx} \beta_y$. Then 
\[
\Pr[\beta | X=x] = p^{|\calx|} e^{-(\frac{1}{2} + \frac{1}{2} S(\beta) - \beta_x ) \leps}.
\]
\end{restatable}
Suppose that the noisy data consists of $n$ bit vectors $\{\beta^i : i \in[n]\}$. 
Then the observation probability matrix $\mg$ has $|\calx|$ rows and $n$ columns   
with $\mg_{x i}=\Pr[\beta^i | X=x]$. Notice from \autoref{lemma:probrappor} that the entries of every column in the matrix $\mg$ are restricted to two distinct 
values, corresponding to the cases $\beta_x=0$ and $\beta_x=1$. 
This allows us to construct a binary matrix $B$ from $\mg$ and translate 
the strict concavity conditions (stated by \autoref{thm:strictconcavity}) on 
$\mg$ into equivalent conditions on $B$. 
Satisfying these conditions depends on the rank of $B$ in $\reals$, which is related to 
its rank in the \emph{binary field}
\footnote{The binary field $\binfield$ is defined on the symbols $\{0,1\}$, 
with the \emph{addition} and \emph{multiplication} defined as the XOR and AND respectively.}
$\binfield$ by the following lemma.   
\begin{restatable}{lemma}{binmatrix}
\label{lemma:binarymatrix}
Let $B$ be a binary matrix, i.e., having entries in $\{0,1\}$. 
Then $B$ has full rank in $\reals$ if it has full rank in $\binfield$.  
\end{restatable}
Now, from Condition \ref{thm_item_3} of \autoref{thm:strictconcavity} together with \autoref{lemma:probrappor} and \autoref{lemma:binarymatrix},   
the following theorem follows. 
\begin{restatable}{proposition}{strictlyConcaverappor}
\label{prop:strictlyConcaverappor}
Consider a basic \textsc{Rappor} mechanism that works on an alphabet $\calx$ and has parameter $\leps>0$. 
Let $p = e^{\leps/2}/(1+e^{\leps/2})$. Suppose that a set of $n \geq |\calx|$ original data points 
are obfuscated by the mechanism. 
Then $L(\cdot)$ is strictly concave with probability at least $\prod_{k=1}^{|\calx|} \max\{0, (1- p^n\, 2^{k-1})\}$ . 
\end{restatable}
The above theorem describes a lower bound for the probability that $L(\cdot)$ is strictly concave after obfuscating 
the original data of $n$ users. Figure \ref{fig:probability_of_s_concavity_rappor} plots this lower bound 
for $|\calx|=10$, different numbers $n$ of users, and different values of $\leps$. Notice that this bound approaches 
probability one after relatively few observations. For instance, with $\leps=1$ the log-likelihood of the basic \textsc{Rappor} 
is strictly concave on $\calc$ with probability almost \emph{one} when $n\geq25$.  

\section{Uniqueness of the MLE}\label{sec:uniqueness}
\begin{figure}[t]
\centering 
\subfigure[$L(\cdot)$ with $\{\it{2},\it{2},\it{2},\it{2}\}$ ]{
      \label{fig:mle_unique_likelihood_not_s_concave_1}
      \includegraphics[width=0.24\textwidth]{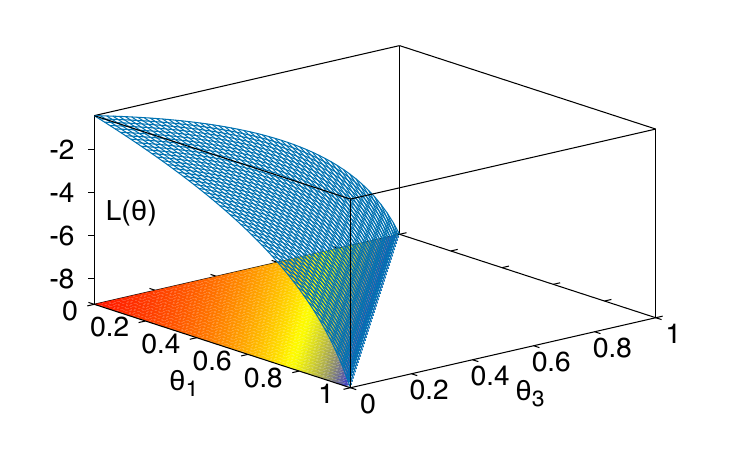}
      }
\hspace{-20pt}
\subfigure[$L(\cdot)$ with $\{\it{2},\it{2},\it{2},\it{2},\it{1},\it{3}\}$]{
      \label{fig:mle_unique_likelihood_not_s_concave_2}
      \includegraphics[width=0.24\textwidth]{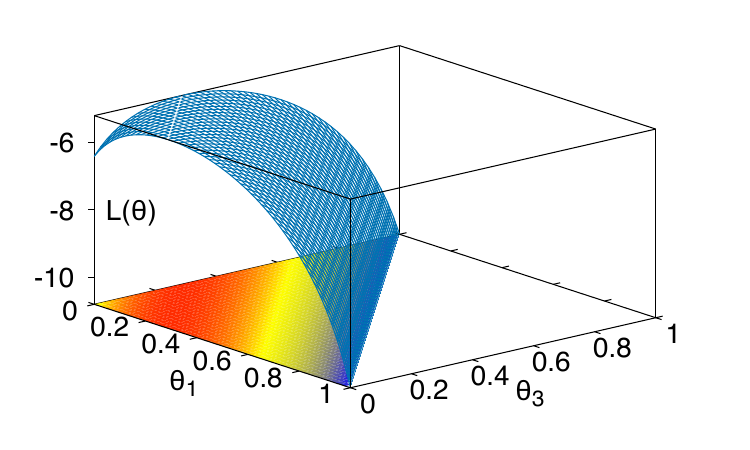}
      }
\caption{The log-likelihood $L(\cdot)$ with the 
mechanism~\eqref{eq:mle_unique_likelihood_not_s_concave} and two sets of observations shown in (a) and (b). 
$L(\cdot)$ is plotted for every distribution $\vtheta$ on the alphabet $\{\it{1},\it{2},\it{3}\}$ 
where $\vtheta$ is fully described by its two components $\theta_1, \theta_3$ since the remaining 
one $\theta_2=1- \theta_1-\theta_3$.
}
\label{fig:uniqueness_vs_sconcavity}
\end{figure}

The uniqueness of the MLE computed by \ibu{} is determined by
the log-likelihood function $L(\cdot)$, which, as shown earlier, 
depends on the underlying privacy mechanism and the noisy data. 
Precisely, this property is linked to the strict concavity of $L(\cdot)$ as follows. 
\begin{restatable}{proposition}{mleunique}
\label{prop:uniqueness}
Suppose that 
$L(\vphi)>-\infty$ for some $\vphi\in\calc$ and 
$L(\cdot)$ is strictly concave on $\calc$. Then the MLE is unique. 
\end{restatable}
While the strict concavity of $L(\cdot)$ implies MLE uniqueness, as stated by 
\autoref{prop:uniqueness}, these two properties are \emph{not equivalent}. 
This is important because, unlike strict concavity that guarantees the MLE's convergence 
to the real distribution (given other conditions described later in Section \ref{sec:conv_real}), 
the MLE's uniqueness does not provide that guarantee.   
To show the nonequivalence between those two properties, 
consider the alphabet $\calx=\{\it{1},\it{2},\it{3}\}$ 
and the following mechanism that maps $\calx$ to itself.
\begin{equation}\label{eq:mle_unique_likelihood_not_s_concave}
    \bbordermatrix{   & \it{1}    & \it{2}    & \it{3}    \cr 
                    \it{1} & 0.45 & 0.10 & 0.45 \cr  
                    \it{2} & 0.05 & 0.90 & 0.05 \cr
                    \it{3} & 0.45 & 0.10 & 0.45 \cr}
\end{equation}
Figure~\ref{fig:mle_unique_likelihood_not_s_concave_1} shows the 
plot of $L(\cdot)$ when the observations are $\{\it{2},\it{2},\it{2},\it{2}\}$. 
Notice that the MLE in this case, $(0,1,0)$, is unique while $L(\cdot)$ 
is \emph{not strictly concave}. 
For instance, all convex combinations of the distributions $(1,0,0)$, $(0,0,1)$
yield the same value of $L(\cdot)$.  
If the above mechanism reports $\it{1}$ and $\it{3}$ in addition to the previously observed data,  
$L(\cdot)$ then turns to have many MLEs, as in Figure~\ref{fig:mle_unique_likelihood_not_s_concave_2}, 
and the MLE never converges to the true distribution, regardless of the amount of noisy data. 
This shows an important fact regarding the consistency of \ibu{}, which computes the MLE: 
uniqueness of the MLE for \emph{some} noisy distribution $\vq$, as in 
Fig.~\ref{fig:mle_unique_likelihood_not_s_concave_1}, does not guarantee \ibu{}'s consistency. 
Alternatively, one might conjecture that \ibu{} is consistent if the MLE is unique for \emph{every} 
noisy distribution.  
Regardless of whether this conjecture holds, it is of limited practical value, as it
excludes some mechanisms that nonetheless allow \ibu{}'s consistency, 
such as the one described by~\eqref{eq:sconcavity_depends_on_observations}.
Thus, the MLE uniqueness alone is unreliable to show \ibu{}'s consistency. 
%

\section{Consistency of \ibu{}}\label{sec:conv_real}
We describe the conditions under which the MLE, which is the limit of the ideal \ibu{} iterative process, 
is consistent, and we analyze them for the mechanisms described in Section~\ref{sec:strictconcavity}.
As explained in Section~\ref{sec:local_privacy_model}, the 
random variables $(Z^i)_{i\in [n]}$, representing the reported noisy observations, are i.i.d. 
Let $\hat\vtheta^n$ be the MLE evaluated from these observations. Then, consistency means that 
$\hat\vtheta^n$ converges to the true distribution as $n \to \infty$.
From the general theory of the MLE \cite{Whitney:94:BOOK}, we can state sufficient conditions for the above convergence. 
\begin{restatable}{theorem}{mlconsistencygen}[consistency of the MLE~\cite{Whitney:94:BOOK}]
\label{thm:mlconsistencygen}
Let $(Z^i)_{i \in [n]}$ be i.i.d. 
random variables having the probability distribution $\Pr[\cdot | \vtheta]$. Suppose that $\vtheta \in \calc$, which is compact. Suppose also that the identification, continuity, and dominance properties 
(\autoref{def:id_cont_dom}) are satisfied. Then $\hat\vtheta^n \to \vtheta$ in probability. 
\end{restatable}
The compactness condition is satisfied because the domain of distributions $\calc$ 
on the alphabet $\calx$ is compact. In the following, we address the remaining conditions of
\autoref{thm:mlconsistencygen}.

\subsection{The identification condition}\label{sec:identification}
The identification is not only part of the set of 
sufficient conditions in \autoref{thm:mlconsistencygen}; it is also necessary, 
as shown in \cite{Whitney:94:BOOK}.
To inspect this condition, 
consider a set of observations and let $\mg$ be their probability matrix 
(see \autoref{def:probability_matrix}). Then the probabilities of these observations 
are given by the vector $\vtheta\, \mg$, where $\vtheta$ is the true distribution 
on $\calx$. The identification (\autoref{def:id_cont_dom}\ref{def:id_cont_dom:id}) 
then means that $\vtheta\, \mg \neq \vphi\, \mg$ for 
some set of observations and all $\vphi \neq \vtheta$. 
By \autoref{thm:strictconcavity}, this is equivalent to the 
strict concavity of $L(\cdot)$ for some observations.  
It is also equivalent to the linear independence of the mechanism's columns as follows. 
\begin{restatable}[Identification]{theorem}{identification}
\label{thm:identification}
Given a mechanism $\mech$ with input alphabet $\calx$, the following conditions are equivalent. 
\begin{enumerate}[label=(\roman*)]
\item\label{prop_id_item_1} $\mech$ satisfies  identification.
\item\label{prop_id_item_2} $L(\cdot)$ is strictly concave for some set of observables.
\item\label{prop_id_item_3} $\mech$ has $|\calx|$ linearly independent columns. 
\end{enumerate}
\end{restatable}

\subsection{The continuity condition}
\label{sec:continuity}
The continuity condition (\autoref{def:id_cont_dom}\ref{def:id_cont_dom:cont}) 
refers to the function $\vphi\mapsto\log\Pr[Z\mid\vphi]$, where $Z$ is a random noisy output. 
It requires that this random function be continuous at every $\vphi\in\calc$ with probability~$1$.

A sufficient condition is that
$\Pr[z|\vphi]$ is strictly positive for every $z\in\calz$ and every $\vphi\in\calc$. 
Since $\Pr[z|\vphi]=\vphi\mech_{*z}$, it is sufficient that every entry of the column $\mech_{*z}$ 
be strictly positive. In this case, $\Pr[z|\vphi]$ remains positive for every $\vphi\in\calc$, 
including points on the boundary of the simplex, so $\log\Pr[z|\vphi]$ is finite and continuous. 
This condition is satisfied by all mechanisms considered in this paper. 
For mechanisms outside this scope, however, continuity must be verified separately.

\subsection{The dominance condition}
\label{sec:dominance}
The dominance condition (\autoref{def:id_cont_dom}\ref{def:id_cont_dom:dom}) can be written as follows: 
there is a function $d:\calz \to \reals^+$ such that 
for every observable $z\in\calz$, the value $d(z)$ is an upper bound on 
$| \log \Pr[z | \vphi] |$ 
for all distributions $\vphi$ over $\calx$, and $d(Z^i)$ has a finite mean, 
i.e., $\E [ d(Z^i) ] < \infty$.  
%
This depends on the mechanism matrix $\mech$; hence, it must be checked on a per-case basis as follows.

\subsection{Consistency for particular mechanisms}  
We will use \autoref{thm:mlconsistencygen} to show the 
consistency of \ibu{} for certain mechanisms. From \autoref{thm:identification}, 
the identification condition is equivalent to the strict concavity of the log-likelihood function 
for some observations. Therefore, we can use the results of Section \ref{sec:strictconcavity}, 
concerning the strict concavity, to ensure the identification condition for particular mechanisms. 
We can also show that the dominance condition holds when these mechanisms are used. 
Thus, we obtain the following results concerning the convergence of the MLE. 
\begin{restatable}{corollary}{consistency1} 
\label{cor:consistency1}
The MLE obtained under $k$-RR, truncated linear geometric, truncated taxicab, and basic \textsc{Rappor} mechanisms 
converges in probability to the real distribution. 
\end{restatable}
The proof of the above theorem relies on taking the dominating function as  
$d(z) = - \log (\inf_{x,y} \mech_{xy})$, where $\mech$ is the mechanism matrix.
\begin{restatable}{corollary}{consistency2} 
\label{cor:consistency2}
Consider a finite alphabet $\calx\subset\integers$. 
Then the MLE obtained under the linear geometric mechanism $\mech: \calx \to \integers$
converges in probability to the real distribution. 
\end{restatable}
Our proof uses, in this case, $d(z) =  \lambda\,s ( | z - x_1 | + x_2 - x_1 ) - \log a$, 
where $x_1, x_2$ are the smallest and largest elements in $\calx$. 
$\lambda, s$ are the parameters of the mechanism, and $a$ is the constant 
given by~\eqref{eq:linearGeometric}.
\begin{restatable}{corollary}{consistency3} 
\label{cor:consistency3}
Consider a finite alphabet $\calx\subset\integers^2$. 
Then the MLE obtained under the taxicab mechanism $\mech:\calx \to \integers^2$  
converges in probability to the real distribution. 
\end{restatable}
We take the dominating function for the taxicab mechanism as 
$d( (\mu,\nu) ) = 
\lambda\,s ( | \mu - u_1 | + | \nu - v_1 | +(u_2-u_1)+(v_2-v_1) ) - 2\,\log a$
where $u_1, u_2$ are the smallest and largest values of the first coordinate of 
any $(u,v) \in \calx$, and $v_1, v_2$ are the smallest and largest values of 
the second coordinate. The detailed proofs of the above theorems are given in 
the supplemental material. 

\subsection{An example of non-consistency}\label{sec:nonconvergence}
In the following, we show an example that does not satisfy the conditions of 
\autoref{thm:mlconsistencygen} and we discuss the consequences.  
We consider a planar region ($5$km $\times$ $5$km) in San Francisco, discretized 
into a grid of $5\times5$ cells with side length $1$km. 
Each location in this region is approximated by the center of its enclosing cell. 
Let the alphabet $\calx$ be the set of these centers,  
and let the true distribution $\vtheta$ over $\calx$ be derived from the Gowalla dataset~\cite{Gowalla:190518}. 
We apply a planar geometric mechanism
\cite{Chatzikokolakis:17:POPETS} 
(see also Appendix \ref{sec:plan_geom})
with $\geps = 0.5$, followed by truncating the noisy data to a smaller grid of $4\times4$ cells.
The matrix $\mech$ of this overall mechanism consists of $25$ rows and $16$ columns, 
therefore has fewer than $|\calx|$ linearly independent columns. This violates Condition \ref{prop_id_item_3}
of \autoref{thm:identification}, which characterizes the identification property, 
hence violating the first condition of the MLE convergence in \autoref{thm:mlconsistencygen}.  

\begin{figure}
\centering 
 \subfigure[Original distribution]{
      \label{fig:ex_original}
      \includegraphics[width=0.18\textwidth]{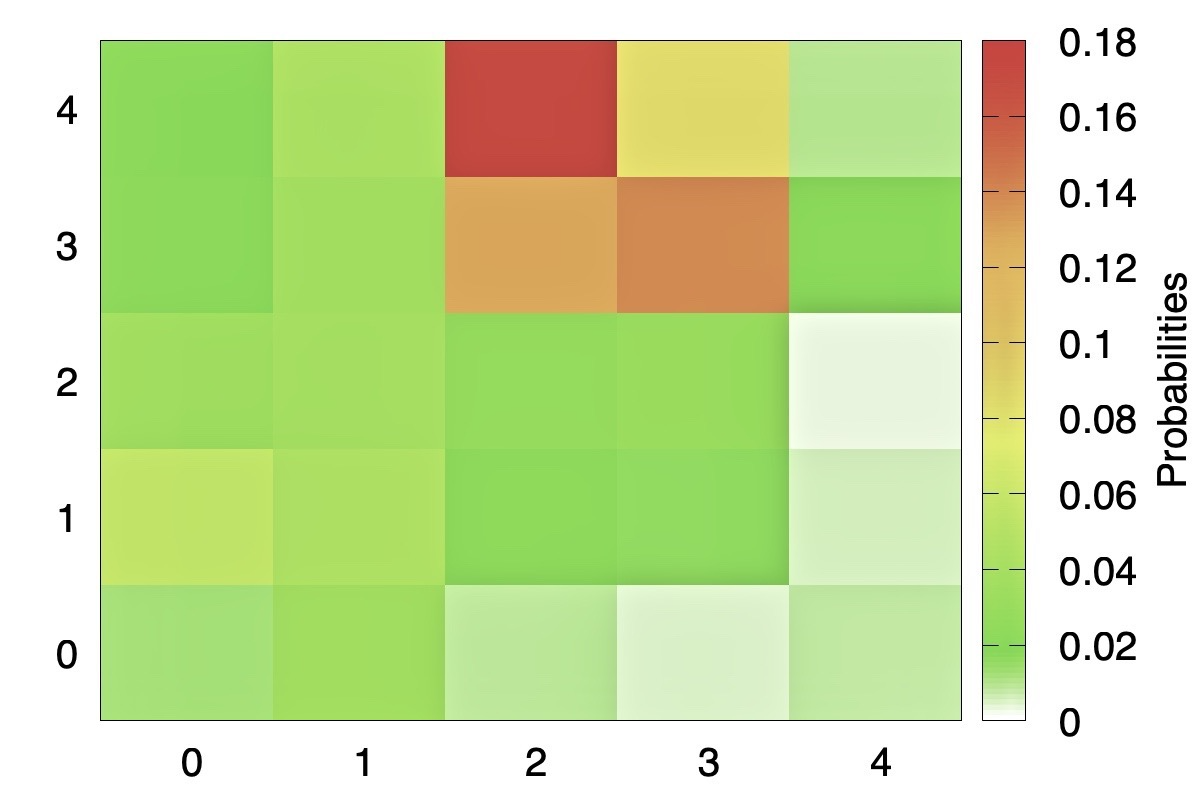}
      }
\subfigure[]{
      \label{fig:ex_mle1}
      \includegraphics[width=0.18\textwidth]{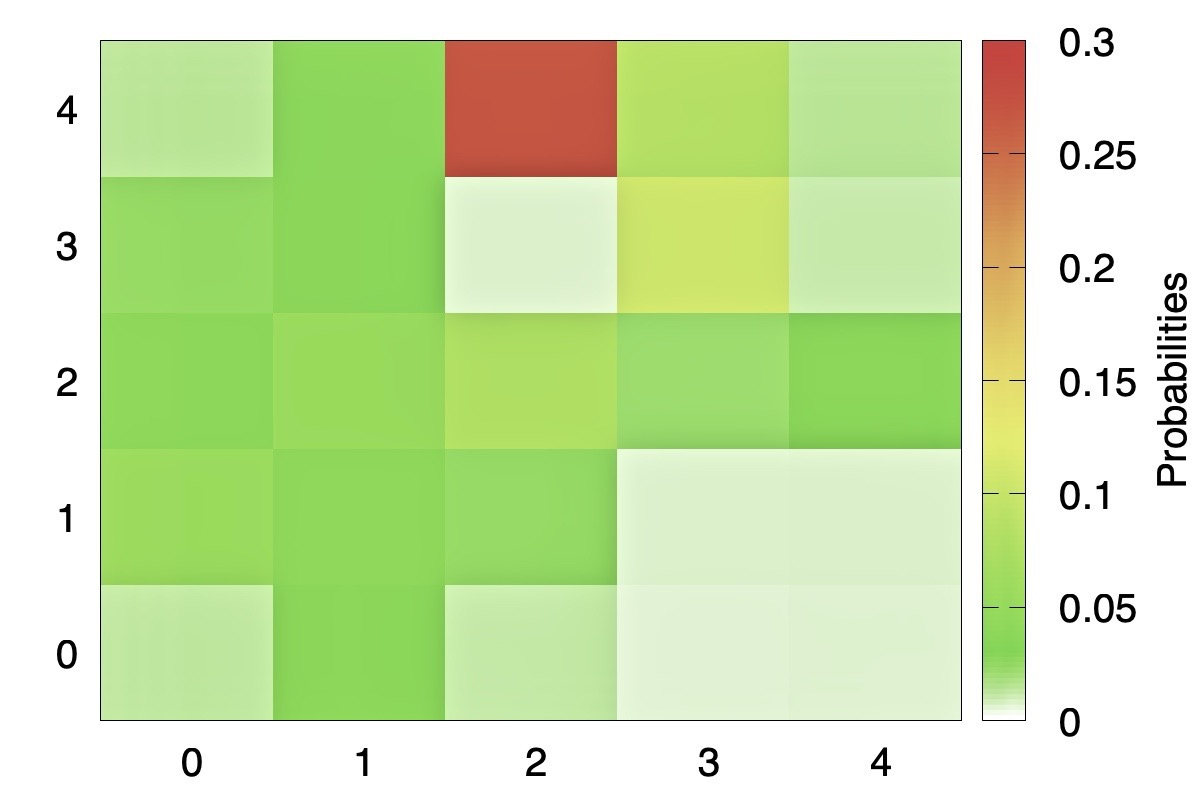}
      }
\subfigure[]{
      \label{fig:ex_mle2}
      \includegraphics[width=0.18\textwidth]{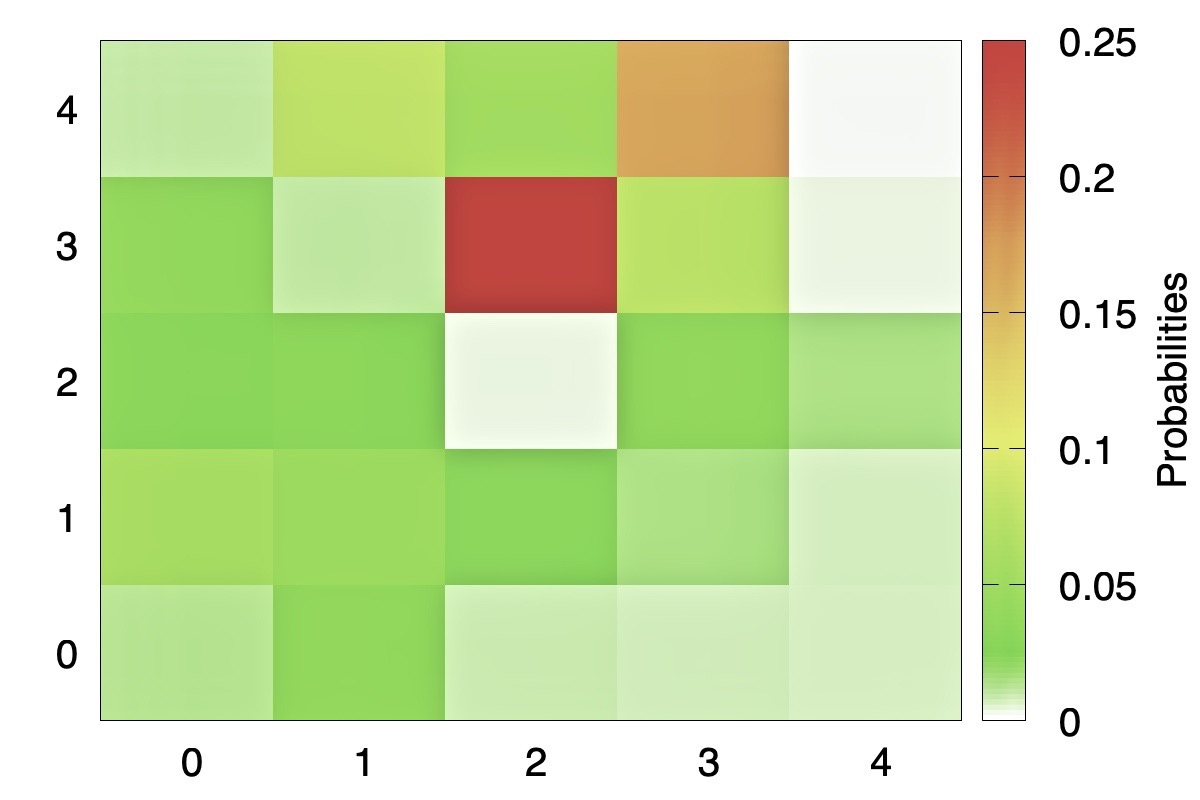}
      }
\subfigure[]{
      \label{fig:ex_mle3}
      \includegraphics[width=0.18\textwidth]{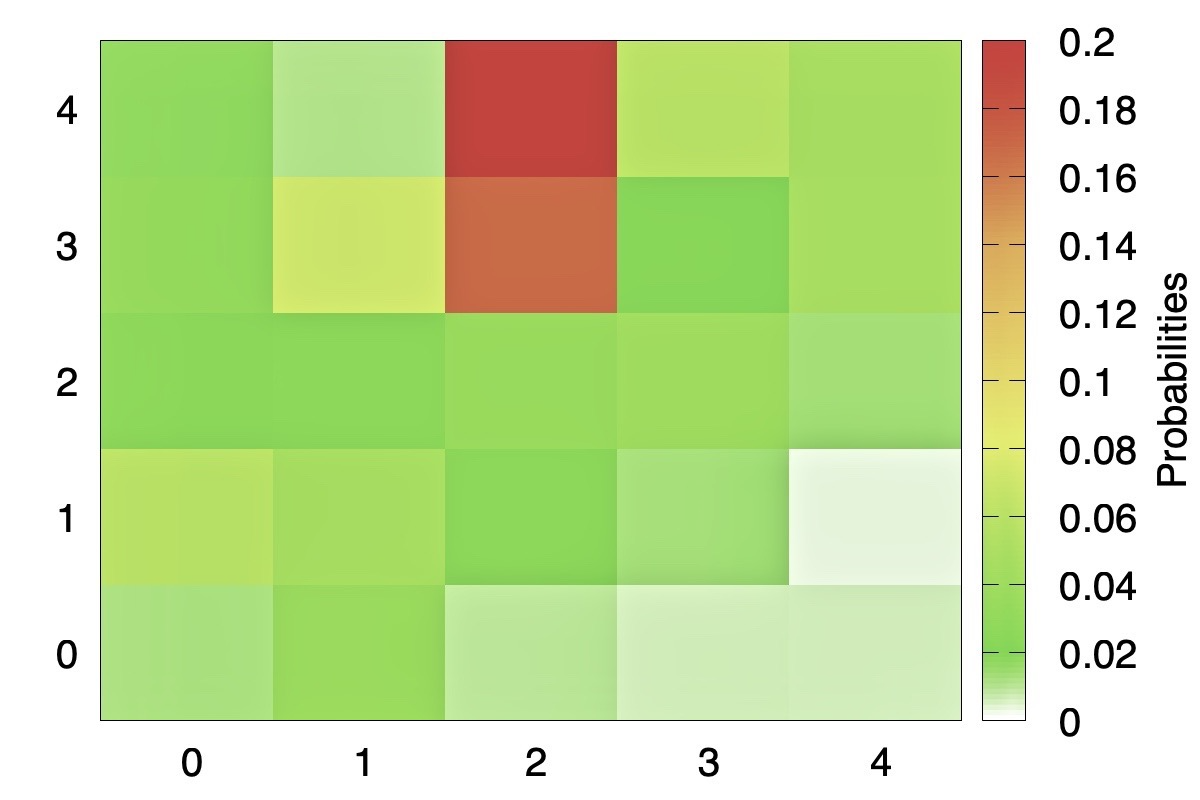}
      }
\caption{
The original distribution obtained from the Gowalla dataset (a), and three MLEs
(b), (c), (d) obtained by \ibu{} using the noisy data provided by a privacy mechanism that violates consistency conditions. } 
\label{fig:multiple_mles}
\end{figure}

To see experimentally that in the above case, the \ibu{}'s estimate (which is an MLE) does not 
converge to $\vtheta$, we let the noisy distribution $\vq$ be the asymptotic one,
equivalently assuming that infinitely many noisy data are available. By running 
\ibu{} on $\vq$ (and $\mech$), we find that different distributions are fixed points 
and therefore may be produced by \ibu{}. Figure~\ref{fig:multiple_mles} shows three such 
estimated distributions, together with the original one $\vtheta$. This means that even with 
infinitely many noisy data, \ibu{}'s estimate is not necessarily the original distribution $\vtheta$.
\subsection{Consistency of \ibu{} on partitions of the alphabet}
While \ibu{}'s estimate may fail to converge to the original distribution $\theta$ on $\calx$, it may still converge to the true distribution over partitions of $\calx$. 
This situation arises when some symbols in $\calx$ are equivalent in 
the sense that they induce identical distributions over the mechanism observables $\calz$, 
i.e., when they correspond to identical rows in the mechanism matrix $\mech$. 
More formally, define the equivalence relation $\sim$ on $\calx$ by writing  
$x \sim x'$ if the two rows $\mech_{x*}$ and $\mech_{x'*}$ are equal. 
Let $\caly$ be the quotient set of equivalence classes, i.e., $\caly = \calx/{\sim}$, 
and define the quotiented mechanism $\tilde{\mech}: \caly \to \calz$ as  
$\tilde{\mech}_{y*} = \mech_{x*}$ for every $y \in \caly$ and $x \in y$; 
this is well-defined since the the rows $\mech_{x*}$ for all $x \in y$ are identical.
Therefore $\Pr[z | y] = \Pr[z | x]$ for all $z\in \calz$ and all $x, y$ with $x\in y$. 
Thus, if the original mechanism $\mech$ satisfies the regularity and dominance 
conditions of \autoref{thm:mlconsistencygen}, and in addition, the quotiented mechanism 
$\tilde{\mech}: \caly \to \calz$ is full-rank, then \ibu{} is consistent on $\caly$.

\section{Experimental evaluation}\label{sec:experiments}
We evaluate the estimation performance of the \ibu{} algorithm in comparison with other estimation methods 
in the literature: the matrix inversion~\cite{Agrawal:05:ICMD,Kairouz:16:ICML} and the \textsc{Rappor}'s estimator~\cite{Erlingsson:14:CCS}. 

{\bf Obfuscation Mechanisms}. We consider three classes of obfuscation mechanisms. 
The first class includes those that work on a linear (one-dimensional) alphabet and satisfy $\lambda\dist$-privacy~\cite{Chatzikokolakis:13:PETS}, 
namely the linear geometric, Laplace, and exponential mechanisms. The second class includes the adaptations
of these mechanisms to planar (two-dimensional) alphabets, i.e., locations, thereby satisfying 
$\geps$-geo-indistinguishablity~\cite{Andres:13:CCS}.
The linear geometric mechanism~\cite{Ghosh:09:STOC}, and its adaptations to planar alphabets, 
namely, the planar geometric~\cite{Chatzikokolakis:17:POPETS} and taxicab mechanisms~(Section \ref{sec:taxicab}), provide privacy protection at a small cost of obfuscation loss.   
The Laplace mechanism provides the same protection, with a minimal penalty on the obfuscation loss, on 
continuous linear~\cite{natasha:21:LICS} and continuous planar~\cite{ElSalamouny:18:JIS} alphabets. 
The exponential mechanism~\cite{Chatzikokolakis:17:POPETS} is flexibly applicable to 
arbitrary finite alphabets of discrete data.
The final class of mechanisms includes those that satisfy $\leps$-LDP, 
namely the $k$-RR and the basic \textsc{Rappor} mechanisms. 
\\[1ex]
{\bf Datasets}. Concerning the users' original data points, we use two real datasets 
that contain different kinds of data. 
The first is the UCI Adult dataset~\cite{adult:1996}, where we focus on the users' ages. 
These data are \emph{one-dimensional}; hence, we use mechanisms suitable for this kind. 
The second is the Gowalla dataset~\cite{Gowalla:190518}, which contains the geographic locations 
of the users. In this case, the data are \emph{planar}, i.e., two-dimensional; hence, we 
use the planar mechanisms. 

\autoref{tab:experiments} lists, in the first column, the mechanisms used in our experiments. 
The second column shows the privacy notion guaranteed by each mechanism, along with the parameter values used. 
The third and fourth columns specify the dataset to which each mechanism is applied and whether the data alphabet is linear or planar. We use truncated versions of the geometric, Laplace, and taxicab mechanisms, which are indicated in the table by ``(T)''.

In our experiments, we measure the performance by the 
Earth Mover’s Distance (EMD) between the original and estimated distributions on the alphabet $\calx$, 
because, as argued by \cite{Alvim:18:CSF}, it is a main notion of utility for data domains equipped with a metric\footnote{The EMD is the minimum cost to transform one distribution into another one by moving probability masses between cells.}. 
In 
Appendix \ref{sec:total_variation}, 
we report, for the same experiments, the estimation errors measured using the total variation. We also perform our experiments on other datasets and report the results in 
Appendices \ref{sec:labTests} and \ref{sec:paris}.

%
\begin{table}[h!]
  \begin{center}
    \caption{Mechanisms and datasets used in the experiments} 
    \label{tab:experiments} 
    \bgroup 
    \begin{tabular}{ l  |  l  |  l  |  l } 
    \toprule 
    \textbf{mechanisms} & 
    \textbf{privacy levels} & 
    \textbf{dataset} & 
    \textbf{type} \\
    \midrule 
    (T)~linear geometric & 
    \multirow{3}{2.2cm}{$\lambda\dist$-privacy with $\lambda$ from 0.01 to 0.3} & 
    \multirow{3}{1.5cm}{UCI Adult (ages)~\cite{adult:1996}} & 
    \multirow{3}{*}{linear}  \\  
    (T)~linear Laplace &&& \\
    exponential        &&& \\
    \midrule
    (T)~taxicab & 
    \multirow{4}{2.2cm}{ $\geps$-geo-indistin-guishability with $\geps$ from 0.2 to 3.0} & 
    \multirow{4}{1.5cm}{Gowalla~\cite{Gowalla:190518} (locations)} & 
    \multirow{4}{*}{planar} \\  
    (T)~planar geometric    &&& \\
    (T)~planar Laplace      &&& \\
    exponential             &&& \\
    \midrule
    \multirow{4}{*}{$k$-RR} & 
    \multirow{4}{2.2cm}{$\leps$-LDP with $\leps$ from 0.5 to 4.0}  & 
    \multirow{2}{1.5cm}{UCI Adult (ages)} & 
    \multirow{2}{*}{linear} \\ 
    &&& \\
    \cmidrule{3-4}
    && \multirow{2}{1.5cm}{Gowalla (locations)} & 
    \multirow{2}{*}{planar} \\ 
    &&& \\
    \midrule
    \multirow{2}{*}{\textsc{Rappor}} & 
    \multirow{2}{2.2cm}{$\leps$-LDP with $\leps$ from 0.1 to 5.0}  & 
    \multirow{2}{*}{Synthetic} & 
    \multirow{2}{*}{linear} \\
    &&& \\
    \bottomrule 
    \end{tabular}
    \egroup
  \end{center}
\end{table}

\subsection{Estimation under metric privacy mechanisms}\label{sec:metric_mechanisms}
In the following, we evaluate the estimation quality of  
the \ibu{} algorithm, \invn{}, and \invp{} 
after applying the metric privacy mechanisms and the 
datasets listed in \autoref{tab:experiments}. 

We start by letting the alphabet $\calx$ be the age values between $0$ and $99$, 
with spacing $s=1$ (years) between successive age values. 
Then we apply the truncated linear geometric mechanism with $\lambda=0.05$.
Finally, we apply the above methods on the noisy data to estimate the original distribution over age values. 
The original, noisy, and estimated distributions on $\calx$ are shown in Figure~\ref{fig:geometric_on_ages} 
in the introduction.
It is clear that the \inv{} methods perform poorly, while the \ibu{}'s estimate is quite precise. The poor estimation performance of \inv{} in this case is formally explained in Section~\ref{sec:inv_error_bounds} via \autoref{prop:invTLinGeomBound}.

\begin{figure}[t]
\centering
\includegraphics[width=0.25\textwidth]{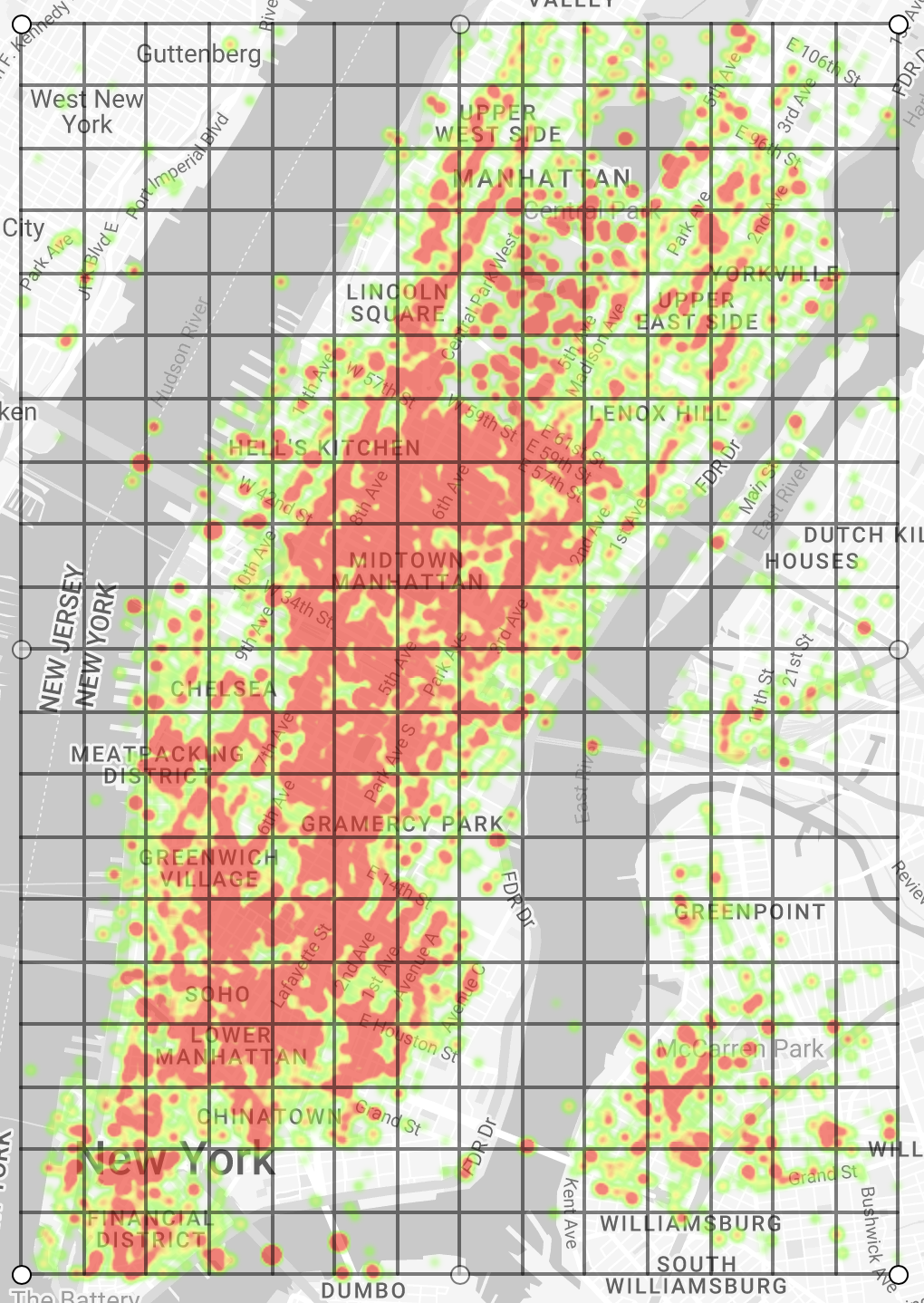}
\caption{The region of Manhattan, discretized in a grid of $20\times14$ cells, 
covering an area of 10km$\times$7km. The   
heatmap is based on the check-ins from the Gowalla dataset.}
\label{fig:manhattan_map}
\end{figure}

We repeat the above experiment, but this time with $\calx$ being 
a set of locations in the planar space. Precisely, we focus on a region 
in Manhattan, bounded by the latitudes $40.7044, 40.7942$, and 
the longitudes $-74.0205, -73.9374$, covering an area of $10$km $\times$ $7$km. 
We discretize this region into a grid of $20\times 14$ square cells of side length $0.5$km 
(see Figure~\ref{fig:manhattan_map}). We approximate every location by the center of 
its enclosing cell, and then we define the alphabet $\calx$ to be the set of these centers. 
Therefore, $\calx$ is regarded as a subset of $\integers^2$ with spacing $s=0.5$km.  
Now we use the check-ins in the Gowalla dataset, restricted to this region ($98060$ check-ins) 
and approximated to the elements of $\calx$, as the users' original data. 
We obfuscate these data using the truncated taxicab mechanism 
(see Section~\ref{sec:taxicab}) with $\lambda = 1.0/\sqrt{2}$, 
hence satisfying $(1.0)$-geo-indistinguishability (see \autoref{prop:geoindTaxicab}).  
\begin{figure}[t]
\centering 
\subfigure[\text{Original}]{
      \label{fig:original}
      \includegraphics[width=0.22\textwidth]{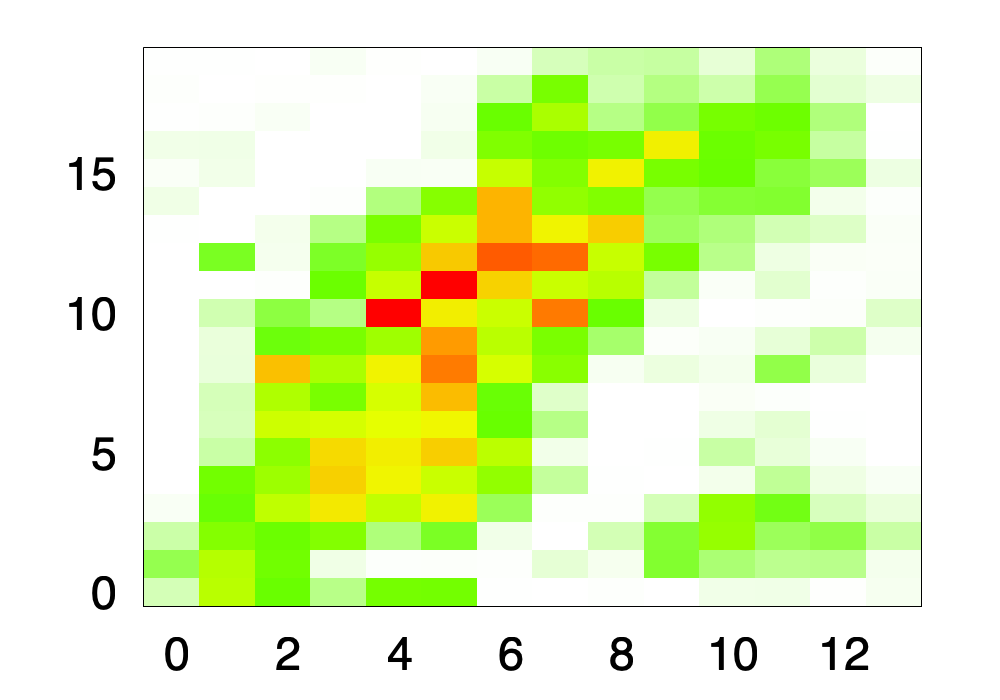}
      }
\hspace{-20 pt}
\subfigure[Noisy  (EMD = 0.8719)]{   
      \label{fig:geom_grid_noisy}
      \includegraphics[width=0.264\textwidth]{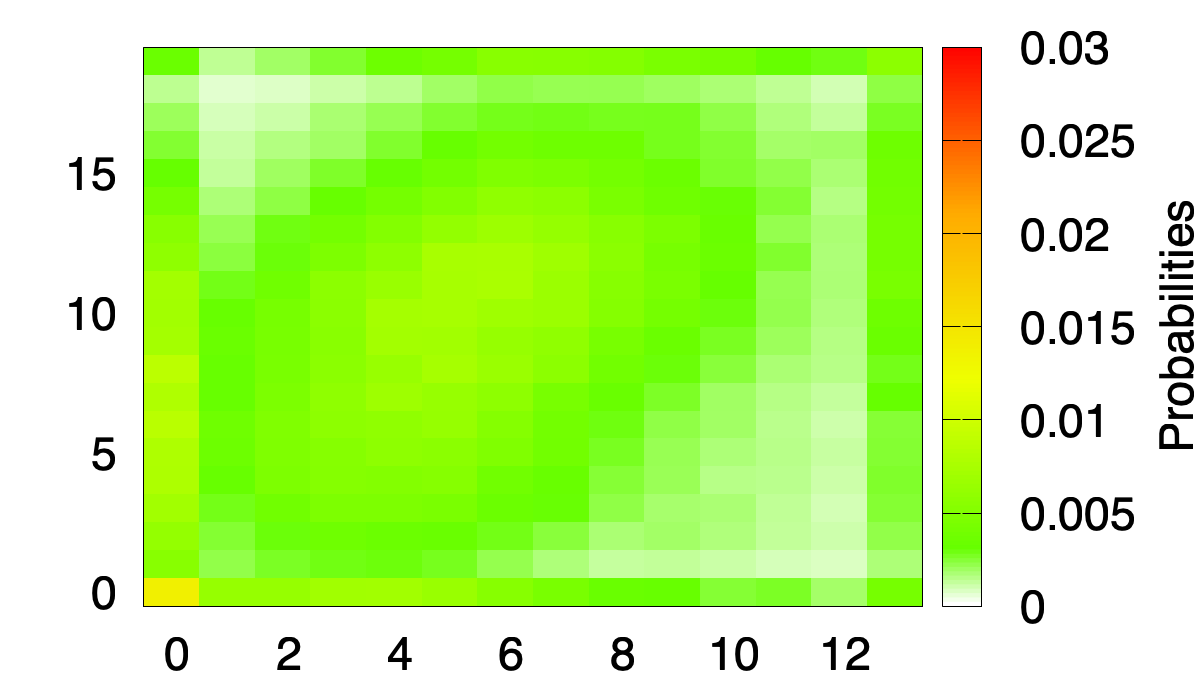}
      }
\subfigure[\invp{} (EMD = 0.9017)]{           
      \label{fig:geom_grid_estimated_invp}
      \includegraphics[width=0.22\textwidth]{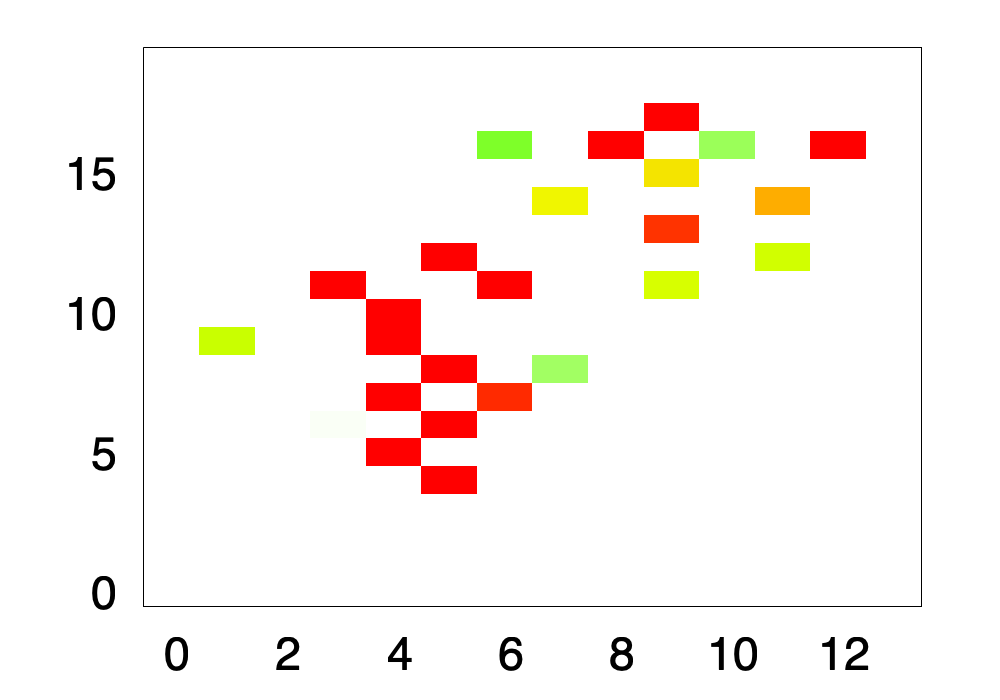}
      }
\hspace{-20 pt}
\subfigure[\invn{} (EMD = 0.7963 )]{        
      \label{fig:geom_grid_estimated_invn}
      \includegraphics[width=0.264\textwidth]{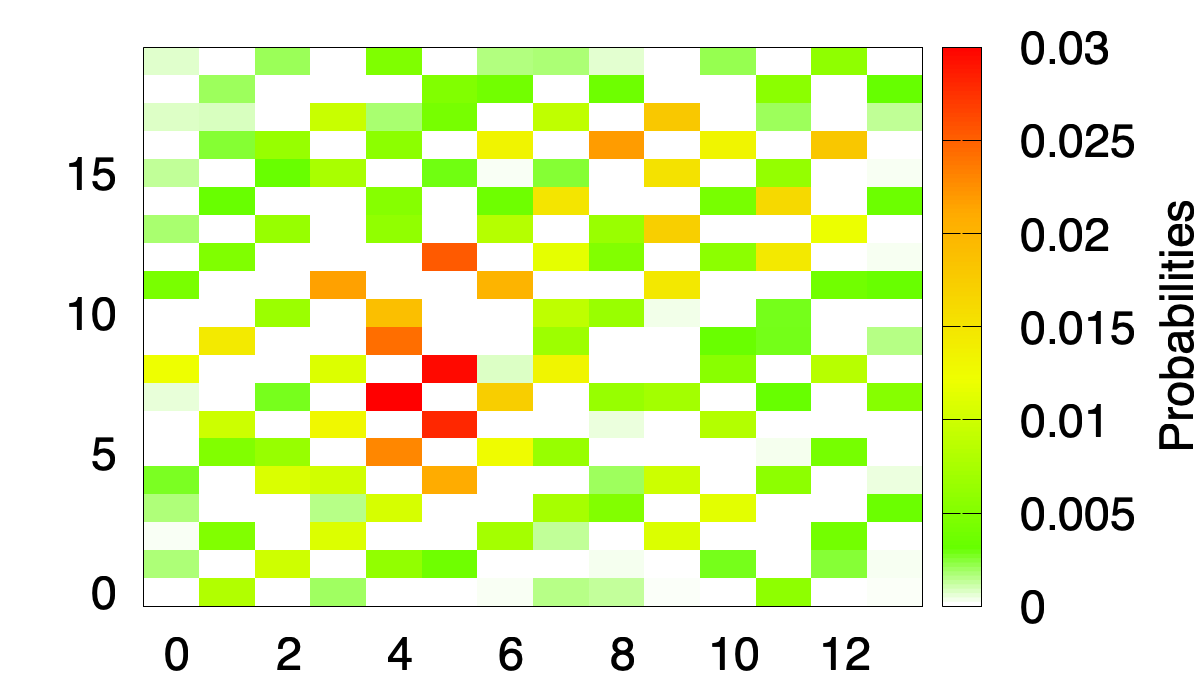}
      }
\subfigure[\ibu{} (EMD = 0.1587)]{    
      \label{fig:geom_grid_estimated_ibu}
      \includegraphics[width=0.264\textwidth]{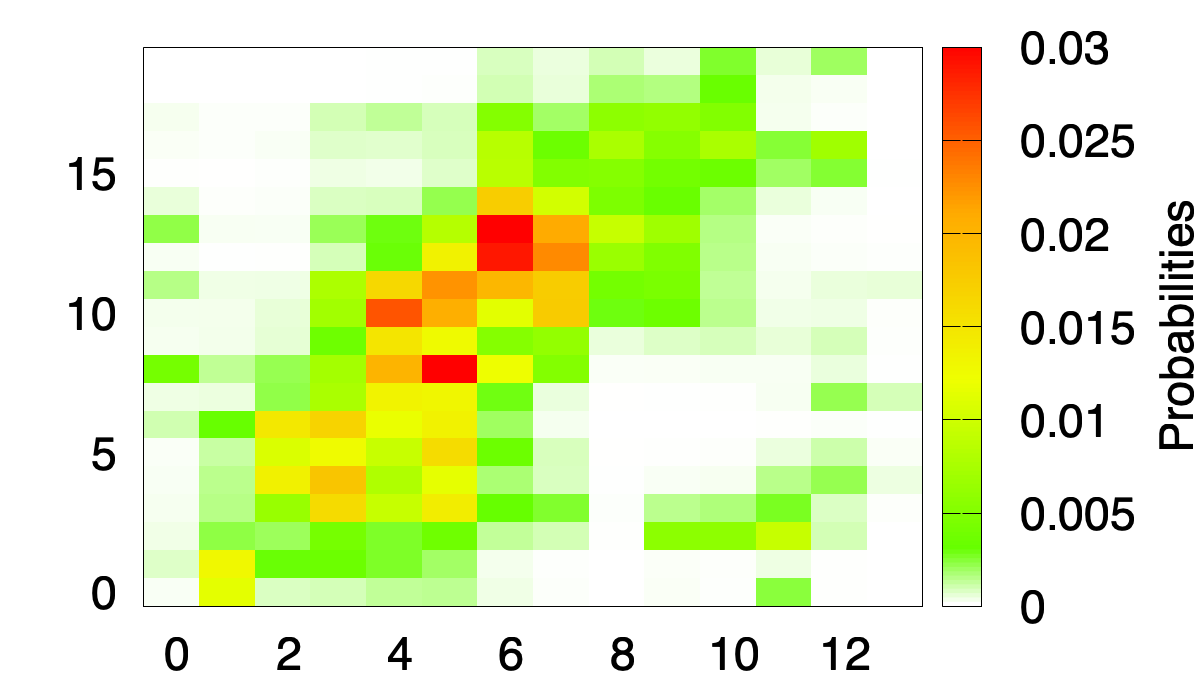}
      }
\caption{Estimation of the Gowalla original distribution. The noisy data are 
produced by a truncated taxicab mechanism that satisfies $(1.0)$-geo-indistinguishability. 
The estimation quality is shown as the EMD between the original and estimated distributions.}
\label{fig:taxicab_on_grid}
\end{figure}
Figure~\ref{fig:taxicab_on_grid} shows the original and noisy distributions on the grid, 
and also the distributions estimated by \invn{}, \invp{}, and the \ibu{} algorithm. 
The figure also shows the estimation quality measured by the EMD between the original 
and estimated distributions. Similar to the case of the one-dimensional alphabet, 
we observe in this experiment that \ibu{} substantially outperforms both \invn{} and \invp{}. 
The poor estimation of the \inv{} methods in this setting will be explained later 
via \autoref{prop:invTTaxicabBound}. 

\begin{figure}[t]
\centering 
\subfigure[obfuscation using truncated \newline linear geometric mechanism]{
      \label{fig:box_ages_vlgeom_emd}
      \includegraphics[width=0.23\textwidth]{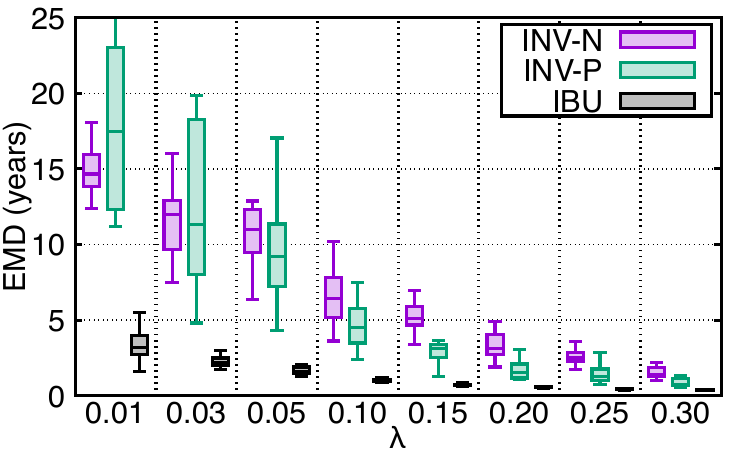}
      }
\subfigure[obfuscation using truncated \newline linear Laplace mechanism]{
      \label{fig:box_ages_vllap_emd}
      \includegraphics[width=0.23\textwidth]{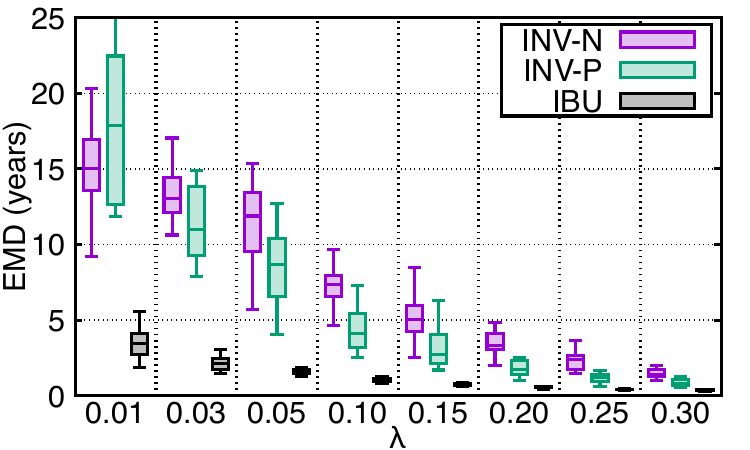}
      }
\subfigure[obfuscation using \newline exponential mechanism]{
      \label{fig:box_ages_exp_emd}
      \includegraphics[width=0.23\textwidth]{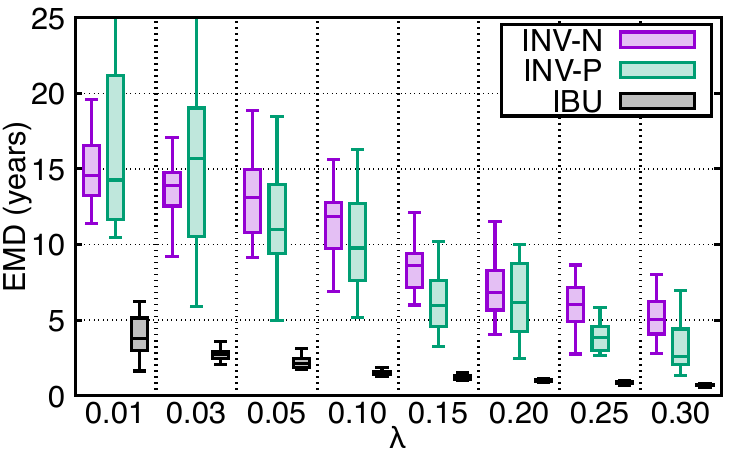}
      }
\caption{The EMD between the original and estimated distributions using the \ibu{} algorithm, 
\invn{}, and \invp{}. Noisy data are produced using different mechanisms applied to the age data 
in the UCI adult dataset.}
\label{fig:box_estimators_dprivacy}
\end{figure}

\begin{figure}[t]
\centering 
\subfigure[obfuscation using truncated \newline taxicab mechanism]{
      \label{fig:box_taxicab_emd}
      \includegraphics[width=0.23\textwidth]{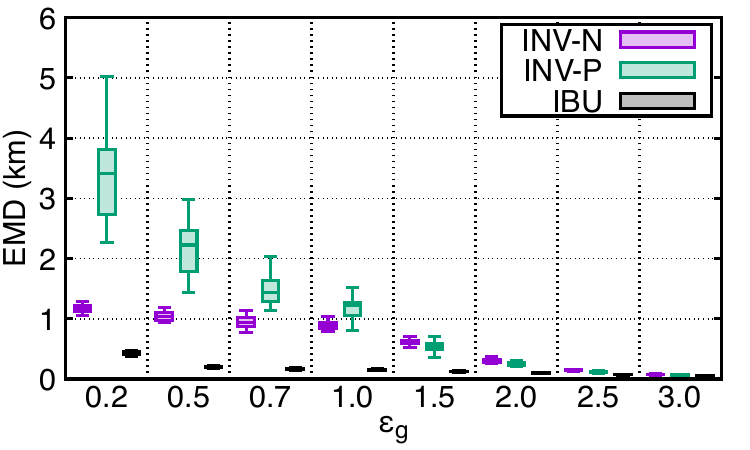}
      }
\subfigure[obfuscation using truncated \newline planar geometric mechanism]{
      \label{fig:box_planar_pgeom_emd}
      \includegraphics[width=0.23\textwidth]{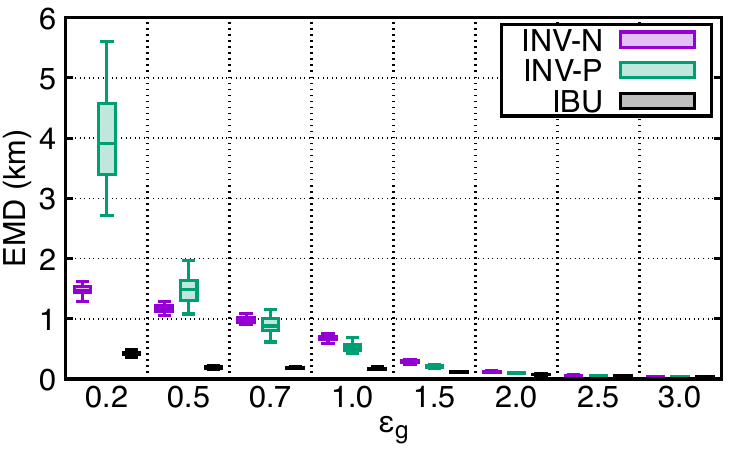}
      }
\subfigure[obfuscation using truncated\newline planar Laplace mechanism]{
      \label{fig:box_planar_plap_emd}
      \includegraphics[width=0.23\textwidth]{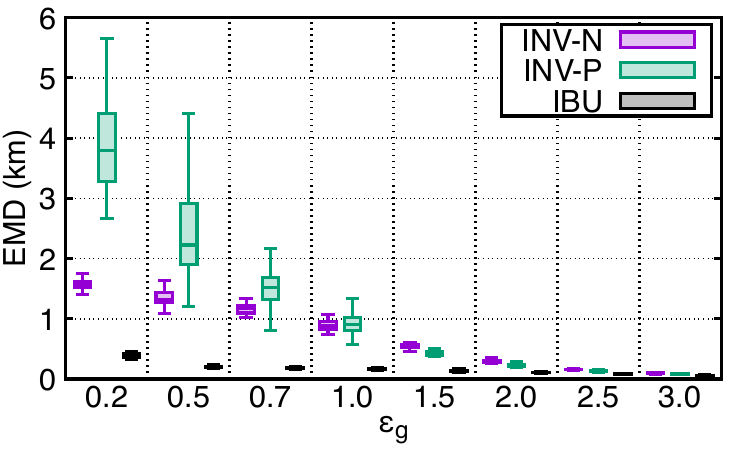}
      }
\subfigure[obfuscation using \newline exponential mechanism]{
      \label{fig:box_planar_exp_emd}
      \includegraphics[width=0.23\textwidth]{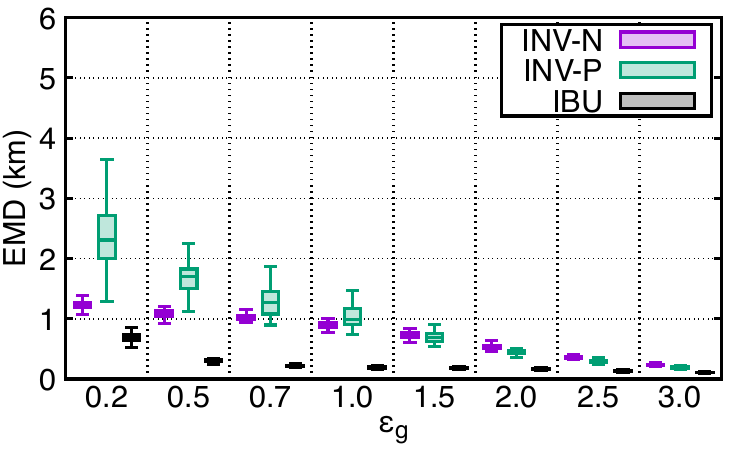}
      }

\caption{The EMD between the original and estimated distributions using \emph{the \ibu{} algorithm}, \invn{}, and \invp{}. 
Noisy data are produced using 
different mechanisms applied to location data (in the Manhattan region) from the Gowalla dataset.}
\label{fig:box_estimators_geind}
\end{figure}

Finally, we evaluate the quality of the estimators under varying levels of privacy. 
To this end, we repeat the obfuscation-estimation procedure from the previous two experiments 
multiple times and record the EMD between the original and estimated distributions in each run.
Figure~\ref{fig:box_estimators_dprivacy} shows the results for linear mechanisms satisfying 
$(\lambda \dist)$-privacy applied to the UCI ages dataset, with the privacy parameter $\lambda$ 
ranging from $0.01$ to $0.3$. Similarly, Figure~\ref{fig:box_estimators_geind} reports 
the results for planar mechanisms satisfying $\geps$-geo-indistinguishability applied to 
the location dataset, with the privacy parameter $\geps$ ranging from $0.2$ to $3.0$.

It is clear from the two figures that the estimation quality of all methods improves with 
larger values of $\lambda$ and $\geps$, corresponding to introducing less noise. 
Furthermore, we observe that \ibu{} is more accurate than \invn{} and \invp{}, 
especially when $\lambda$ and $\geps$ are small (strong levels of privacy). 
This empirically confirms the poor estimation of \inv{} in our earlier two experiments 
with the truncated linear geometric mechanism (\autoref{fig:geometric_on_ages}), 
and the truncated taxicab mechanism (\ref{fig:taxicab_on_grid}).  
and extends this observation to other metric privacy mechanisms (both linear and planar). 

For the truncated linear geometric and planar taxicab mechanisms, 
we will present later \autoref{prop:invTLinGeomBound} and \autoref{prop:invTTaxicabBound} 
which show lower bounds for the error of \inv{}, hence explaining its poor estimation in these cases. 


\subsection{Estimation after protection by k-RR}\label{sec:var_priv-krr}

\begin{figure}[t]
\centering 
\subfigure[age data sanitized by $k$-RR]{
      \label{fig:box_ages_krr_emd}
      \includegraphics[width=0.23\textwidth]{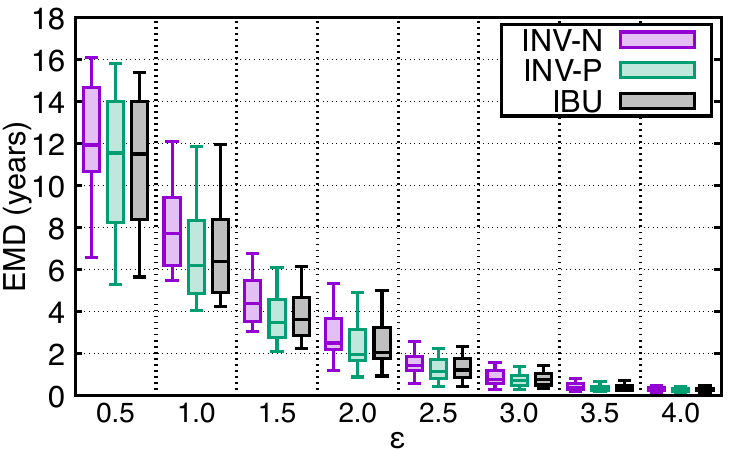}
      }
\subfigure[location data sanitized by $k$-RR]{
      \label{fig:box_planar_krr_emd}
      \includegraphics[width=0.23\textwidth]{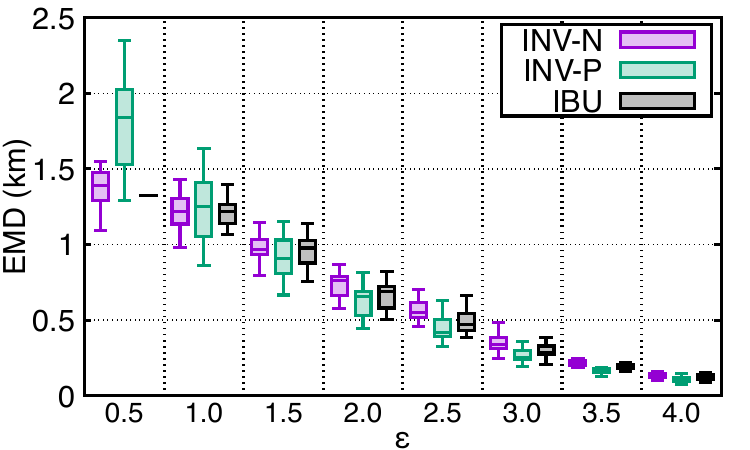}
      }
\caption{The EMD between the original and estimated distributions using the indicated methods. 
Noisy data are produced using a $k$-RR mechanism with the shown values of $\leps$.}
\label{fig:box_estimators_krr_emd}
\end{figure}

Now we use the $k$-RR mechanism (described in Section \ref{sec:strictlyConcave:krr}) to obfuscate the real data, 
with the privacy level $\leps$ varying from $0.5$ to $4.0$, and plot the estimation error for \ibu{} 
and the \inv{} variants at each level. Figure~\ref{fig:box_estimators_krr_emd} shows the results for
the (one-dimensional) UCI age data and the (planar) location data in Manhattan. 

Unlike the earlier experiments, where metric privacy mechanisms were used, 
we observe that under $k$-RR the estimation quality of the \inv{} methods 
is comparable to that of \ibu{}. 
Later in Section \ref{sec:inv_error_bounds}, we will show via \autoref{prop:invKrrBound} that the 
\inv{} estimation error is low in this setting.  

\subsection{Estimation after protection by \textsc{Rappor} mechanism}\label{sec:exp_rappor}

\begin{figure}[t]
\centering 
\subfigure[Binomial distribution]{
      \label{fig:binomial_rappor_em}
      \includegraphics[width=0.23\textwidth]{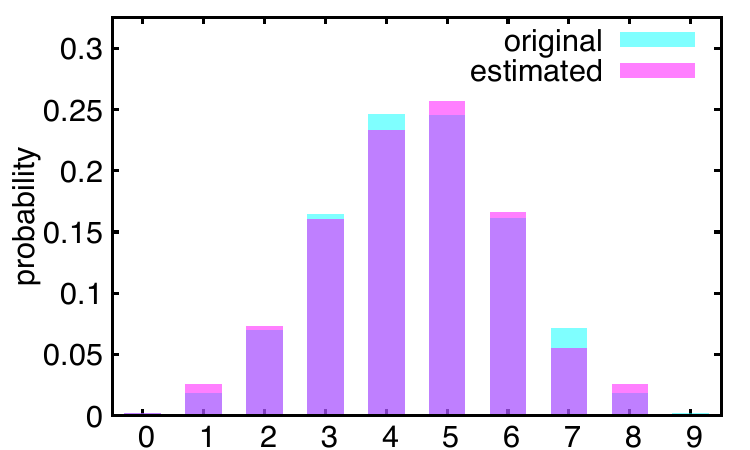}
      }
\subfigure[Uniform on $\{3,4,5,6\}$]{
      \label{fig:uniform_rappor_em}
      \includegraphics[width=0.23\textwidth]{fixed_privacy/line_binomial_samples100000_BOTRAPPOR_eps0.5_IBU.pdf}
      }
\caption{Using the \ibu{} algorithm to estimate the original distributions on $\{0,1,\dots, 9\}$ from noisy data produced by a \textsc{Rappor} mechanism with $\leps=0.5$. 
In (a), the real distribution is binomial. In (b), the distribution sets $\Pr[x] =1/4$ for every $x\in \{3,4,5,6\}$ 
and $\Pr[x]=0$ otherwise.}
\label{fig:rappor_on_line_em}
\end{figure}

We consider obfuscating the users' data using the basic \textsc{Rappor} mechanism 
(Section~\ref{sec:rappor}). 
In this experiment, we consider an alphabet $\calx= \{0,1, \dots, 9\}$ and two 
synthetic data sets, each consisting of $10^5$ data points. The first dataset is 
obtained by sampling its data points from the binomial distribution $B(9, 0.5)$; 
and the other dataset is generated from a distribution that assigns uniform probabilities to 
$\{3,4,5,6\}$ and probability $0$ to the other elements of $\calx$. 
In the two cases, we obfuscate the data using a \textsc{Rappor} mechanism with $\leps=0.5$, 
and then apply the \ibu{} algorithm to estimate the original distribution. 
The results, depicted in Figure~\ref{fig:rappor_on_line_em}, show that \ibu{} 
produces precise estimates in the two cases. 

\begin{figure}[t]
\centering 
\subfigure[original data sampled from\newline binomial distribution]{
      \label{fig:box_rappor_binomial}
      \includegraphics[width=0.23\textwidth]{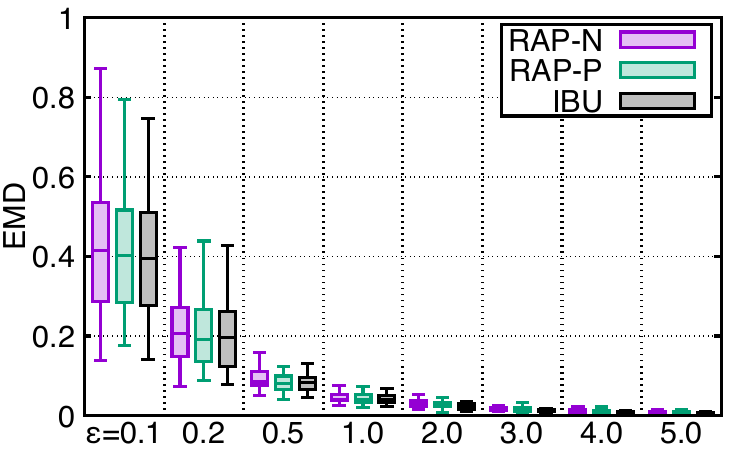}
      }
\subfigure[original data sampled from\newline uniform distribution]{
      \label{fig:box_rappor_uniform}
      \includegraphics[width=0.23\textwidth]{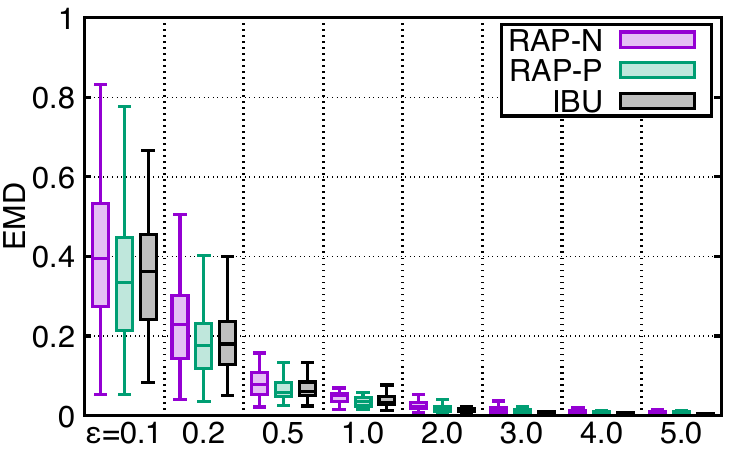}
      }
\caption{The EMD between the original and estimated distributions using the \ibu{} algorithm, \rapn{}, and \rapp{},  
after applying \textsc{Rappor} mechanism for obfuscation. Original data of users are sampled from a Binomial distribution (a), 
and from uniform distribution on $\{3,4,5,6\}$ (b).}
\label{fig:box_estimators_rappor}
\end{figure}

Now, we consider two variants of the \textsc{Rappor} estimator, 
denoted by \rapn{} and \rapp{}, obtained by using \invn{} and \invp{}, respectively, 
to estimate the individual components of the original distribution (see Section~\ref{sec:rappor}).
We evaluate the performance of these two estimators relative to 
\ibu{} by repeatedly running the latter experiment with the three estimators. 
Figure~\ref{fig:box_estimators_rappor} reports the results over a range of $\leps$.
Notice that \rapn{} and \rapp{} give as precise estimates as \ibu{}. 
This observation is anticipated from the similarly accurate estimation of the \inv{} methods under  
the $k$-RR mechanism (cf. Figure~\ref{fig:box_estimators_krr_emd}), 
since in both cases the obfuscation uses the idea of randomized response, and the estimation is based on 
matrix inversion. 
\subsection{Estimation accuracy with varying data sizes}\label{sec:var_datasize}
\begin{figure}[t]
\centering 
\subfigure[$\lambda = 0.05$]{
      \label{fig:box_var_data_lgeom_0.05}
      \includegraphics[width=0.23\textwidth]{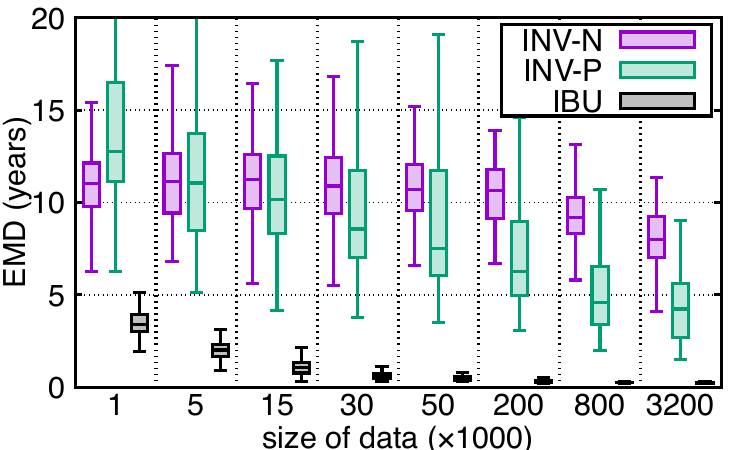}
      }
\subfigure[$\lambda = 0.1$]{
      \label{fig:box_var_data_lgeom_0.1}
      \includegraphics[width=0.23\textwidth]{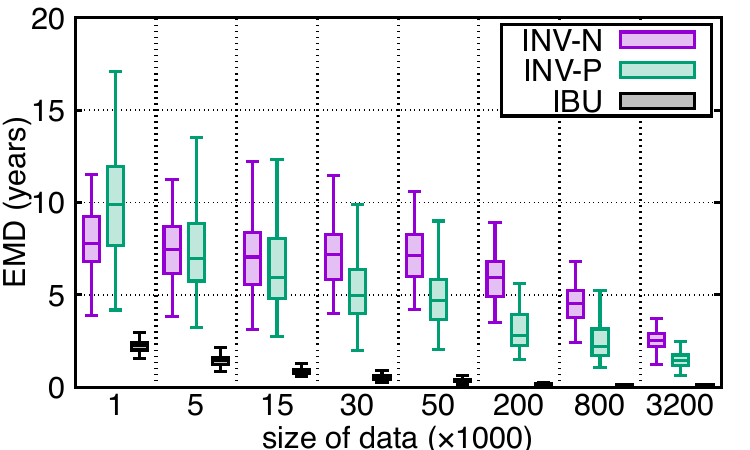}
      }
\caption{The estimation error of the \ibu{} algorithm, \invn{}, and \invp{} 
applied to noisy age data of various sizes. The noisy data are produced by 
the truncated linear geometric mechanism that satisfies $\lambda \dist$-privacy.}
\label{fig:box_var_data_lgeom_emd}
\end{figure}

We continue to compare the estimation performance of \ibu{}, \invn{}, and \invp{}, but this time we vary the size of the data used for the estimation, i.e., the number of users $n$. 
We start by considering
the one-dimensional alphabet of `ages', and sample $n$ data points from the
\emph{population} distribution obtained from the frequencies of the age values in the UCI Adult dataset. We regard
the $n$ data points as real data of the users and obfuscate them with a truncated linear geometric mechanism, 
which offers $\lambda \dist$-privacy, to produce the corresponding noisy data. Finally, we run the estimators on the available noisy data and measure the EMD between the original and estimated distributions of the $n$ data.   
Figure \ref{fig:box_var_data_lgeom_emd} shows the results for $n$ growing from $10^3$ to $3200\times10^3$.  
We also use two strong privacy levels $\lambda$, namely $0.05$ (\ref{fig:box_var_data_lgeom_0.05}) 
and $0.1$ (\ref{fig:box_var_data_lgeom_0.1}).
We note that in the two settings, both \invn{} and \invp{} are very inaccurate compared to \ibu{}, 
and do not approach $0$ EMD  with the given amount of data, while \ibu{} converges to nearly 
$0$ EMD at about $n= 200\times10^3$. 
Indeed, their estimation errors decline much more slowly than \ibu{}. 
In particular, with $\lambda=0.1$, they need about $3200\times10^3$ data points to attain the error level $2.5$, 
which is reached by \ibu{} using only $10^3$ data points. The situation is even worse with $\lambda=0.05$, where 
\invn{} entirely fails to reach the error that \ibu{} attains using only $10^3$ data points. 

\begin{figure}[t]
\centering 
\subfigure[$\geps=0.5$]{
      \label{fig:box_var_data_taxicab_0.5}
      \includegraphics[width=0.23\textwidth]{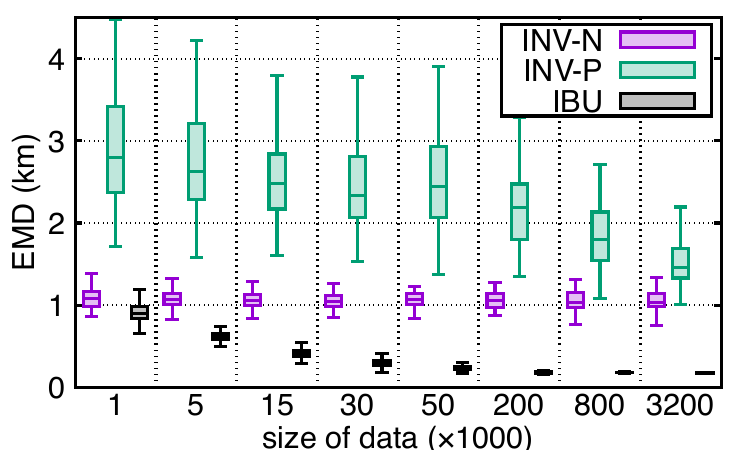}
      }
\subfigure[$\geps = 1.0$]{
      \label{fig:box_var_data_taxicab_1.0}
      \includegraphics[width=0.23\textwidth]{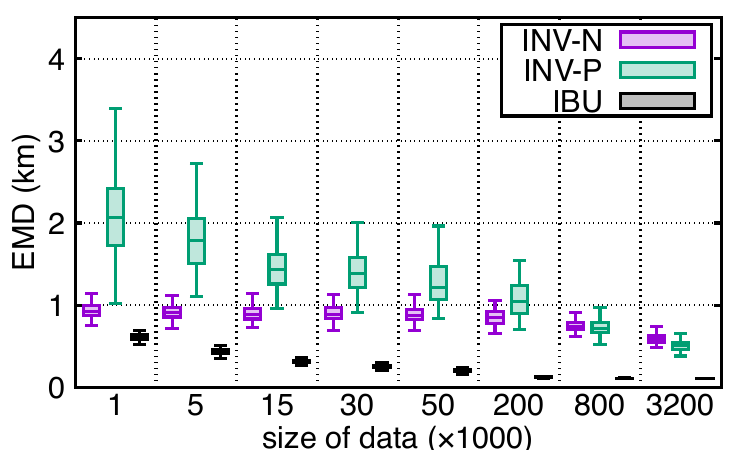}
      }
\caption{
The estimation error of \ibu{} algorithm, \invn{}, and \invp{} 
applied to noisy location data of various sizes. 
The noisy data are produced by a taxicab mechanism that 
satisfies $\geps$-geo-indistinguishability.}
\label{fig:box_var_data_taxicab_emd}
\end{figure}

We redo the above experiment, but now on the two-dimensional (planar) alphabet described in 
Section \ref{sec:metric_mechanisms}. 
Similar to the above experiment, we calculate a population distribution on this alphabet (locations), 
but from the Gowalla dataset.
From this distribution, $n$ data points are sampled and regarded as the original users' data. 
These data points are obfuscated by a taxicab mechanism, which offers $\geps$-geo-indistinguishability. 
Then, following the steps of the previous experiment, we 
run the three estimators on the noisy data and measure their estimation error. 
Figure \ref{fig:box_var_data_taxicab_emd} shows 
the results with $n$ growing from $10^3$ to $3200\times10^3$, 
and with two strong privacy levels $\geps$, namely $0.5$
(\ref{fig:box_var_data_taxicab_0.5}) and $1.0$ (\ref{fig:box_var_data_taxicab_1.0}). 
Similar to the previous experiment, we 
observe that both \invn{} and \invp{} are very inaccurate with the considered values of $n$, 
and their errors decline slowly compared to \ibu{}. 
In particular,  neither \invn{} nor \invp{} approach $0$ EMD  with the given amount of data, 
while \ibu{} again converges 
to nearly $0$ EMD at about $n= 200\times10^3$. 
Furthermore, with $\geps=1.0$, \invn{} and \invp{}'s errors decline at $n=3200\times10^3$ to nearly $0.6$, 
which is reached by \ibu{} with much fewer data points (at $n=10^3$). With $\geps=0.5$, 
they decline more slowly due to 
the larger amount of noise added by the mechanism in this case. 

\begin{figure}[t]
\centering 
\subfigure[Ages]{
      \label{fig:box_var_data_krr_ages}
      \includegraphics[width=0.23\textwidth]{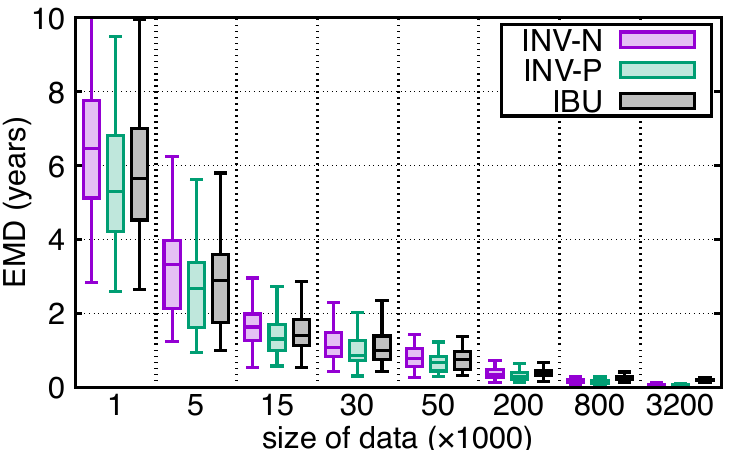}
      }
\subfigure[Locations]{
      \label{fig:box_var_data_krr_locations}
      \includegraphics[width=0.23\textwidth]{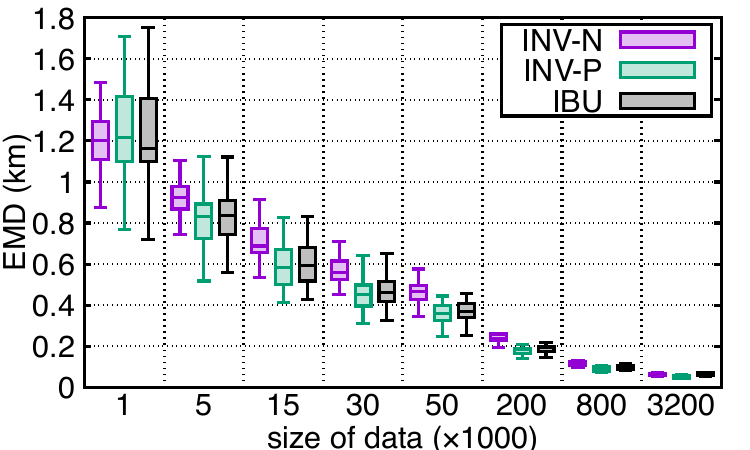}
      }
\caption{The estimation error of \ibu{} algorithm, \invn{}, and \invp{} 
applied to various data sizes of ages (a), and locations (b). 
These data are produced using a $k$-RR mechanism that satisfies 
$3.0$-LDP.}
\label{fig:box_var_data_krr_emd}
\end{figure}

Finally, we repeat the above two experiments, but this time we replace the linear geometric and 
the taxicab mechanisms with
a $k$-RR mechanism that offers $(3.0)$-LDP. Figure \ref{fig:box_var_data_krr_emd} shows the result of this 
experiment for the cases of linear and planar alphabets. It can be seen that the three estimators are 
comparable when
the $k$-RR is used for obfuscation, even with various data sizes. This confirms our earlier results of 
Section \ref{sec:var_priv-krr} in which the estimates were comparable given various privacy levels and a fixed data size.  

\subsection{On the estimation error of \inv{}}\label{sec:inv_error_bounds}
From our experiments, \ibu{} may be seen as a \emph{baseline} for the estimation quality 
of the other estimators. In particular, we noticed that the estimation error of \inv{} 
coincides with this baseline when the $k$-RR mechanism is used (Figures~\ref{fig:box_estimators_krr_emd} 
and \ref{fig:box_var_data_krr_emd}), while the \inv{}'s estimation error is way above the baseline when 
the metric privacy mechanisms are used (Figures~\ref{fig:box_estimators_dprivacy}, 
\ref{fig:box_estimators_geind},\ref{fig:box_var_data_lgeom_emd}, and \ref{fig:box_var_data_taxicab_emd}). 
We recall that the \inv{}'s estimate is obtained by evaluating the vector $\vw = \vq \mech^{-1}$, 
and then post-processing $\vw$ to produce a distribution. 
We note that the mean of $\vw$ is the original distribution 
(See~\autoref{lem:q_unbiased} in Appendix~\ref{sec:proofs_experiments}), 
hence summing the variances of its components yields the 
mean squared error of $\vw$. Using this observation and the inverse matrix $\mech^{-1}$ we show, via the 
following propositions, bounds for this mean squared error in the cases when the $k$-RR, truncated linear geometric, and truncated taxicab mechanisms are used. 
These bounds are independent of the original distribution over $\calx$. 

For $k$-RR, we can derive a relatively small upper bound on the mean squared error of the \inv{} estimate (before post-processing) $\vw$, 
which justifies the good performance of \inv{} in this case.
\begin{restatable}[\inv{} with $k$-RR]{proposition}{invkrrbound}
\label{prop:invKrrBound}
Consider any alphabet $\calx$ of size $k$, and let $\vtheta$ be any original distribution on $\calx$. 
Let $n$ data points be sampled from $\calx$ and then obfuscated by a $k$-RR mechanism with local privacy parameter $\leps$. 
Then the \inv{}-estimated vector $\vw$ satisfies  
$\E[ {\| \vw - \vtheta \|}_2^2 ] \le \frac{1}{n}\left( \frac{e^\leps + k -1}{e^\leps -1} \right)^2$. 
\end{restatable}
For the truncated linear geometric and taxicab mechanisms, on the contrary, 
we derive a relatively large lower bound for the mean squared error of the \inv{} estimate 
(before post-processing), which justifies the poor performance of \inv{} in these cases.

\begin{restatable}[\inv{} with truncated linear geometric]{proposition}{invgeombound}
\label{prop:invTLinGeomBound}
Consider the alphabet $\calx=\{ 0,1,\cdots, k\}$ with spacing $s$, and any original distribution $\vtheta$ over $\calx$.
Let $n$ data be sampled from $\calx$ and then obfuscated by the truncated linear geometric mechanism $\mech:\calx\to\calx$ 
with parameter $0< \lambda \leq 0.94/s$. Let $\alpha = e^{-\lambda s}$ and $\beta = 1/(1- e^{-\lambda s})$.
Then the \inv{}-estimated vector $\vw$ satisfies  $\E[ {\| \vw - \vtheta \|}_2^2 ]\geq \frac{1}{n}(\beta^3 - 2 \alpha \beta^2 - 2)$. 
\end{restatable}
\begin{restatable}[\inv{} with truncated taxicab mechanisms]{proposition}{invTTaxicabBound}
\label{prop:invTTaxicabBound}
For two finite ranges $\calu, \calv$ of integers, consider the planar alphabet $\calx = \calu\times\calv$ with spacing $s$, 
and any original distribution $\vtheta$ over $\calx$. Let $n$ data points be sampled from $\calx$ and then obfuscated by the 
truncated taxicab mechanism $\mech: \calx\to\calx$ with parameter $0< \lambda \leq 1.09/s$. 
Let $\alpha = e^{-\lambda s}$ and $\beta = 1/(1- e^{-\lambda s})$. 
Then the \inv{}-estimated vector $\vw$ satisfies $\E[ {\| \vw - \vtheta \|}_2^2 ]\geq \frac{1}{n}(\beta^5 - 4 \alpha \beta^3 - 4)$. 
\end{restatable}
The error bounds given by \autoref{prop:invTLinGeomBound} and \autoref{prop:invTTaxicabBound} 
hold for broad, practical ranges of privacy. In particular, the constraint $\lambda \leq 0.94/s$ in \autoref{prop:invTLinGeomBound} is satisfied for all values of $\lambda$ in our experiments 
with the truncated linear geometric mechanism ($s=1.0$; Fig.~\ref{fig:box_ages_vlgeom_emd}). 
Similarly, the constraint $\lambda \leq 1.09/s$ in \autoref{prop:invTTaxicabBound}, 
which is equivalent to geo-indistinguishability $\geps \leq 1.54/s$, holds for all levels 
$\geps$ used in our experiments with the truncated taxicab mechanism on location data 
($s=0.5$ km; Fig.~\ref{fig:box_taxicab_emd}).
It is evident from these experiments that for the (weak) privacy levels beyond these 
constraints, both \inv{} and \ibu{} incur only negligible errors.

To see the impact of the given error bounds, consider, 
for instance, the alphabet of ages (described in Section \ref{sec:metric_mechanisms}) which has size $k=100$, and let us 
compare between the \inv{} error when the original data are 
obfuscated using a $k$-RR mechanism with $\leps=2$ and when they are obfuscated instead using a linear geometric mechanism with $\lambda=0.05$. 
These two values ($\leps=2$ and  $\lambda=0.05$) give approximately the same utility for the $k$-RR and linear geometric mechanisms when \ibu{} is used for estimation  
(See Figures \ref{fig:box_ages_krr_emd} and \ref{fig:box_ages_vlgeom_emd}), hence providing a common ground for comparing the two mechanisms. 
However, if we use \inv{} for estimation instead of \ibu{} we find from \autoref{prop:invKrrBound} that in the case of $k$-RR 
mechanism the error is below $227.28/n$ while for the linear geometric mechanism (\autoref{prop:invTLinGeomBound}), the mean error overshoots above $7238/n$. 

Similarly, for the alphabet of locations, we can see from Figures \ref{fig:box_planar_krr_emd}, \ref{fig:box_taxicab_emd} 
that obfuscating the users' real locations using a $k$-RR mechanism with $\leps=2.5$ gives approximately the same utility as
obfuscating those data points using a truncated taxicab mechanism that satisfies $0.2$-geo-indistinguishability (i.e. $\lambda = 0.2/\sqrt{2}$) 
when \ibu{} is used for the estimation in the two cases. However, if \inv{} is used for estimation instead of \ibu{}, we find 
from \autoref{prop:invKrrBound} that the mean error in the case of $k$-RR is below $678.04/n$, while for the taxicab mechanism (\autoref{prop:invTTaxicabBound}), the mean error overshoots above $662648.6/n$. 

While the estimation error of \inv{} is practically reduced by 
post-processing $\vw$ (via normalization or projection), this post-processing is merely 
an ad-hoc correction to obtain a valid distribution, with no guarantee of approximating the MLE. 
Therefore, the resulting estimate is still significantly large compared to \ibu{}. 
\subsection{The impact of the parameters of the \ibu{} algorithm}\label{sec:parameters}
In this section, we experimentally examine how sensitive the outcome of the \ibu{} algorithm is to the choice of the initial distribution $\theta^0$ and the threshold $\delta$, which determines the stopping condition by limiting the difference between two consecutive iterations.
For these experiments, we consider an alphabet of 20 symbols, $\calx=\{0,1,\ldots,19\}$, with unit spacing. We assume that the data follow a binomial distribution with parameter $p_\calx = 0.2$ and we generate a dataset of $10^5$ samples from it. These data points are then obfuscated using the geometric mechanism with different values of $\lambda$ ($0.1, 0.3$, and $0.9$). We apply the \ibu{} algorithm to the obfuscated data and compute the EMD between the estimate and the original true distribution. 
We considered three values of $\delta$ ($1.0$, $10^{-1}$, and $10^{-2}$), 
which are small compared to 
the absolute value of the log-likelihood $L(\cdot)$, whose magnitude scales with $n$, the number of data points (see Equation~\eqref{eq:L_ibu}). Namely, $L(\cdot)$ is the sum of the log-probabilities $L_i(\cdot)$, 
over the individual noisy data points, $i\in[n]$. We also considered 
several starting priors $\theta^0$. Each $\theta^0$ is a binomial distribution, and its parameter $p$ ranges from $0.1$ to $0.8$. For every combination of $\lambda$, $\delta$, and $p$, we repeat the experiment 30 times, always using the same original dataset but reapplying the (random) geometric obfuscation each time.
\begin{figure}[t]
\centering 
\subfigure{
      \label{fig:01-10^-5}
      \includegraphics[width=0.23\textwidth]{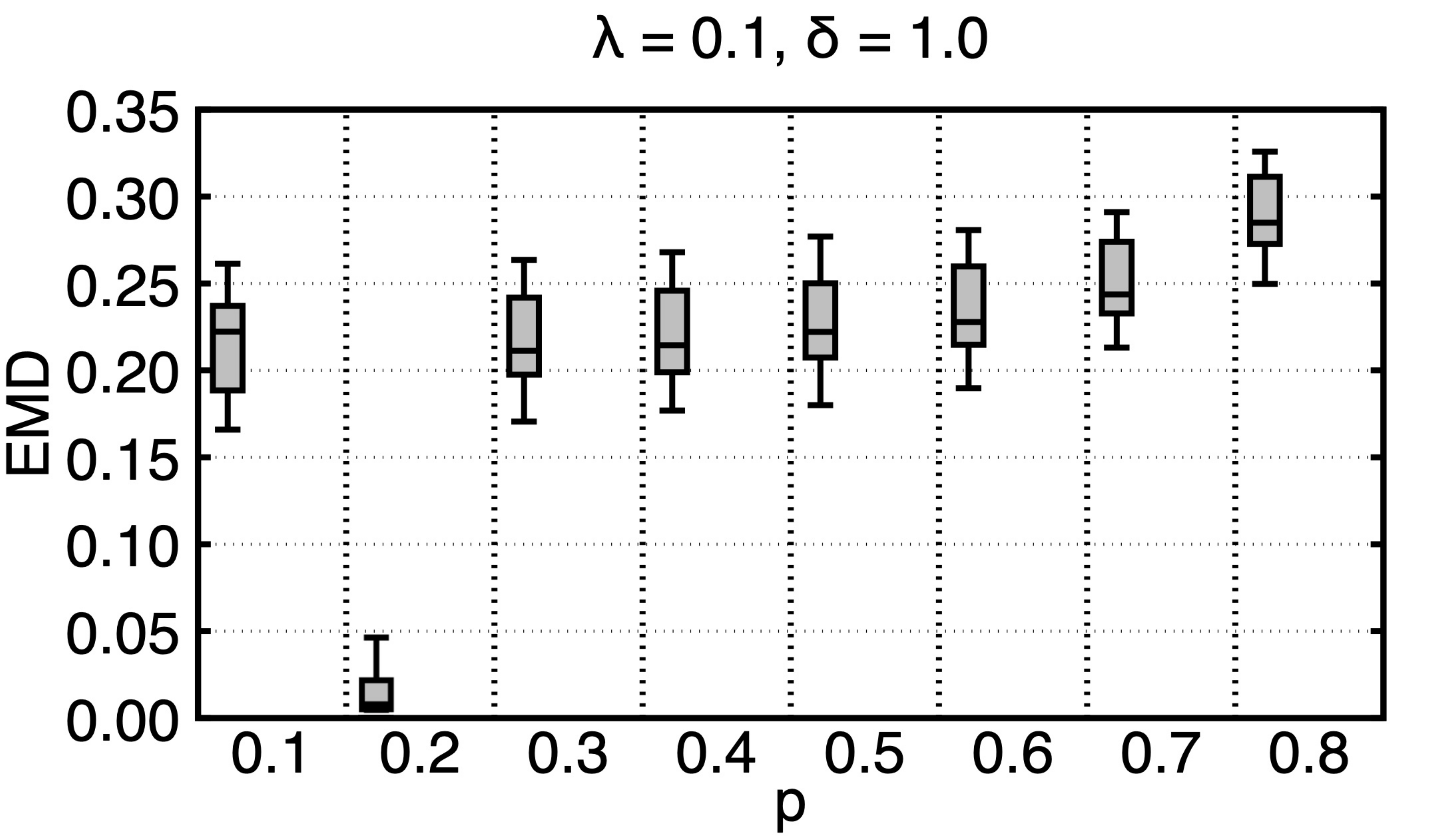}
      }
\vspace{-10pt}
\subfigure{
      \label{fig:0.1-10^-6}
      \includegraphics[width=0.23\textwidth]{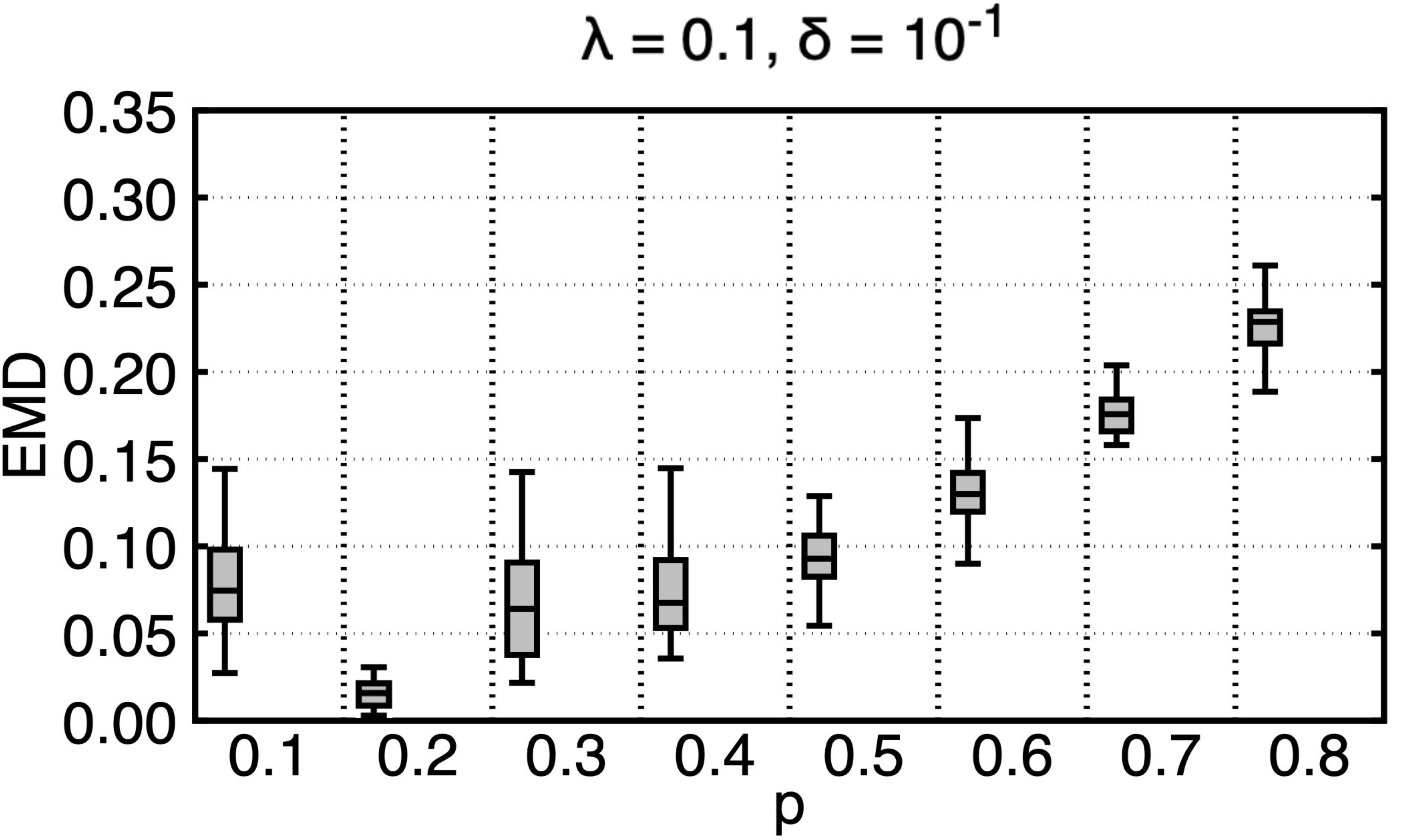}
      }
\vspace{-10pt}
\subfigure{
      \label{fig:0.1-10^-7}
      \includegraphics[width=0.23\textwidth]{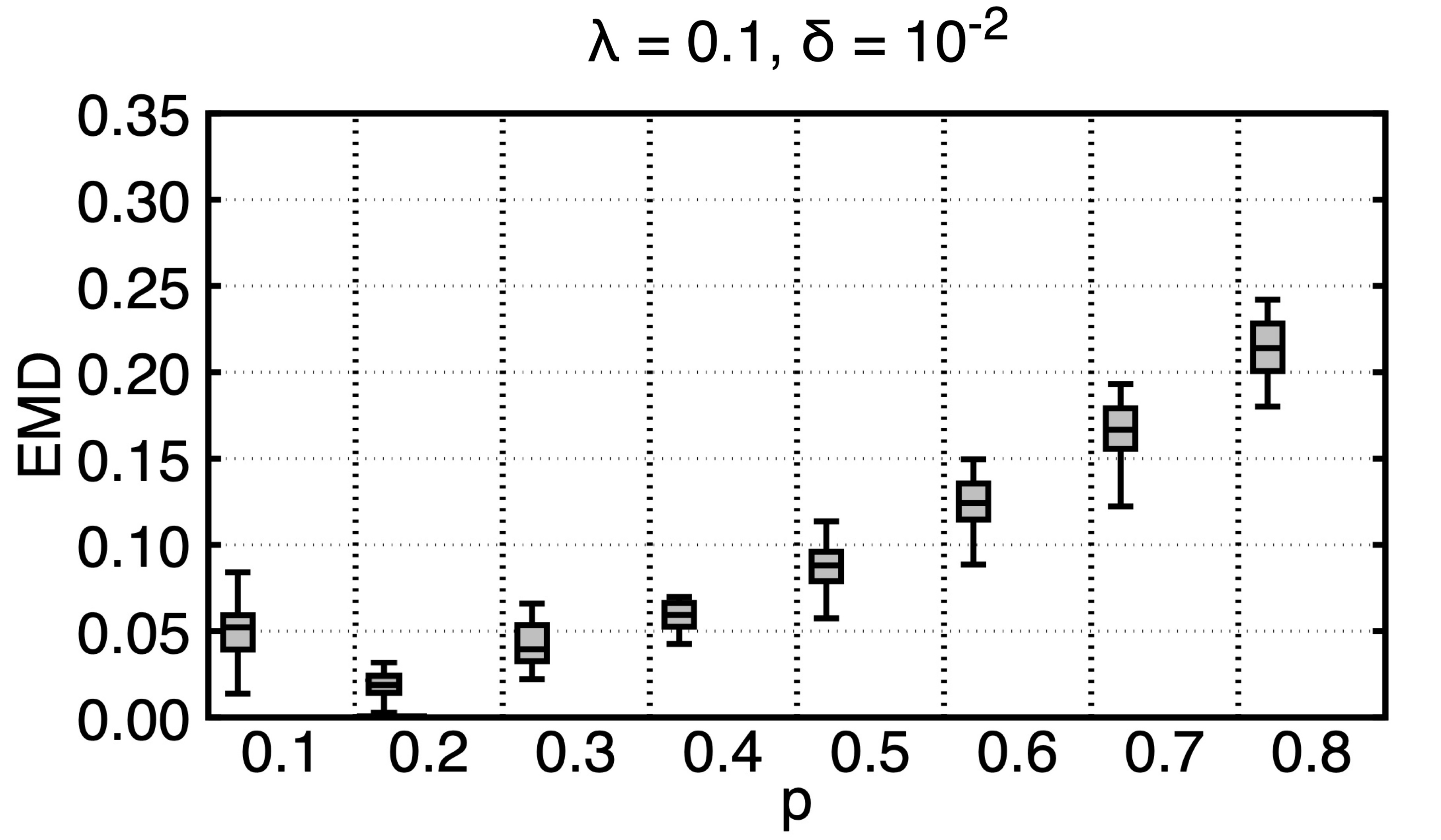}
      }
\subfigure{
      \label{fig:0.3-10^-5}
      \includegraphics[width=0.23\textwidth]{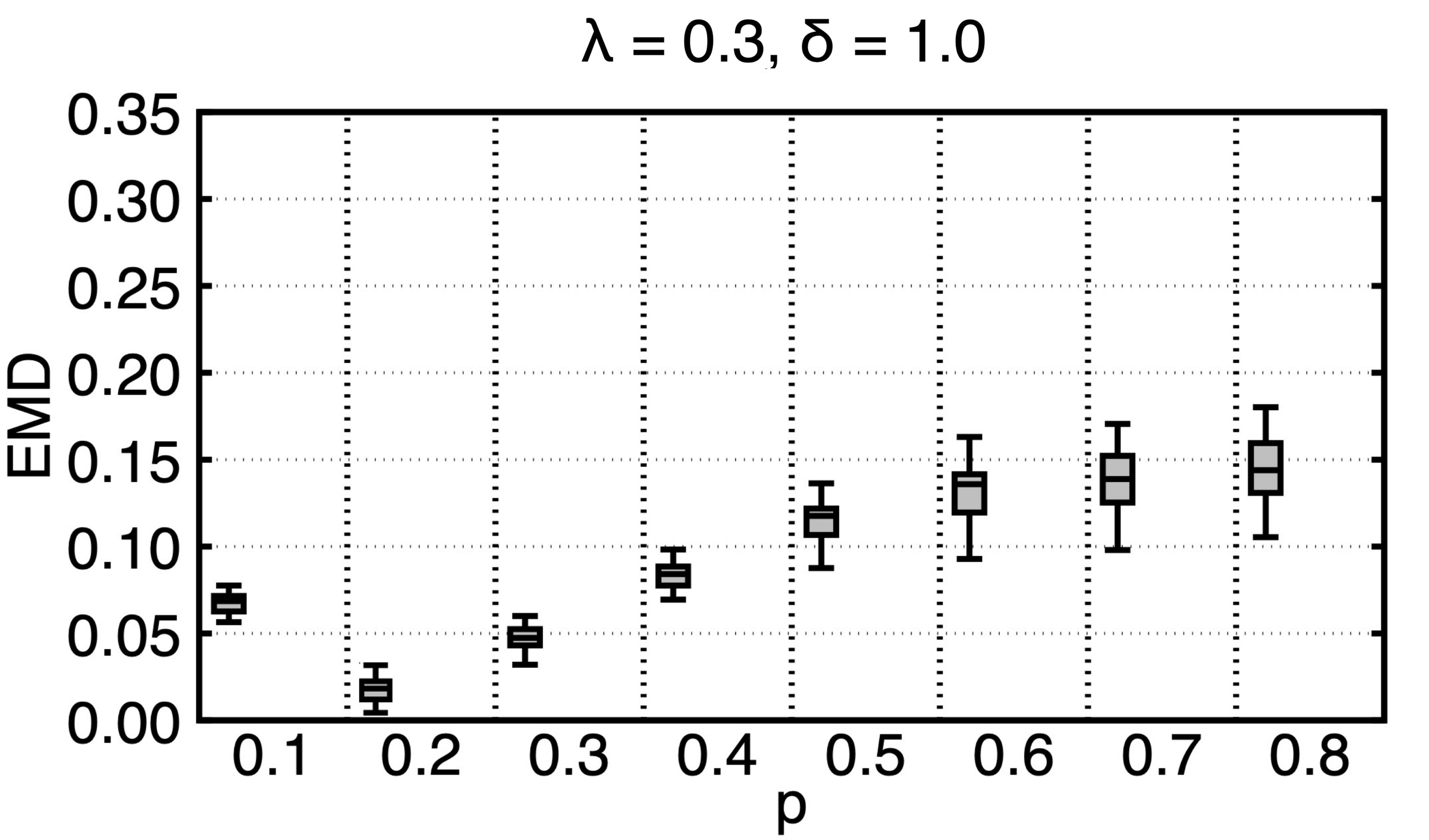}
      }
\vspace{-10pt}
\subfigure{
      \label{fig:0.3-10^-6}
      \includegraphics[width=0.23\textwidth]{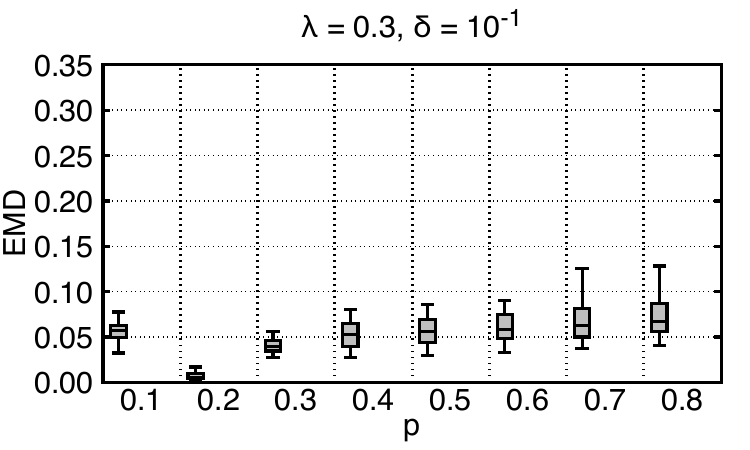}
      }
\subfigure{
      \label{fig:0.3-10^-7}
      \includegraphics[width=0.23\textwidth]{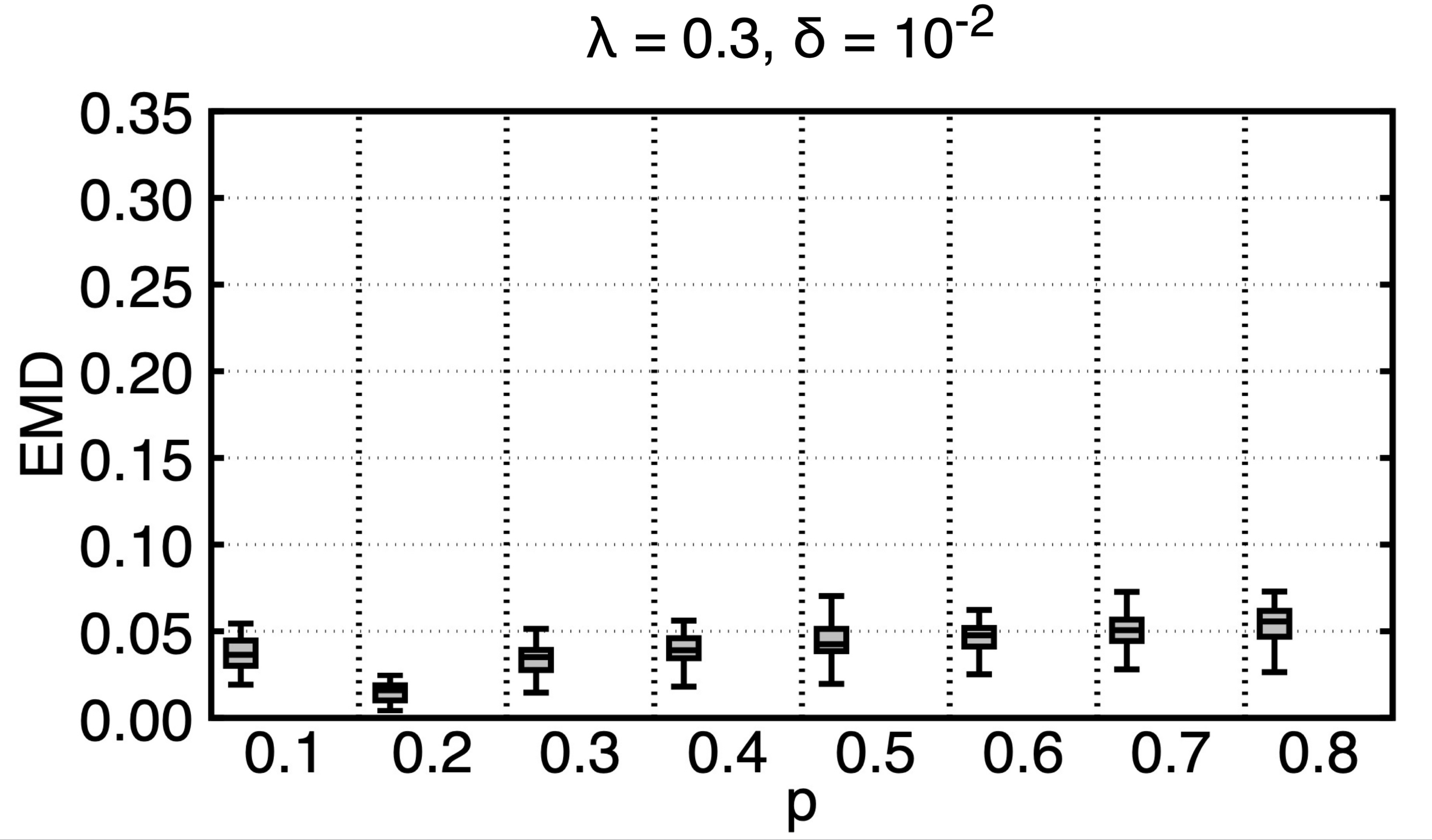}
      }
\vspace{-10pt}
\subfigure{
      \label{fig:0.9-10^-5}
      \includegraphics[width=0.23\textwidth]{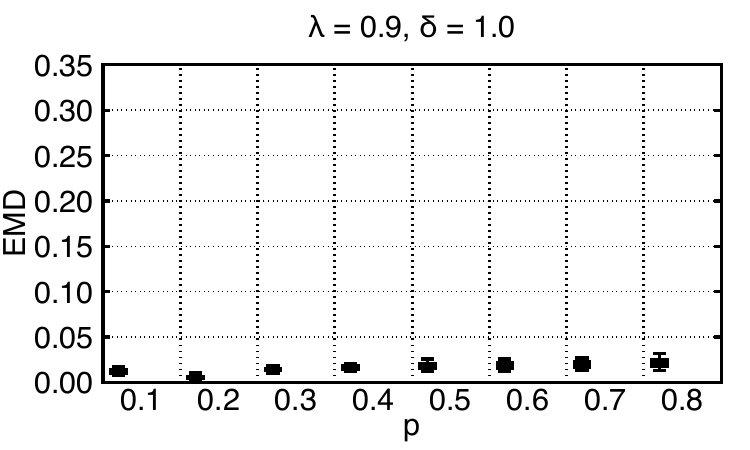}
      }
\subfigure{
      \label{fig:0.9-10^-6}
      \includegraphics[width=0.23\textwidth]{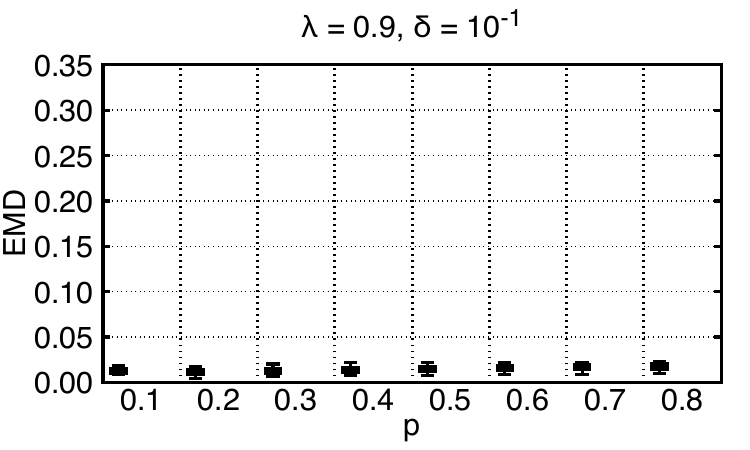}
      }
\caption{The EMD between the the \ibu{} estimate and the true distribution for different parameters $p$ of the starting distribution $\theta^0$.}
\label{fig:EMD_ini_delta}
\end{figure}

The results of the experiments are illustrated in Figure~\ref{fig:EMD_ini_delta}. As expected, the estimates become more precise as $\lambda$ increases (the noise is reduced) and as $\delta$ decreases. The choice of the starting distribution $\theta^0$ also matters: the closer it is to the true distribution, the more accurate the estimate. In particular, when we initialize \ibu{} algorithm with the exact same distribution as the real one ($p = p_\calx = 0.2$), we clearly achieve higher precision, especially in the high privacy regime (small $\lambda$) and large $\delta$. However, as $\delta$ decreases and $\lambda$ increases, the influence of the choice of $\theta^0$ gradually fades. This is important because we assume no prior knowledge of the true distribution, and therefore have no guidance on how to choose the starting point.

\begin{figure}[t]
\centering 
\subfigure{
      \includegraphics[width=0.23\textwidth]{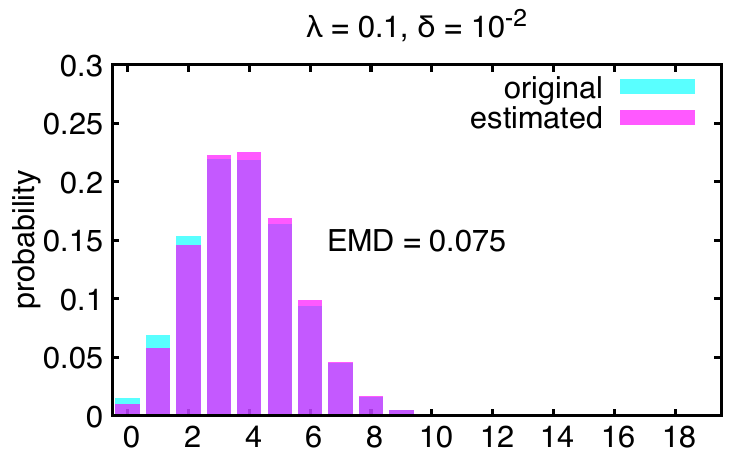}
      }
\subfigure{
      \includegraphics[width=0.23\textwidth]{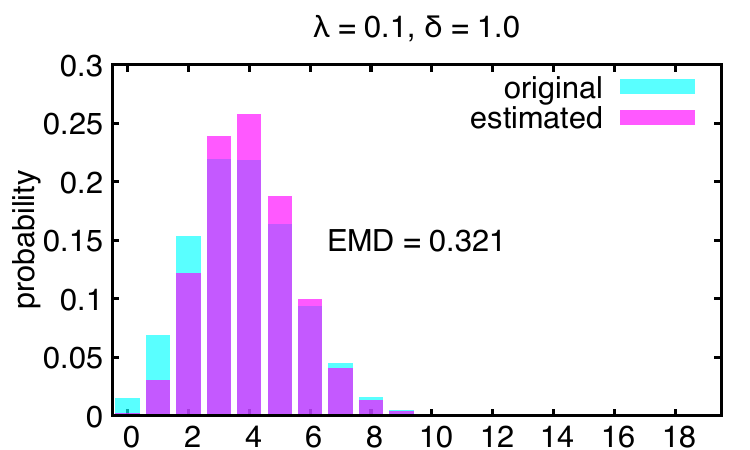}
            }
\caption{Two \ibu{} estimates with different EMD from the true distribution. In both plots, the true distribution is a binomial with parameter $p_\calx=0.2$.}
\label{fig:distributions}
\end{figure}

We remark that, although the differences in estimate accuracy for different parameters may look substantial in Figure~\ref{fig:EMD_ini_delta}, they are in fact rather minor, since all corresponding EMDs reflect only a small divergence. For instance, for $\lambda = 0.1$, one of the best configurations in our experiments is $ \delta = 10^{-2}$ and $p=0.3$, where a particular run produced EMD $= 0.075$, whereas one of the worst configurations is $\delta = 1.0$ and $p=0.8$, where a particular run produced EMD $= 0.321$ (almost $5$ times larger!). If in the first case the estimate is almost perfect, in the latter case it is still reasonably precise, see Figure~\ref{fig:distributions}. This suggests that, in the context of our experiments, \ibu{} does not seem to be highly sensitive to the specific choice of parameters.

It is important to highlight that the number of iterations performed by the \ibu{} algorithm can fluctuate significantly depending on the selected parameters. For instance, for $\lambda = 0.1$ and $\delta = 10^{-2}$, the algorithm required only $10$ iterations when initialized with a $\theta^0$ with $p=0.2$, but it took $296$ iterations when started from a $\theta^0$ with $p=0.8$. This suggests that imposing a fixed upper bound on the number of iterations would not constitute a dependable stopping criterion.

\subsection{On the computational cost of \ibu{} algorithm} and \inv{}
\ibu{} and \inv{} also differ significantly in computational cost. The \ibu{} algorithm 
is an iterative procedure whose 
runtime depends on the sizes of the input and output alphabets, $\calx, \calz$ of the mechanism, 
as well as the number of iterations $I$ required for convergence, yielding a total complexity of $O(|\calx|\,|\calz|\,|I|)$ (see \autoref{alg:ibu} -- note that the denominator appearing in the algorithm’s update rule can be precomputed once for each $z\in\calz$).
In contrast, \inv{} is a one-shot estimator, which relies 
on multiplying the noisy data distribution $\vq$ by the 
mechanism matrix inverse $\mech^{-1}$, yielding a complexity 
of $O(|\calx|^2)$. The postprocessing operations on \inv{} do not increase the complexity, because the projection is  $O(|\calx| \log |\calx|)$, and the normalization is $O(|\calx|)$. Consequently, \inv{} is substantially faster in practice, 
while \ibu{} offers greater flexibility and accuracy at the expense of higher computational overhead.

\begin{figure}[t]
\footnotesize
\setlength{\tabcolsep}{2pt} 
\begin{tabular}{@{} c c}
 $\delta=10^{-2}$ & $\delta=10^{-3}$\\[0em]
\mysubfig[]{fig:time-10^-2}{0.23\textwidth}{varying_alphabet/varying_alphabet_time_th0.01.pdf}&
\mysubfig[]{fig:time-10^-3}{0.23\textwidth}
{varying_alphabet/varying_alphabet_time_th0.001.pdf}\\
\mysubfig[]{fig:distance-10^-2}{0.23\textwidth}
{varying_alphabet/time_vs_distance_th0.01.pdf}&
\mysubfig[]{fig:distance-10^-3}{0.23\textwidth}
{varying_alphabet/time_vs_distance_th0.001.pdf}
\end{tabular}
\caption{Execution time of \invn{}, \invp{}, and \ibu{} algorithm. 
The true distribution is a binomial with $p_\calx=0.5$, and \ibu{} 
is initialized with the uniform distribution.}
\label{fig:time_alphabet_delta}
\end{figure}

We complete the analysis of the computational cost of the \ibu{} algorithm and \inv{} 
by performing some experiments. 
We consider alphabets varying in size $k$ from 
$100$ to $800$, and a binomial distribution on the data with parameter $p_\calx = 0.5$. 
We collect $10^5$ samples, obfuscate them using a geometric mechanism with parameter $\lambda\in\{0.01,0.03,0.05\}$, 
and finally run \ibu{} algorithm and \inv{} for estimation.  
For \ibu{} we use the uniform distribution for initialization and 
$\delta\in\{10^{-2}, 10^{-3}\}$ for termination.
We repeat the experiment $30$ times.
The experiments are performed using our C++ implementation, executed on 
a MacBook Pro with an Apple M2 processor (8-core CPU) and 24 GB of RAM, running macOS.

The first row of Figure~\ref{fig:time_alphabet_delta} illustrates the execution time of 
\ibu{} for each combination of $k$, $\delta$ and $\lambda$, 
and also \inv{} whose time depends mainly on the alphabet size $k$. 
As expected, the execution time of the \ibu{} algorithm is larger than that of \inv{}, 
and both times increase with the size of the alphabet. 
Moreover, the execution time of \ibu{} increases as $\delta$ decreases, 
because the stopping condition is met later. Conversely, it decreases as $\lambda$ increases, because the empirical data more closely match the ``ideal'' output distribution, which speeds up convergence.
To complement the execution-time comparison, the second row of Figure~\ref{fig:time_alphabet_delta} 
plots the execution time against the estimation error, measured by the EMD, for each run in the above experiment. The results clearly illustrate the trade-off between computational efficiency and estimation accuracy: while the \ibu{} algorithm consistently produces more accurate estimates than \invn{} and \invp{}, this improvement comes at the cost of higher computational overhead.

\section {Obtaining MLEs for infinite alphabets}\label{sec:infinite_alphabet}
In many situations, the alphabet $\calx$ can be very large or even infinite.
This alphabet may correspond, for example, to a linear attribute such as the annual income, 
internet usage, or others for which we don't have specific bounds. In this case, 
$\calx$ can be described as a discretized version of $\reals^+$, which is infinite. 
The collected data, however, are always confined within a bounded range.  
Another example is an online location-based service that wants to estimate the distribution of 
visited locations in a large geographic region based on sanitized check-ins collected from individuals. 
Then $\calx$ may be seen as a grid covering the entire region, hence consisting of a large number of cells. 
The collected (sanitized) check-ins may, however, be situated in a few cells or within a small area inside this region. 

Applying a privacy mechanism on a large alphabet, such as those in the above examples, 
is relatively simple: there are mechanisms such as the geometric, Laplace, and taxicab mechanisms, 
which can be applied in the above cases because they work on both finite and infinite alphabets 
of sensitive data. However, applying \ibu{} to estimate the original distribution in these cases 
is not practical because in every iteration $t$, an estimate $\theta^t_x$ for every $x\in\calx$ 
has to be computed (See~\autoref{alg:ibu}). 
In this section, we address this problem by showing that, given a privacy mechanism and some noisy data, 
we can obtain from the original large alphabet a smaller subset which we refer to as ``likely subset''. 
As shown later via \autoref{thm:likelysubset}, computing an MLE on this likely subset (via \ibu{}) provides 
the required MLE on the original alphabet $\calx$. We will derive the likely set for linear and planar alphabets 
when particular privacy mechanisms are used (\autoref{prop:likelylinear}, \autoref{prop:likelyplanar}, and \autoref{prop:likelykrr}).

\subsection{Likely subsets of the alphabet}
Consider noisy data produced by a mechanism $\mech$. 
Given these data and the mechanism, we call an element $x'\in\calx$ 
\emph{unlikely} if there exits another element $x\in\calx$ that is at least as likely as $x'$ 
to produce every observation in the noisy data, and \emph{strictly more} likely than $x'$ 
for one of these observations. 
In terms of the conditional probabilities of $\mech$, this can be formalized as follows. 
\begin{definition}[unlikely element]\label{def:unlikely}
Consider a privacy mechanism $\mech$ that works on an alphabet $\calx$. Consider also a set of observations $\{z^i: i\in[n]\}$ 
produced by $\mech$. Then an element $x' \in \calx$ is unlikely for the mechanism 
and the observations if there is another $x \in \calx$ such that $\mech_{x' z^i} \leq \mech_{x z^i}$ for all $i \in [n]$, and this 
inequality is strict for some $i$. 
\end{definition}
From the above definition, whether an element $x'$ is unlikely depends on 
the observed noisy data 
and the mechanism that produced them. The following lemma shows that 
any MLE on $\calx$ must assign probability $0$ to the unlikely elements. 
\begin{restatable}{lemma}{unlikely}
\label{lem:unlikely}
Consider a mechanism that works on an alphabet $\calx$ and consider a set of observations. 
If $\hat\vtheta$ is an MLE on $\calx$ with respect to the mechanism and the observations, 
then it must hold that $\hat{\theta}_{x'}=0$ for every unlikely $x'\in\calx$. 
\end{restatable}
By removing any number of unlikely elements from $\calx$, we obtain a smaller set which we refer to as a \emph{likely} subset of $\calx$ 
and denote by $\hat\calx$. More formally, we define the likely subsets of $\calx$ as follows. 
\begin{definition}[likely subsets]\label{def:likely}
Consider a privacy mechanism that works on an alphabet $\calx$, and consider a set of observations. 
Then any subset $\hat\calx \subseteq \calx$ is likely with respect to the mechanism and observations if 
every $x' \in (\calx \setminus \hat\calx)$ is unlikely. 
\end{definition}
Likely subsets of $\calx$ are useful because the required MLEs on the large $\calx$ can be obtained more 
efficiently by computing the MLE only on a (smaller) likely subset of $\calx$ as shown by the following theorem.  
\begin{restatable}{theorem}{likelysubset}
\label{thm:likelysubset} 
Consider a mechanism that works on an alphabet $\calx$ and consider a set of observations. 
Let $\hat\calx$ be a likely subset of $\calx$. 
Then, with respect to the mechanism and observations, 
a distribution $\hat{\vtheta}$ is an MLE on $\calx$ if and only if 
$\hat{\theta}_x=0$ for all $x \not\in \hat\calx$, and the distribution $(\hat\theta_x : x\in \hat\calx)$
is an MLE on $\hat\calx$. 
\end{restatable}
\autoref{thm:likelysubset} establishes an equivalence between the MLEs on $\calx$ and 
those on the smaller likely subset $\hat\calx$. This equivalence allows us to restrict the computation of 
the MLE to $\hat\calx$ with no loss of generality. In the following, we describe typical combinations of 
the alphabets and mechanisms, and we identify in each case a finite likely subset $\hat\calx$. 
\subsection{Linear alphabets}
To satisfy a reasonable trade-off between privacy and utility, many mechanisms that work on linear numeric
alphabets generate an observable $z$ from a true value $x$ with a probability depending 
on the difference between $x$ and $z$. Specifically, for a mechanism $\mech: \calx\to\calz$, 
with $\calx,\calz \subset \reals$, 
the above property means that the closer $x,z$ are, the larger $\mech_{xz}$ is. 
That is for any $x,x'\in \calx$ and an observable $z\in\calz$, we have 
$\mech_{x z} > \mech_{x' z}$ whenever $| x - z | < | x' - z |$.
Many classical mechanisms such as the linear geometric~\cite{Ghosh:09:STOC}~\eqref{eq:linearGeometric},  
Laplace~\cite{Dwork:06:TCC}, and exponential~\cite{mcsherry:2007} mechanisms satisfy this property. 
Given a mechanism $\mech$ of this class and a set of observations $\{z^i: i\in [n]\}$, we identify via the following proposition a likely subset of $\calx$.
Define $x_\textit{min} = \sup \{x: x\in \calx, x \leq z^i, i \in[n] \}$ and $x_\textit{max} = \inf \{x: x\in \calx, x \geq z^i, i \in[n] \}$. 
\begin{restatable}[A linear likely subset]{proposition}{likelylinear}
\label{prop:likelylinear}
Suppose that $\calx \subset \reals, \calz \subset \reals$. Consider the mechanism $\mech:\calx\to\calz$ 
that satisfies $\mech_{x z} > \mech_{x' z}$ whenever $| x - z | < | x' - z |$ for all $x,x'\in\calx$ and $z\in\calz$. 
Consider also the observations $\{z^i : i\in[n] \}$ produced by $\mech$. 
Then $[x_\textit{min},x_\textit{max}] \cap \calx$ is a likely subset of $\calx$. 
\end{restatable}
The above proposition is demonstrated in Figure~\ref{fig:convex_likely_linear}.  
The numbers $\{z^1, \dots, z^6\}$ are reported by a mechanism that satisfies the condition of 
Proposition \ref{prop:likelylinear}. The points $x_\textit{min}$ and $x_\textit{max}$ which 
are elements of $\calx$ are also shown. Note that they are respectively lower and upper bounds 
for $z^i$. Then the likely subset given by \autoref{prop:likelylinear} consists exactly of $x_\textit{min}, x_\textit{max}$ 
and all elements of $\calx$ between them. 
To justify this, consider any $x'\in\calx$ that is larger than $x_\textit{max}$. 
Then $x'$ is clearly farther away from every reported $z^i$ compared to $x_\textit{max}$. 
This makes $x'$ unlikely in the sense of Definition \ref{def:unlikely}, compared to $x_\textit{max}$, to produce any 
observable $z^i$. A similar argument holds for all elements of $\calx$ that are smaller than $x_\textit{min}$ 
making them unlikely too. 
\begin{figure}
\centering
\begin{tikzpicture}
\node (mininf) at (-1, 0) {$-\infty$};
\node (plusinf) at (7, 0) {$+\infty$};
\coordinate [label={above:$z^1$}] (z1) at (3, 0);       
\coordinate [label={above:$z^2$}] (z2) at (4, 0);       
\coordinate [label={above:$z^3$}] (z3) at (2, 0);       
\coordinate [label={above:$z^4$}] (z4) at (2.5, 0);    
\coordinate [label={above:$z^5$}] (z5) at (0.5, 0);    
\coordinate [label={above:$z^6$}] (z6) at (1, 0);       

\fill (z1)  circle[radius=1.5pt];
\fill (z2)  circle[radius=1.5pt];
\fill (z3)  circle[radius=1.5pt];
\fill (z4)  circle[radius=1.5pt];
\fill (z5)  circle[radius=1.5pt];
\fill (z6)  circle[radius=1.5pt];

\def\xshift{0.75}

\coordinate [label={below:$x_\textit{min}$}] (xmin) at (0, 0);   
\coordinate [label={below:$x_\textit{max}$}] (xmax) at ($(xmin)+(6*\xshift,0)$);   
\coordinate [label={below:$x'$}] (xprime) at (5.5, 0);       

\draw[thick] ($(xmin)+(0,0.15)$) -- +(0,-0.30);
\node[scale=1.0] at ($(xmin)+(1*\xshift,0)$) {+};
\node[scale=1.0] at ($(xmin)+(2*\xshift,0)$) {+};
\node[scale=1.0] at ($(xmin)+(3*\xshift,0)$) {+};
\node[scale=1.0] at ($(xmin)+(4*\xshift,0)$) {+};
\node[scale=1.0] at ($(xmin)+(5*\xshift,0)$) {+};
\draw[thick] ($(xmax)+(0,0.15)$) -- +(0,-0.30);
\node[scale=1.0] at (xprime) {+};

\draw[dashed,thick,->] (xmin) -- (mininf); 
\draw[thick] (xmin) -- (xmax) ;
\draw [dashed,thick] (xmax) -- (xprime);
\draw [dashed,thick,->] (xprime) -- (plusinf);

\end{tikzpicture}
\caption{The plot of observations $\{z^1,\dots, z^6\}$ generated by a mechanism that satisfies \autoref{prop:likelylinear}.
Note that $| x_\textit{max} - z^i | < | x' - z^i |$ for every $x'  > x_\textit{max}$, 
and $| x_\textit{min} - z^i | < | x' - z^i |$ for every $x'  < x_\textit{min}$. 
Then $[x_\textit{min} , x_\textit{max}] \cap \calx$ is a likely subset.} 
\label{fig:convex_likely_linear}
\end{figure}
\subsection{Planar alphabets}
We consider the case when the sensitive data points of the users are their locations 
in the two-dimensional (planar) space. 
In this case, the alphabet $\calx$ consists of discrete points approximating 
the planar space $\reals^2$. A typical way to define $\calx$ is to consider a grid as in 
Figure~\ref{fig:likely_planar:1} and let $\calx$ be the set of the centers of the grid cells. 
A variety of mechanisms may be used to obfuscate locations, such as the
planar Laplace~\cite{Andres:13:CCS}, planar geometric~\cite{Chatzikokolakis:17:POPETS}, 
and the taxicab (Section~\ref{sec:taxicab}), 
which work on infinite $\calx$, and the Tight-Constrained and exponential 
mechanisms~\cite{Chatzikokolakis:17:POPETS} which work on finite but arbitrarily large $\calx$.  
These mechanisms tend to report points that are relatively close to the real location in a similar 
sense to the property stated in \autoref{prop:likelylinear}. Based on this 
feature, we identify a bounded likely subset of $\calx$ as follows. 
Depending on the specification of $\calx$, we define the discretization error $\delta$ 
to be the maximum distance between any point in $\reals^2$ and its nearest point in 
$\calx$. For instance in Figure~\ref{fig:likely_planar:1}, $\delta = h/\sqrt{2}$ where 
$h$ is the width of each cell. From the reported points $\{z^i: i\in[n]\}$, we also define
\begin{equation}\label{eq:extension}
\delta' = (\delta^2 + 2\delta\,\max_{i,j \in [n]} \dist(z^i,z^j))^{1/2},
\end{equation}
where $\dist(\cdot,\cdot)$ denotes the Euclidean distance between two points. 
Let $\calb$ be the convex hull of the reported points $\{z^i : i\in[n]\}$, and 
$T_{\delta'}(\calb)$ be a larger region constructed by extending $\calb$ in all directions 
by distance $\delta'$ as shown in Figure~\ref{fig:likely_planar:2}. 
More formally 
\[
T_{\delta'}(\calb)=  \{v \in \reals^2: \exists u\in \calb, \dist(v,u)\leq \delta' \}.
\] 
Then $T_{\delta'}(\calb)\cap\calx$ is a likely subset of $\calx$ as follows.  
\begin{restatable}[A planar likely subset]{proposition}{likelyplanar}
\label{prop:likelyplanar}
Suppose that $\calx \subset \reals^2$, $\calz \subset \reals^2$. Consider the mechanism
$\mech: \calx \to \calz$ that satisfies $\mech_{x z} > \mech_{x' z}$ whenever $\dist(x, z) < \dist(x',z)$ 
for every $x,x'\in \calx$ and $z\in\calz$. Let $\calb\subset\reals^2$ be the convex hull of 
the reported points $\{z^i: i\in[n]\}$.
Then $T_{\delta'}(\calb) \cap \calx$ is a likely subset of $\calx$.  
\end{restatable}

To demonstrate the above proposition, suppose that the real locations of the users are 
obfuscated to yield the points shown in Figure~\ref{fig:likely_planar:2}. 
Then the inner region $\calb$ is a convex hull that contains all 
observed points. The outer region is $T_{\delta'}(\calb)$ which results from 
extending $\calb$ with distance $\delta'$, given by~\eqref{eq:extension}, in all directions.  

\begin{figure}
\newcommand\cellside{0.3}  
\FPeval{\del}{round(cellside/(2^0.5),2)}  
\centering
\subfigure[The points of $\calx$. Here $\delta=\del$ km.] 
{
\begin{tikzpicture}[scale=0.8]
\draw[step=1,gray,very thin] (-0.5,-0.5) grid (1.5,1.5);
\foreach \px in {-0.5,0.5,1.5}  \foreach \py in {-0.5,0.5,1.5}  \node[scale=0.7] at (\px,\py) {+};  
\draw[dashed,<->] (0,2) -- node [above] {\footnotesize$h=\cellside$km} (1,2);
\draw[dashed,<->] (0,0) -- node [above, sloped, xshift=2pt,yshift=-2pt] {\footnotesize $\delta$} (0.5,0.5);
\draw[dotted,thick] ($(-0.5,-0.5) - (0.2,0.2)$) -- +(-0.2,-0.2);   
\draw[dotted,thick] ($(-0.5,0.5) - (0.2,0)$) --  +(-0.2,0);
\draw[dotted,thick] ($(-0.5,1.5) - (0.2,-0.2)$)--  +(-0.2,+0.2);
\draw[dotted,thick] ($(1.5,-0.5) + (0.2,-0.2)$) -- +(0.2,-0.2);   
\draw[dotted,thick] ($(1.5,0.5) + (0.2,0)$) --  +(0.2,0);
\draw[dotted,thick] ($(1.5,1.5) + (0.2,+0.2)$)--  +(0.2,+0.2);
\draw[dotted,thick] ($(0.5,-0.5) + (0,-0.2)$) -- +(0,-0.2);   
\draw[dotted,thick] ($(0.5,1.5) + (0,0.2)$) -- +(0,0.2);   
\end{tikzpicture}
\label{fig:likely_planar:1}
}
\hspace{2pt}
\newcommand{\dmaxval}{8.25}   
\FPeval{\delp}{round(  (del*del + 2*del*dmaxval)^0.5,2)}     
\subfigure[The inner region $\calb \subset \reals^2$ contains observed points, and $\dmax = \dmaxval$ km. 
Then $\delta'= \delp$ km and $T_{\delta'}(\calb)$ is shown.] 
{
\begin{tikzpicture}[scale=0.3]
\draw[step=\cellside,gray,very thin] (-2,-2) grid (10,6);
%
\def\drawArcBetweenPoints[#1](#2)(#3)(#4)  
{
 \tkzFindSlopeAngle(#3,#2) \tkzGetAngle{sA};    
 \tkzFindSlopeAngle(#3,#4) \tkzGetAngle{fA};      
 \tikzmath{ if (\sA>\fA) then {\sA=\sA-360;}; } 
 \draw[#1] (#2) arc (\sA: \fA: \delp);   
 }

\coordinate (z1) at (0, 0);      
\coordinate (z2) at (6, 0);      
\coordinate (z3) at (8, 2);       
\coordinate  (z4) at (5, 4);       
\coordinate  (z5) at (5, 3);       
\coordinate  (z6) at (1, 3);       
\coordinate  (z7) at (2, 2.4);    

\foreach \p in {(1,1),(2,2),(1.5,2), (3,2),(3,0.2),(4,2.5),(4,3),(6,3),(6,2),(5,0.5),(5.5,0.5) } \fill \p circle[radius=3pt];

\draw[dashed,<->] (z1)  -- node [above, sloped] {\footnotesize $d_{\textit{max}}$}  (z3);   
\fill (z1)  circle[radius=3pt];
\fill (z2)  circle[radius=3pt];
\fill (z3)  circle[radius=3pt];
\fill (z4)  circle[radius=3pt];
\fill (z5)  circle[radius=3pt];
\fill (z6)  circle[radius=3pt];
\fill (z7)  circle[radius=3pt];
\draw[dashed,thick]  (z1) -- (z2) --(z3) --(z4) --node [below,yshift=1pt] {$\calb$} (z6) -- cycle;

\path (z6) -- (z1) -- ([turn] -90:\delp) coordinate (x1);
\path (z2) -- (z1) -- ([turn] 90:\delp) coordinate  (x2); 
\drawArcBetweenPoints[thick](x1)(z1)(x2)

\path (z1) -- (z2) -- ([turn] -90:\delp) coordinate (x3);
\path (z3) -- (z2) -- ([turn] 90:\delp) coordinate  (x4); 
\drawArcBetweenPoints[thick](x3)(z2)(x4)

\path (z2) -- (z3) -- ([turn] -90:\delp) coordinate (x5);
\path (z4) -- (z3) -- ([turn] 90:\delp) coordinate (x6); 
\drawArcBetweenPoints[thick](x5)(z3)(x6)

\path (z3) -- (z4) -- ([turn] -90:\delp) coordinate (x7);
\path (z6) -- (z4) -- ([turn] 90:\delp) coordinate (x8); 
\drawArcBetweenPoints[thick](x7)(z4)(x8)

\path (z4) -- (z6) -- ([turn] -90:\delp) coordinate (x9);
\path (z1) -- (z6) -- ([turn] 90:\delp) coordinate (x10); 
\drawArcBetweenPoints[thick](x9)(z6)(x10)

\draw[thick] (x2)--(x3);
\draw[thick] (x4)--(x5);
\draw[thick] (x6)--(x7);
\draw[thick] (x8)--(x9);
\draw[thick] (x10)--(x1);

\coordinate (ex_p) at ($(z6)!1.7!(z4)$);
\draw[dashed] (z4)--(ex_p);
\draw[dashed] (x9) -- ($(x9)!1.6!(x8)$); 
\draw[dashed,<->] (ex_p) -- node [right] {\footnotesize $\delta'$} ($(x9)!(ex_p)!(x8)$);  
\end{tikzpicture}
\label{fig:likely_planar:2}
}
\caption{An infinite planar alphabet $\calx$ is shown in (a). A (finite) likely subset of $\calx$ is $T_{\delta'}(\calb) \cap \calx$ 
which is bounded by the outer region in (b). 
} 
\label{fig:likely_planar}
\end{figure}

\subsection{Alphabets under k-RR mechanisms}
Consider an alphabet $\calx$ that is finite but arbitrarily large. 
Suppose that a $k$-RR mechanism~\eqref{eq:krr}
is used to sanitize the users' original values from $\calx$. Recall that both the inputs and outputs of the $k$-RR are elements of $\calx$. 
Since $\calx$ is finite, \ibu{} may be used to compute an MLE directly for $\calx$. 
However, this would be costly when $\calx$ is large, since \ibu{} has to estimate the 
probability of each $x\in\calx$. 
By \autoref{thm:likelysubset}, the required estimate can be obtained more efficiently 
by running \ibu{} only on a likely subset, which turns out in this case to be exactly the set of reported observations. 

\begin{restatable}[likely subset under $k$-RR]{proposition}{likelykrr}
\label{prop:likelykrr}
Consider any finite alphabet $\calx$, and suppose that a $k$-RR mechanism is used 
to obfuscate the users' data from $\calx$ to produce the observations $\calx' = \{z^i : i\in[n]\}$. 
Then $\calx'$ is a likely subset of $\calx$. 
\end{restatable}

Suppose, for example, that $\calx$ consists of the cells that cover a certain geographic region, 
and the users apply a $k$-RR mechanism to sanitize their locations (cells). Then, according to \autoref{prop:likelykrr}, the likely set of $\calx$ consists only of the reported cells. 
If the original users' check-ins do not occupy the entire region, for example, due to the existence of lakes, mountains, etc., the noisy dataset may be distributed over only a small number of cells, 
allowing us to use \ibu{} with a lower computational cost.  

\section{Conclusion and future work}\label{sec:conclusion}
In this paper, we analyzed the consistency of the iterative Bayesian update (\ibu{}). 
In other words, we precisely described the conditions that assure the convergence of 
\ibu{}'s estimate to the original distribution. These are the identification, 
dominance, and other regularity conditions. 
We characterized the `identification' in terms of the mechanism matrix, and proved that it 
is equivalent to the strict concavity of the log-likelihood function for some noisy data. 
We also characterized the strict concavity, for a given mechanism and noisy 
data, and showed that it implies (but is not equivalent to) the uniqueness of the \ibu{}'s estimate. 
We showed that $k$-RR, linear geometric, taxicab, and \textsc{Rappor}'s mechanisms satisfy 
the consistency conditions.  
We also presented a mechanism that violates these conditions, and we showed that \ibu{}'s 
estimate does not converge to the real distribution in this case.  
We experimentally compared \ibu{} and other estimators concerning estimation performance 
and showed that \ibu{} is significantly superior for metric privacy mechanisms, and gave 
a mathematical reason for the suboptimal performance of \inv{} in this case. 
Finally, we extended the 
application of \ibu{} to the case when the alphabet of the sensitive attribute is unbounded. 

In future work, we plan to inspect the consistency of \ibu{} when the other mechanisms are used, 
such as the planar geometric and exponential ones, and also the one resulting from the Blahut-Arimoto 
construction algorithm~\cite{Oya:17:CCS}, which satisfies geo-indistinguishability. 
In addition, we plan to inspect the consistency and estimation performance of \ibu{} when the 
users apply different mechanisms, the situation in which we use an extension of \ibu{} called \gibu{}, which 
was presented in \cite{ehab:eurosp:2020}. 

\bibliographystyle{ieeetran}
\bibliography{short}


%

\ifCLASSOPTIONcompsoc
  \section*{Acknowledgments}
\else
  \section*{Acknowledgment}
\fi
This work was funded by the European Research Council (ERC) under the European Union’s Horizon 2020 research and innovation programme. Grant agreement № 835294.

\ifCLASSOPTIONcaptionsoff
  \newpage
\fi

\appendices

%
\section{Proofs of Sections \ref{sec:strictconcavity} and \ref{sec:uniqueness}}\label{sec:proofs_sconcavity}

%
{\bf \autoref{lemma:concavity}}. 
{\it
For a given observation probability matrix $\mg$ and two distributions $\vtheta, \vphi \in \calc$, 
it holds for every $0<\gamma<1$ that 
\[
L(\gamma \vtheta + (1-\gamma) \vphi) \geq \gamma L(\vtheta) +(1-\gamma) \, L(\vphi)
\]
where the equality holds if and only if $(\vtheta-\vphi)\, \mg = \vzeros$. 
}

\begin{proof}
From~\eqref{eq:L_ibu} we can write for every $i\in[n]$ 
\begin{flalign*}
&L_i(\gamma \theta + (1-\gamma) \phi) = \log \sum_{u \in \calx} (\gamma \theta_u + (1-\gamma) \phi_u) \mg_{u i} \\ 
          &\qquad\qquad= \log \left( \gamma \sum_{u \in \calx} \theta_u \mg_{u i} + (1-\gamma)\sum_{u \in \calx} \phi_u \mg_{u i} \right) \\
          &\qquad\qquad\geq \gamma \log \sum_{u \in \calx} \theta_u \mg_{u i} +  (1-\gamma) \log \sum_{u \in \calx} \phi_u \mg_{u i}\\
          & \qquad\qquad = \gamma \, L_i(\theta) +(1-\gamma) \, L_i(\phi).
\end{flalign*}
The above inequality follows from Jensen's inequality and the concavity of  
$\log(\cdot)$. Since the latter function is not linear on $\reals$, the above inequality turns into equality if and only if
\begin{equation}\label{thm:unique:eq:2}
\sum_{u \in \calx} \theta_u \mg_{u i} = \sum_{u \in \calx} \phi_u \mg_{u i}.
\end{equation}
Since $L(\cdot) = \sum_{i=1}^n L_i(\cdot)$ it follows that 
\begin{equation}\label{thm:unique:eq:3}
L(\gamma \vtheta + (1-\gamma) \vphi) \geq \gamma L(\vtheta) +(1-\gamma) \, L(\vphi),
\end{equation}
where the equality holds if and only if \eqref{thm:unique:eq:2} holds for all $i\in [n]$, i.e., $(\vtheta-\vphi)\, \mg = \vzeros$. 
\end{proof}

%
{\bf \autoref{thm:strictconcavity}} {\rm (strict concavity)}.
{\it
For a given mechanism and noisy data, let $\mg$ be the corresponding
observation probability matrix. Then the following conditions are equivalent.
\begin{enumerate}[label=(\roman*)]
\item $L(\cdot)$ is strictly concave on $\calc$.
\item All distributions $\vtheta, \vphi$ with $\vtheta\neq\vphi$
satisfies $\vtheta \,\mg \neq \vphi \,\mg$.   
\item Every nonzero vector $\vectr{w}\in \reals^{|\calx|}$ with $\sum_{x\in\calx} w_x=0$
satisfies $\vectr{w} \,\mg \neq \vzeros$. 
\end{enumerate}
}

\begin{proof}
\ref{thm_item_1} $\iff$ \ref{thm_item_2}:
holds directly by the definition of strict concavity and \autoref{lemma:concavity}. 

\ref{thm_item_2} $\implies$ \ref{thm_item_3}: 
Consider a nonzero vector $\vectr{w}$ with $\sum_{x\in\calx} w_x = 0$. Clearly there are distinct $\vtheta, \vphi \in \calc$ 
such that $\vtheta - \vphi = \tau \, \vectr{w}$ for some scalar $\tau>0$. Now since $\vtheta\,\mg \neq \vphi\,\mg$ by \ref{thm_item_2}
we get $\tau \, \vectr{w} \, \mg \neq \vzeros$, hence $\vectr{w} \, \mg \neq \vzeros$.  

\ref{thm_item_3} $\implies$ \ref{thm_item_2}:
Consider any distinct $\vtheta, \vphi \in \calc$ and let $\vectr{w}= \vtheta - \vphi$. Note that $\vectr{w}\neq\vzeros$. 
Also $\sum_{x\in\calx} w_x = 0$ because $\vtheta,\vphi$ are distributions. 
Now since $\vectr{w}\,\mg \neq \vzeros$ by \ref{thm_item_3} we get $(\vtheta-\vphi)\,\mg \neq \vzeros$. 
\end{proof}

{\bf \autoref{cor:strictlyConcave_one_mech}}.
{\it
Let $\mech$ be a mechanism that works on an alphabet $\calx$.
Suppose that the noisy data contains a set $\tilde\calz$ of 
$|\calx|$ distinct observables such that the columns $\{\mech_{* z} : z\in \tilde\calz\}$ are linearly independent. 
Then $L(\cdot)$ is strictly concave on $\calc$.  
}
\begin{proof}
Suppose that the condition in \autoref{cor:strictlyConcave_one_mech} holds. Then the observation probability matrix $\mg$ 
contains the columns $\{\mech_{* z} : z\in \tilde\calz\}$ which are linearly independent. Since the number of these columns is $|\calx|$ the matrix $\mg$ is full-rank and 
the equation $\vectr{w}\, \mg = \vzeros$ is satisfied only by $\vectr{w}=\vzeros$. Thus by \autoref{thm:strictconcavity} $L(\cdot)$ is strictly concave on $\calc$. 
\end{proof}

\begin{lemma}\label{lem:krrLinIndependence}
Consider a finite alphabet $\calx$, and a $k$-RR mechanism $\mech:\calx \to \calx$.  
Then the columns $\{\mech_{*z} : z\in\calx \}$ are linearly independent. 
\end{lemma}
\begin{proof}
To show the linear independence we equivalently show in the following that 
for a row vector $\vw$, the equation $\vw \, \mech =\vzeros$ has only 
the trivial solution $\vw=\vzeros$. 

Let $\alpha = e^\leps/(1-k+ e^\leps)$, $\beta = 1/(1-k+ e^\leps)$, and 
$\mech_{* z}$ be the column of $\mech$ corresponding to the observable $z\in\calx$. Then by \eqref{eq:krr}
\[
\mech_{* z} = [\beta,\beta,\dots,\beta, \alpha, \beta,\beta,\dots, \beta]^T
\]
in which the $z$-th entry is $\alpha$ and each other entry is $\beta$. 
Now suppose that there is a row vector $\vw$ satisfying $\vw \mech = \vzeros$. Then for every $z\in\calx$, it must hold
$\vw \, (\mech_{* z} - \beta \sum_{u\in\calx} \mech_{* u}) = 0$. 
Since by the stochasticity of $\mech$, $\sum_{u\in\calx} \mech_{* u} = \vones$, 
it must follow that $w_z = 0$ for every $z$, i.e. $\vw=\vzeros$. 
\end{proof}

\begin{lemma}\label{lem:linearGeometricLinIndependence}
Consider a finite alphabet $\calx\subset\integers$, and a linear geometric mechanism $\mech:\calx \to \integers$ \eqref{eq:linearGeometric}. 
Then the columns $\{\mech_{*z} : z\in\calx \}$ are linearly independent. 
\end{lemma}
\begin{proof}
Let $\tilde \mech$ be the square matrix consisting of the columns $\{\mech_{*z} : z\in\calx \}$. 
Then the linear independence of these columns is equivalent to saying that $\tilde \mech$ is invertible.  
This is also equivalent to saying that for a variable row vector $\vw$, 
the equation $\vw \, \tilde \mech =\vzeros$ has only the trivial solution $\vw=\vzeros$. 
We write $\calx = \{x_1,x_2,\dots, x_k\}$ and we assume with no loss of generality that 
$x_i < x_{i+1}$ for all $i<k$. Then by \eqref{eq:linearGeometric}, and letting $\alpha= e^{-\lambda\,s}$
we have 
\[
\tilde \mech =a\, \begin{bmatrix}
    1                                     & \alpha^{x_2 - x_1}        & \alpha^{x_3 - x_1}             &\dots        & \alpha^{x_k - x_1}  \\
    \alpha^{x_2 - x_1}          & 1                                   & \alpha^{x_3 - x_2}             &\dots        & \alpha^{x_k - x_2}  \\
    \alpha^{x_3 - x_1}          & \alpha^{x_3 - x_2}         & 1                                       &\dots        & \alpha^{x_k - x_3}    \\
    \vdots                             & \vdots                            & \vdots                               &\ddots       &\vdots                       \\
    \alpha^{x_k - x_1}         & \alpha^{x_k - x_2}        & \alpha^{x_k - x_3}           &\dots         & 1
\end{bmatrix}
\]
We show that any row vector $\vw$ satisfying $\vw\,\tilde \mech = \vzeros$ must be equal to $\vzeros$. This is a system 
of $k$ linear equations in which the $i$-th one is $\vw \, \tilde \mech_{* x_i} = 0$. 
It is easy to see that $\vw \, (\tilde \mech_{* x_1} - \alpha^{(x_2 - x_1)} \tilde \mech_{* x_2})= 0$ 
yields $w_1=0$. Using this together with $\vw \, (\tilde \mech_{* x_2} - \alpha^{(x_3-x_2)} \tilde \mech_{* x_3}) = 0$ yields $w_2=0$. 
By repeating this procedure inductively on every successive two columns, we get that $w_i=0$ for all $i \leq k$, that is $\vw = \vzeros$.
\end{proof}

\begin{lemma}\label{lem:TLinearGeometricLinIndependence}
Let $\calz$ be a finite range of integers. Consider an alphabet $\calx\subseteq\calz$,  
and a truncated linear geometric mechanism $\mech:\calx \to \calz$ \eqref{eq:TLinearGeometric}. 
Then the columns $\{\mech_{*z} : z\in\calx \}$ are linearly independent. 
\end{lemma}
\begin{proof}
Similar to the proof of \autoref{lem:linearGeometricLinIndependence} we show that the matrix $\tilde \mech$ 
consisting of the columns $\{\mech_{*z} : z\in\calx \}$ is invertible. We proceed with the same proof lines
except replacing the constant $a$ in every entry of $\tilde \mech$ with $a(z)$ in~\eqref{eq:TLinearGeometric}. 
\end{proof}

\begin{lemma}\label{lem:taxicabLinIndependence}
Consider a finite alphabet $\calx\subset\integers^2$, and a taxicab mechanism $\mech:\calx \to \integers^2$.  
Then the columns $\{\mech_{*z} : z\in\calx \}$ are linearly independent. 
\end{lemma}
\begin{proof}
Let $\tilde \mech$ be the square matrix consisting of the columns $\{\mech_{*z} : z\in\calx \}$. 
Then, the linear independence of these columns is equivalent to saying that $\tilde \mech$ is invertible. 
Notice first that every entry of $\tilde \mech$ has the form $\tilde \mech_{(u,v) (\mu,\nu)}$, and is exactly  
the probability of producing $(\mu,\nu)$ from $(u,v)$. Let $\mathcal{U} = \{u_1, u_2, \dots, u_{k_1}\}$ be the 
set of all horizontal coordinates in the elements of $\calx$. 
Similarly, let $\mathcal{V} = \{v_1, v_2, \dots, v_{k_2}\}$ be the set of all vertical coordinates in the elements of $\calx$. 
Define an alphabet $\calx' = \mathcal{U} \times \mathcal{V}$, and let $K$ be a taxicab mechanism that works on $\calx'$ (Note that $\calx\subseteq\calx'$).  
Let $\tilde K$ be the submatrix of $K$ consisting only of 
columns that correspond to the elements of  $\calx'$.  
That is $\tilde K_{(u,v), (\mu, \nu)} = \Pr[(\mu, \nu) | (u,v)]$ for all $(u,v) ,(\mu, \nu) \in\calx'$. We show in the following that $\tilde K$ is invertible. 

Let $U$ be a linear geometric mechanism that works on $\mathcal{U}$ with parameter $\lambda$, 
and let $\tilde U$ be the submatrix of $U$ consisting only of the columns that correspond to $\mathcal{U}$. 
Similarly, let $V$ be a linear geometric mechanism that works on $\mathcal{V}$ with parameter $\lambda$, 
and let $\tilde V$ be the submatrix of $V$ consisting only of the columns that correspond to $\mathcal{V}$.
From \eqref{eq:taxicab}, $\tilde K$ is the Kronecker product of the matrices $\tilde U$ and $\tilde V$ which are 
invertible because their respective columns are linearly independent (by \autoref{lem:linearGeometricLinIndependence}). 
Thus $\tilde K$ is invertible too. 

Finally, for every $(u,v) \in \calx'\setminus \calx$, remove from the matrix $\tilde K$ the $(u,v)$-th row and also the $(u,v)$-th column. 
This yields the original (and smaller) matrix $\tilde \mech$ which is indexed exactly by $\calx\times\calx$ (instead of $\calx'\times\calx'$). 
Since $\tilde K$ is invertible, $\tilde \mech$ is also invertible. 
\end{proof}

\begin{lemma}\label{lem:TTaxicabLinIndependence}
Let $\calz_1, \calz_2$ be two finite ranges of integers. Consider an alphabet $\calx\subseteq(\calz_1\times\calz_2)$,  
and a truncated taxicab mechanism $\mech:\calx \to (\calz_1\times\calz_2)$. 
Then the columns $\{\mech_{*z} : z\in\calx \}$ are linearly independent. 
\end{lemma}
\begin{proof}
The proof follows the same lines of that of \autoref{lem:linearGeometricLinIndependence}, with the following differences: $K$ is a truncated 
taxicab mechanism; $U,V$ are truncated linear geometric mechanisms with parameter $\lambda$; Equation \eqref{eq:truncatedTaxicab} 
shows that the matrix $\tilde K$ is the Kronecker product of matrices $\tilde U$ and $\tilde V$, and \autoref{lem:TLinearGeometricLinIndependence} 
is used to show the invertibility of both $\tilde U$ and $\tilde V$. 
\end{proof}
%

{\bf \autoref{cor:strictlyConcave-krr}}.
{\it
Consider a finite alphabet $\calx$, and a $k$-RR mechanism $\mech:\calx \to \calx$.  
Then $L(\cdot)$ is strictly concave for any noisy data that contains all the elements of $\calx$. 
}
\begin{proof}
The noisy data contains the set $\calx$, whose size is $|\calx|$. Also  
the columns $\{\mech_{*z} : z\in\calx\}$ are linearly independent by \autoref{lem:krrLinIndependence}. 
Thus, it follows from \autoref{cor:strictlyConcave_one_mech} that $L(\cdot)$ is strictly concave.
\end{proof}

{\bf \autoref{cor:strictlyConcaveLinearGeometric}}.
{\it
Consider a finite alphabet $\calx\subset\integers$, and a linear geometric mechanism $\mech:\calx \to \integers$.  
Then $L(\cdot)$ is strictly concave for any noisy data that contains all the elements of $\calx$. 
} 
\begin{proof}
The noisy data contains the set $\calx$, whose size is $|\calx|$. Also  
the columns $\{\mech_{*z} : z\in\calx\}$ are linearly independent by \autoref{lem:linearGeometricLinIndependence}. 
Thus it follows from \autoref{cor:strictlyConcave_one_mech} that $L(\cdot)$ is strictly concave.
\end{proof}

{\bf \autoref{cor:strictlyConcaveTLinearGeometric}}.
{\it
Let $\calz$ be a finite range of integers. Consider an alphabet $\calx\subseteq\calz$,  
and a truncated linear geometric mechanism $\mech:\calx \to \calz$.  
Then $L(\cdot)$ is strictly concave for any noisy data that contains all the elements of $\calx$. 
}
\begin{proof}
The noisy data contains the set of $\calx$, whose size is $|\calx|$. 
Since also the columns $\{\mech_{*z} : z\in\calx\}$ are linearly independent by \autoref{lem:TLinearGeometricLinIndependence}, 
it follows from \autoref{cor:strictlyConcave_one_mech} that $L(\cdot)$ 
is strictly concave.
\end{proof}

{\bf \autoref{prop:geoindTaxicab}}.
{\it
The truncated and non-truncated taxicab mechanisms with parameter $\lambda$   
satisfy $(\sqrt{2}\, \lambda)$-geo-indistinguishability. 
}
\begin{proof}
For all $(u,v),(u', v')\in \calx$ and $(\mu, \nu)\in\integers^2$ we have from \eqref{eq:taxicab} (and \eqref{eq:truncatedTaxicab} for the truncated mechanism)
\begin{align}
\frac{\Pr[(\mu,\nu) | (u,v)]}{\Pr[(\mu,\nu) | (u',v')]} &= e^{ \lambda s ( |\mu - u' | - |\mu -  u |  ) +  \lambda s( |\nu - v' | - |\nu - v | ) }\nonumber \\
   &\leq e^{ \lambda s (| u' - u | ) +  \lambda s (| v' - v | ) } \label{prop:geoindTaxicab:eq1}
\end{align}
where the last inequality holds because, from the reverse triangle inequality on real numbers, we have $|\mu - u' | - |\mu -  u | \leq | u' - u |$ 
and $|\nu - v' | - |\nu - v | \leq | v' - v |$. 
Now let $\ell = \dist((u, v), (u', v')) $, i.e. the Euclidean distance between $(u, v)$ and $(u', v')$, 
and suppose that the line segment connecting these two points makes an angle $0\leq\omega\leq\pi/2$ with the horizontal axis; 
then we have 
\[
s | u' - u | = \ell \cos(\omega) \quad\mbox{and} \quad s | v' - v | = \ell \sin(\omega). 
\] 
With fixed $\ell$, define $f(\omega) = s (| u' - u |) + s (| u' - u |) = \ell (\cos(\omega) + \sin(\omega))$. 
Then we want to get the maxima of $f(\omega)$ by inspecting its first and second derivatives 
\[
f'(\omega) = \ell (-\sin(\omega) + \cos(\omega)), \quad f''(\omega) = \ell (-\cos(\omega)  - \sin(\omega)).
\]
Clearly at $\omega = \pi/4$, we have $f'(\omega) = 0$ and $f''(\omega) = -\sqrt{2} \ell < 0$. 
Therefore $f(\omega)$ is maximum when $\omega = \pi/4$, in which case   
$f(\omega)$ = $\sqrt{2} \ell$. Thus $s (| u' - u |) + s (| v' - v |) \leq \sqrt{2}\, \dist ((u,v), (u',v'))$. 
Then from \eqref{prop:geoindTaxicab:eq1}, we obtain 
\[
\frac{\Pr[(\mu,\nu) | (u,v)]}{\Pr[(\mu,\nu) | (u',v')]} \leq e^{\sqrt{2}\lambda \,\dist ((u,v), (u',v'))}. 
\]
\end{proof}

{\bf\autoref{cor:strictlyConcaveTaxicab}}
{\it
Consider a finite alphabet $\calx\subset\integers^2$, and a taxicab mechanism $\mech:\calx \to \integers^2$.  
Then $L(\cdot)$ is strictly concave for any noisy data that contains all the elements of $\calx$. 
}
\begin{proof}
The noisy data contains all the elements of $\calx$, whose size is $|\calx|$. Since also  
the columns $\{\mech_{*z} : z\in\calx\}$ are linearly independent by \autoref{lem:taxicabLinIndependence}, 
it follows from \autoref{cor:strictlyConcave_one_mech} that $L(\cdot)$ 
is strictly concave.
\end{proof}

{\bf\autoref{cor:strictlyConcaveTTaxicab}}
{\it
Let $\calz_1, \calz_2$ be two finite ranges of integers. Consider an alphabet $\calx\subseteq(\calz_1\times\calz_2)$,  
and a truncated taxicab mechanism $\mech:\calx \to (\calz_1\times\calz_2)$.  
Then $L(\cdot)$ is strictly concave for any noisy data that contains all the elements of $\calx$. 
}
\begin{proof}
The columns $\{\mech_{*z} : z\in\calx\}$ are linearly independent by \autoref{lem:TTaxicabLinIndependence}, 
Thus, it follows from \autoref{cor:strictlyConcave_one_mech} that $L(\cdot)$ is strictly concave.
\end{proof}

{\bf \autoref{lemma:probrappor}}.
{\it
Consider a basic \textsc{Rappor} mechanism that works on alphabet $\calx$ and has parameter $\leps>0$.
Let $p = e^{\leps/2}/(1+e^{\leps/2})$. For every bit-vector $\beta\in\{0,1\}^{|\calx|}$ and $y\in\calx$, 
let $\beta_y$ be the value of the $y$-th bit and $S(\beta)= \sum_{y\in\calx} \beta_y$. Then 
\[
\Pr[\beta | X=x] = p^{|\calx|} e^{-(\frac{1}{2} + \frac{1}{2} S(\beta) - \beta_x ) \leps}.
\]
}
\begin{proof}
Suppose that the original datum is $x$. 
Therefore $x$ is encoded into a bit-vector $b\in \{0,1\}^{|\calx|}$ where $b_x=1$ and $b_y=0$ for all $y\neq x$. 
Now from $b_x=1$ we get $\beta_x=1$ with probability $\frac{e^{\frac{\leps}{2}  }} {1+ e^{\frac{\leps}{2}}}$ and 
$\beta_x=0$ with probability $\frac{1} {1+ e^{\frac{\leps}{2}}}$.  
Therefore 
\begin{equation}\label{eq:probrappor:1}
\Pr[\beta_x | X=x ] = \frac{e^{\frac{\leps}{2} \beta_x  }} {1+ e^{\frac{\leps}{2}}}. 
\end{equation}
For every $y\in\calx\setminus\{x\}$, we have $b_y=0$. Therefore 
we get $\beta_y=1$ with probability $\frac{1} {1+ e^{\frac{\leps}{2}}}$ and 
$\beta_y=0$ with probability $\frac{e^{\frac{\leps}{2}}} {1+ e^{\frac{\leps}{2}}}$.  
That is 
\begin{equation}\label{eq:probrappor:2}
\Pr[\beta_y | X=x ] = \frac{e^{\frac{\leps}{2} (1-\beta_y)  }} {1+ e^{\frac{\leps}{2}}} \quad\forall y \neq x. 
\end{equation}
The bits of the original vector $b$ are randomized independently to obtain the noisy vector $\beta$. 
Therefore we get from \eqref{eq:probrappor:1} and \eqref{eq:probrappor:2}
\begin{align*}
\Pr[\beta | X=x] &= \frac{e^{\frac{\leps}{2} \beta_x  }} {1+ e^{\frac{\leps}{2}}} \prod_{y \neq x} \frac{ e^{ \frac{\leps}{2} (1-\beta_y) } }{ 1 + e^{\frac{\leps}{2}}    } 
\\
&= \left(\frac{1} {1+ e^{\frac{\leps}{2}}}\right)^{|\calx|}
e^{  \frac{\leps}{2} \beta_x} \,
e^{\frac{\leps}{2} \sum_{y\neq x} (1-\beta_y)}
\\
&= \left(\frac{1} {1+ e^{\frac{\leps}{2}}}\right)^{|\calx|}
e^{\frac{\leps}{2} \left(\beta_x +(|\calx|-1) - \sum_{y\neq x} \beta_x\right)}
\\
&= \left(\frac{e^{\frac{\leps}{2}}} {1+ e^{\frac{\leps}{2}}}\right)^{|\calx|}
e^{\frac{\leps}{2} \left(\beta_x -1 - (S(\beta)-\beta_x)\right)}
\\
&= p^{|\calx|} e^{-(\frac{1}{2} + \frac{1}{2} S(\beta) - \beta_x ) \leps}
\end{align*}
where the last equality follows by substituting 
$p= \frac{e^{\frac{\leps}{2}}} {1+ e^{\frac{\leps}{2}}}$.
\end{proof}

{\bf \autoref{lemma:binarymatrix}}.
{\it
Let $B$ be a binary matrix, i.e., having entries in $\{0,1\}$. 
Then $B$ has full rank in $\reals$ if it has full rank in $\binfield$.  
}
\begin{proof}
Consider a binary matrix $B$ with $m$ rows and $k$ columns. With no loss of generality, we assume that $m\leq k$. 
Then $B$ has full rank if and only if it has $m$ linearly independent columns, i.e. if and only if it has $m$ columns that 
form an $m \times m$ sub-matrix with nonzero determinant. Note also that the determinant of this 
submatrix in $\binfield$ is binary. Therefore, it is enough for the proof to show that every $m \times m$ 
binary matrix $C$ satisfies
\begin{equation}\label{eq:binmatrix}
\textit{det}_{\reals}(C) \mbox{ is odd} \quad\quad \text{iff} \quad\quad \textit{det}_{\binfield}(C)=1.
\end{equation}
We prove \eqref{eq:binmatrix} by induction on the size $m$. 
If $m=1$, i.e. $C$ consists of one entry in $\{0,1\}$, then the statement \eqref{eq:binmatrix} holds 
in this case because $\textit{det}_{\binfield}(C) = \textit{det}_{\reals}(C) \in \{0,1\}$. 
Now suppose that \eqref{eq:binmatrix} holds for every binary matrix of size $m\geq1$, 
and let $C$ be a binary matrix of size $m+1$. Then with any row index $i\in\{1,2,\dots, m+1\}$, 
the determinant of $C$ in $\reals$ is 
\begin{equation}\label{eq:det-R}
\textit{det}_{\reals}(C) = \sum_{j=1}^{m+1} (-1)^{i+j} C_{ij}\, \textit{det}_{\reals}(M_{ij}). 
\end{equation}
where $M_{ij}$ is the $m$-size submatrix resulting from removing the $i$-th row and $j$-th column from $C$.
The determinant of $C$ in the field $\binfield$ is defined exactly as \eqref{eq:det-R}, but with the value 
$-1$ replaced by the value $1$, the summation replaced by $\textit{XOR}$, and the product replaced by  
$\textit{AND}$ (which we write as $\land$). Then with any row index $i\in\{1,2,\cdots, m+1\}$ the determinant 
of $C$ in $\binfield$ is 
\begin{equation}\label{eq:det-b}
\textit{det}_{\binfield}(C) = \textit{XOR}_{j=1}^{m+1}  \left(C_{ij} \land \textit{det}_{\binfield}(M_{ij})\right) 
\end{equation}
By the assumption that \eqref{eq:binmatrix} holds for all binary matrices of size $m$, we have for any fixed $i$ that
the counts
\begin{flalign*}
&\textit{count}\{j : 1\leq j \leq m+1, C_{ij}=1, \textit{det}_{\binfield}(M_{ij})=1\}, \\
&\textit{count}\{j : 1\leq j \leq m+1, C_{ij}=1, \textit{det}_{\reals}(M_{ij}) \mbox{ is odd}\}
\end{flalign*}
are identical. Let $c$ denote these counts. 
From \eqref{eq:det-b} $\textit{det}_{\binfield}(C)=1$ iff $c$ is odd. Notice also from \eqref{eq:det-R} that   
$\textit{det}_{\reals}(C)$ is odd iff $c$ is odd. Thus $\textit{det}_{\binfield}=1$ iff $\textit{det}_{\reals}(C)$ is odd, 
i.e. \eqref{eq:binmatrix} holds for $m+1$. 
\end{proof}

{\bf \autoref{prop:strictlyConcaverappor}}.
{\it
Consider a basic \textsc{Rappor} mechanism that works on an alphabet $\calx$ and has parameter $\leps>0$. 
Let $p = e^{\leps/2}/(1+e^{\leps/2})$. Suppose that a set of $n \geq |\calx|$ original data are obfuscated by the mechanism. 
Then $L(\cdot)$ is strictly concave with probability at least $\prod_{k=1}^{|\calx|} \max\{0, (1- p^n\, 2^{k-1})\}$ . 
}

\begin{proof}
By \autoref{thm:strictconcavity} $L(\cdot)$ is strictly concave on $\calc$ if the equations $\vw \, \mg = \vzeros $ and $\vw \, \vones = 0$ 
are satisfied only by $\vw=\vzeros$. We recall from \autoref{lemma:probrappor} that each entry $\mg_{xi}$ has value  
either $a_i = p^{|\calx|} e^{-(\frac{1}{2} + \frac{1}{2} S(\beta^i) - 1 ) \leps}$ when $\beta^i_x=1$, or $b_i = a_i e^{-\leps}$ when $\beta^i_x=0$, where $\beta^i$ is the $i$-th noisy bit-vector. Then we can combine 
the columns of $\mg$ linearly with the column vector $\vones$ to form a new matrix $B$ whose columns are defined as 
\begin{equation}\label{eq:B}
B_{* i} = (G_{* i} - b_i \vones)/(a_i-b_i). 
\end{equation} 
Then the above condition of \autoref{thm:strictconcavity} is equivalent 
to saying that $\vw B = \vzeros$ and $\vw \, \vones = 0$ 
are satisfied only by $\vw=\vzeros$. 
Therefore $L(\cdot)$ is strictly concave if $B$ has full rank, i.e. has $|\calx|$ linearly independent 
rows (recall that $B$ has $|\calx|$ rows and $n$ columns, and $|\calx| \leq n$). 

Note from \eqref{eq:B} that the matrix $B$ is binary (having entries in $\{0,1\}$) and also $B_{* i} = \beta^i$, i.e. the $i$-th noisy bit-vector.
Clearly, $B$ is random and depends on the original data $D = \{x^i: i\in [n]\}$. Also the entries $B_{xi}$  
(for all $x\in\calx$ and $i\in[n]$) are independent of each other, and 
\begin{equation}\label{eq:B_bounds}
1-p \leq \Pr[B_{xi} = \nu \,|\, D] \leq p \quad \forall \nu\in\{0,1\}. 
\end{equation}
Now since $B$ is binary, it follows from \autoref{lemma:binarymatrix} that $B$ 
has full rank in $\reals$ if it has full rank in the binary field $\binfield$.
Therefore
\[
\Pr[\textit{rank}_\reals(B) = |\calx| \,\big|\, D] \geq \Pr[\textit{rank}_{\binfield}(B)= |\calx| \,\big|\, D].
\]
With no loss of generality let $\calx = \{1,2,\dots, |\calx|\}$, and for every $x\in\calx$ let $\rho_x$ be the 
event that the rows $\{B_{y *} : y \leq x\}$ are linearly independent in $\binfield$. Then 
\begin{align}\label{eq:rank}
&\Pr[\textit{rank}_{\binfield}(B) = |\calx| \,\big|\, D] = \nonumber\\
&\qquad\qquad\qquad \Pr[\rho_1 \,\big|\, D]\, \prod_{x=2}^{|\calx|} \Pr[\rho_x \,\big|\, \rho_{x-1}, D].
\end{align}
The set consisting of the row $B_{1*}$ is linearly dependent if and only if $B_{1*}=\vzeros$, which occurs with probability at most $p^n$ by \eqref{eq:B_bounds}. Therefore $\Pr[\rho_1 | D] \geq 1 - p^n$.  
To evaluate $\Pr[\rho_x | \rho_{x-1}, D]$, suppose that $\rho_{x-1}$ holds, i.e. the rows
$B_{1*}, B_{2*}, \cdots, B_{(x-1)*}$ are linearly independent in $\binfield$. Then they span the vector 
space $\{0,1\}^{x-1}$ which consists of $2^{x-1}$ distinct members (vectors). Now the row $B_{x*}$ is 
linearly dependent on those rows if and only if $B_{x*}$ is a member of the space $\{0,1\}^{x-1}$.  
The probability of this event is at most  $2^{x-1} p^n$ (recall from \eqref{eq:B_bounds} that the maximum 
probability of any value for $B_{x*}$ is $p^n$). 
Therefore, it holds that 
$\Pr[\rho_x | \rho_{x-1}, D] \geq \max\{0, (1- 2^{x-1} p^n)\}$. 
The proof is finally completed by substitution in~\eqref{eq:rank}.  
\end{proof}

{\bf \autoref{prop:uniqueness}}.~{\it
Suppose that 
$L(\vphi)>-\infty$ for some $\vphi\in\calc$ and 
$L(\cdot)$ is strictly concave on $\calc$. Then the MLE is unique. 
}

\begin{proof}
Suppose that there is some distribution $\vphi\in\calc$ such that $L(\vphi)>-\infty$. Then the set 
$\calc'=\{\vtheta\in\calc : L(\vtheta) \geq L(\vphi)\}$
is compact by \cite[Lemma~1]{ehab:eurosp:2020}.
Note also that the function $L(\cdot)$ \eqref{eq:L_ibu} is continuous everywhere in $\calc'$ because $L(\vtheta) \geq L(\vphi) > -\infty$ for every $\vtheta\in\calc'$.
Therefore, by the extreme value theorem, there is 
a maximizer $\hat\vtheta\in\calc'$ for $L(\cdot)$ across the points of $\calc'$. From the definition of $\calc'$,  $\hat\vtheta$ is also a maximizer for $L(\cdot)$ over the entire $\calc$.  

Now suppose that $L(\cdot)$ is strictly concave on $\calc$, which is convex by its definition. 
It is known that if a function is strictly concave on a convex domain where there exists a maximum, then its maximizer is unique 
(see e.g.~\cite[Proposition~2.1.2]{bertsekas:2003:BOOK}). 
Since $L(\cdot)$ is strictly concave on $\calc$, which is convex,
and $L(\cdot)$ has a maximizer $\hat\vtheta\in\calc$, this maximizer is unique in $\calc$, that is, a unique MLE. 
\end{proof}

\section{Proofs of Section~\ref{sec:conv_real}}

{\bf \autoref{thm:identification}}~{\rm (Identification)}.~{\it
Given a mechanism $\mech$ with input alphabet $\calx$, the following conditions are equivalent. 
\begin{enumerate}[label=(\roman*)]
\item $\Pr[\cdot|\vtheta] \neq \Pr[\cdot|\vphi]$ for all distributions $\vtheta\neq\vphi$.
\item $L(\cdot)$ is strictly concave for some set of observables.
\item $\mech$ has $|\calx|$ linearly independent columns. 
\end{enumerate}
}
\begin{proof}
\ref{prop_id_item_1} $\iff$ \ref{prop_id_item_2}:
holds by \autoref{thm:strictconcavity} in which $\vtheta\,\mg$ is a part of the 
distribution $\Pr[\cdot | \vtheta]$. 
\ref{prop_id_item_3} $\implies$ \ref{prop_id_item_2}: Consider the set of observations corresponding to the 
independent columns of $\mech$. Then by \autoref{thm:strictconcavity} $L(\cdot)$ is strictly concave for these observations. 
\ref{prop_id_item_2} $\implies$ \ref{prop_id_item_3}:
Let $\tilde\mech$ be the set of all linearly independent columns in $\mech$. 
By strict concavity and statement \ref{thm_item_3} of \autoref{thm:strictconcavity}, 
the set of columns $\tilde\mech \cup \{\vones\}$ contains $|\calx|$ linearly independent columns. 
By the definition of $\tilde\mech$, both $\sum_{\vectr{v}\in\tilde \mech} \vectr{v}$ and $\sum_{\vectr{v}\in \mech\setminus\tilde\mech} \vectr{v}$ 
are linearly dependent on $\tilde \mech$. Therefore, the vector $\vones$, which is equal to the sum of these two vectors (due to the stochasticity of $\mech$)
is also linearly dependent on $\tilde \mech$. Therefore $\tilde\mech$ contains $|\calx|$ linearly independent columns. 
\end{proof}
%
{\bf \autoref{cor:consistency1}}.~{\it
The MLE obtained under $k$-RR, truncated linear geometric, truncated taxicab, and basic \textsc{Rappor} mechanisms 
converges in probability to the real distribution. 
}
\begin{proof}
From \autoref{cor:strictlyConcave-krr}, \autoref{cor:strictlyConcaveTLinearGeometric}, \autoref{cor:strictlyConcaveTaxicab}, and \autoref{prop:strictlyConcaverappor}
corresponding respectively to the mentioned mechanisms, there are observables for which $L(\cdot)$ is strictly concave, 
hence by \autoref{thm:identification}, the identification condition is satisfied. 
The continuity condition holds by the argument of Section \ref{sec:continuity} and 
that every column of these mechanisms consists of only strictly positive probabilities.
Finally let $d(z) = - \log (\inf_{x,y} \mech_{xy})$. 
It is clear that $| \log \Pr[ z | \vphi] | \leq d(z)$ for all $z\in\calz$ and all $\vphi\in\calc$. 
Note that for all the above mechanisms we have finite $\calz$, and $d(z) < \infty$ for every $z\in\calz$. 
Therefore, we have $\E[d(Z^i)] < \infty$, which satisfies the dominance condition. 
\end{proof}
%
{\bf \autoref{cor:consistency2}}.~{\it
Consider a finite alphabet $\calx\subset\integers$. 
Then the MLE obtained under the linear geometric mechanism $\mech: \calx \to \integers$ converges in probability to the real distribution. 
}
\begin{proof}
From \autoref{cor:strictlyConcaveLinearGeometric} (for the linear geometric) 
there are observables for which $L(\cdot)$ is strictly concave, hence, by \autoref{thm:identification}, the identification condition is satisfied. The continuity condition holds by the argument of 
Section \ref{sec:continuity}, since every entry of the mechanism matrix is strictly positive.
Finally, for the dominance, we recall for the linear geometric mechanism that $\calz = \integers$ and $\calx\subset\integers$. 
Since $\calx$ is finite, we have $x_1 \leq x \leq x_1 +\ell$ for some integers $x_1,\ell$ where $\ell>0$. 
Then by the definition~\eqref{eq:linearGeometric} of geometric mechanism every $z \in \integers$ satisfies

\begin{align} 
%
&\begin{aligned}
a\,e^{-\lambda s(|z-x_1|+\ell)}
&\leq \Pr[z\mid x]
 \leq a\,e^{-\lambda s|z-x_1|},\\[-2pt] 
&\hspace{2cm}\text{if } z\leq x_1
\end{aligned}
\label{eq:consistency2_1}
\\[4pt] 
&\hspace{1.3cm}\begin{aligned} 
a\,e^{-\lambda s\ell}
&\leq \Pr[z\mid x]
 \leq a,\\[-2pt]
&\hspace{2cm}\text{if } x_1\leq z\leq x_1+\ell
\end{aligned}
\label{eq:consistency2_2}
\\[4pt]
&\hspace{0.6cm}\begin{aligned}
a\,e^{-\lambda s|z-x_1|}
&\leq \Pr[z\mid x]
 \leq a\,e^{-\lambda s(|z-x_1|-\ell)},\\[-2pt]
&\hspace{2cm}\text{if } x_1+\ell\leq z
\end{aligned}
\label{eq:consistency2_3}
\end{align}
For any $z\in\integers$, it follows from the left side inqualities of 
\eqref{eq:consistency2_1}-\eqref{eq:consistency2_3} that 
\begin{equation}\label{eq:consistency2_4}
\Pr[z \mid x] \geq a \, e^{- \lambda\,s ( | z - x_1 | + \ell ) } \quad\text{for every $x\in\calx$}. 
\end{equation}
We also have for every $\vphi\in\calc$ that
\[
\Pr[ z \mid \vphi] = \sum_{x\in\calx} \phi_x \Pr[z \mid x] 
\geq a \, e^{- \lambda\,s ( | z - x_1 | + \ell ) }
\]
where the last inequality follows from \eqref{eq:consistency2_4} and 
that $\sum_{x\in\calx} \phi_x =1$.
Therefore $| \log \Pr[ z | \vphi] | \leq d(z)$ 
where $d(z) =  \lambda\,s ( | z - x_1 | + \ell ) - \log a$. Then
from the right side inequalities of \eqref{eq:consistency2_1}-\eqref{eq:consistency2_3}, 
we obtain
\begin{align*}
\E[d(Z^i)]
&\leq
\sum_{z\in\integers\cap[x_1,x_1+\ell]} a\,d(z) \\
&\quad+
\sum_{z\in\integers\setminus[x_1,x_1+\ell]}
a\,e^{-\lambda s(|z-x_1|-\ell)}\,d(z)
<\infty,
\end{align*}
In the above formula, the first sum is finite since the summation range for $z$ is finite. 
In the second sum, $d(z)=O(|z-x_1|)$, 
while the exponential factor provides geometric decay 
in $|z-x_1|$. Thus, both sums are finite, 
yielding $\E[d(Z^i)]<\infty$,
which satisfies the dominance condition of \autoref{thm:mlconsistencygen}. 
\end{proof}

{\bf \autoref{cor:consistency3}}.~{\it
Consider a finite alphabet $\calx\subset\integers^2$. 
Then the MLE obtained under the taxicab mechanism $\mech:\calx \to \integers^2$  
converges in probability to the real distribution.
}
\begin{proof}
From \autoref{cor:strictlyConcaveTaxicab}
there are observables for which $L(\cdot)$ is strictly concave, hence, by \autoref{thm:identification}, the identification condition is satisfied.  
The continuity condition also holds by the argument of Section \ref{sec:continuity}, since the mechanism matrix contains only strictly positive elements. 
Finally, for the dominance condition, we recall that the taxicab mechanism works on a finite set $\calx \subset \integers^2$ and the 
probability of reporting $(\mu, \nu)\in \integers^2$ from $(u,v) \in \calx$ is given as
\[ 
\Pr[ (\mu,\nu) | (u,v)] = \Pr[\mu | u]\, \Pr[\nu | v]
\]
where the probabilities $\Pr[\mu | u]$ and $\Pr[\nu | v]$ are defined by two 
independent linear geometric mechanisms~\eqref{eq:linearGeometric} obfuscating 
the coordinates $u,v$ into $\mu,\nu$ respectively.
Suppose $u_1\leq u\leq u_1+\ell$ and $v_1\leq v\leq v_1+ \ell'$, for some integers $u_1, v_1, \ell,\ell'$ where $\ell,\ell' >0$.  
Using the inequalities~\eqref{eq:consistency2_1} - \eqref{eq:consistency2_3}, 
and letting 
\[
B(\mu,\nu) = a^2 \, e^{- \lambda\,s ( | \mu - u_1 | + | \nu - v_1 | +\ell+\ell' ) }
\]
we obtain the following inequalities for $\Pr[ (\mu,\nu) | (u,v)]$
\begin{align}
&\begin{aligned}
&B(\mu,\nu) \leq \Pr[(\mu,\nu) | (u,v)] \leq a^2\\   
&\hspace{0.5cm}\text{if } \mu\in \integers\cap [u_1, u_1+\ell] \;\mbox{and}\; \nu \in \integers\cap [v_1,  v_1+\ell']
\end{aligned}\label{eq:consistency3_1}
\\[4pt]
&\begin{aligned}
&B(\mu,\nu)\leq \Pr[(\mu,\nu) | (u,v)]\leq a^2\,e^{- \lambda\,s ( | \mu - u_1 | - \ell ) } \\
&\hspace{0.5cm}\text{if } \mu \in \integers\setminus [u_1, u_1+\ell] \quad\mbox{and} \quad \nu \in \integers\cap [v_1, v_1+\ell']
\end{aligned}\label{eq:consistency3_2}
\\[4pt]
&\begin{aligned}
&B(\mu,\nu)\leq \Pr[(\mu,\nu) | (u,v)]     \leq a^2\,e^{- \lambda\,s ( | \nu - v_1 | - \ell' ) } \\
&\hspace{0.5cm}\text{if } \mu \in \integers\cap[u_1, u_1+\ell]  \quad \mbox{and} \quad \nu \in\integers\setminus [v_1, v_1+\ell']
\end{aligned}\label{eq:consistency3_3}
\\[4pt]
&\begin{aligned}
&B(\mu,\nu)     \leq \Pr[(\mu,\nu) | (u,v)]     \leq a^2\,e^{- \lambda\,s ( | \mu - u_1 | + | \nu - v_1 | - \ell - \ell') } \\ 
&\hspace{0.5cm}\text{if } \mu \in\integers\setminus [u_1, u_1+\ell]  \quad \mbox{and} \quad \nu \in\integers\setminus [v_1, v_1+\ell']
\end{aligned}\label{eq:consistency3_4}
\end{align}
For any $(\mu,\nu)\in\integers^2$, we have from the left side inequalities of 
\eqref{eq:consistency3_1}-\eqref{eq:consistency3_4} that 
\begin{equation}\label{eq:consistency3_5}
\Pr[(\mu,\nu) | (u,v)] \geq B(\mu,\nu) \quad\text{for every $(u,v)\in\calx$}. 
\end{equation}
We also have for every $\vphi\in\calc$ that
\[
\Pr[ (\mu,\nu) | \vphi] = \sum_{(u,v)\in\calx} \phi_{(u,v)} \Pr[(\mu,\nu) | (u,v)]
\geq B(\mu,\nu)
\]
where the last inequality follows from \eqref{eq:consistency3_5} and 
that $\sum_{(u,v)\in\calx} \phi_{(u,v)} =1$. Thus from the definition 
of $B(\mu,\nu)$, we obtain
$| \log \Pr[ (\mu,\nu) | \vphi) | \leq d( (\mu,\nu) ]$ where 
\begin{align*}
d( (\mu,\nu) ) &=  - \log B(\mu,\nu) \\
&= \lambda\,s ( | \mu - u_1 | + | \nu - v_1 | +\ell+\ell' ) - 2\,\log a. 
\end{align*} 
Then from the right side inqualities of \eqref{eq:consistency3_1}-\eqref{eq:consistency3_4}, we obtain
\begin{align*}
&\E[d(Z^i)]
\leq
\sum_{\substack{
    \mu \in \integers\cap[u_1,u_1+\ell]\\
    \nu \in \integers\cap[v_1,v_1+\ell']
}}
a^2\,d((\mu,\nu))
\\
&\hspace{0.5cm}+
\sum_{\substack{
    \mu\in\integers\setminus[u_1,u_1+\ell]\\
    \nu\in\integers\cap[v_1,v_1+\ell']
}}
a^2 e^{-\lambda s(|\mu-u_1|-\ell)}\,d((\mu,\nu))
\\
&\hspace{0.5cm}+
\sum_{\substack{
    \mu\in\integers\cap[u_1,u_1+\ell]\\
    \nu\in\integers\setminus[v_1,v_1+\ell']
}}
a^2 e^{-\lambda s(|\nu-v_1|-\ell')}\,d((\mu,\nu))
\\
&\hspace{0.5cm}+
\sum_{\substack{
    \mu\in\integers\setminus[u_1,u_1+\ell]\\
    \nu\in\integers\setminus[v_1,v_1+\ell']
}}
a^2 e^{-\lambda s(|\mu-u_1|+|\nu-v_1|-\ell-\ell')}\,
d((\mu,\nu)).
\end{align*}
The first sum is finite since the summation ranges for $\mu,\nu$ are finite. 
In each of the remaining sums, $d((\mu,\nu))=O(|\mu-u_1|+|\nu-v_1|)$, 
while the exponential factor provides geometric decay 
in $|\mu-u_1|$ and $|\nu-v_1|$. Thus, all four sums are finite, 
yielding $\E[d(Z^i)]<\infty$, 
which satisfies the dominance condition of \autoref{thm:mlconsistencygen}. 
\end{proof}

\section{Proofs of Section \ref{sec:experiments}}\label{sec:proofs_experiments}

\begin{lemma}\label{lem:q_unbiased}
Let a mechanism $\mech$ be applied to data sampled from the real distribution $\vtheta$. 
Let $\vq$ be the noisy distribution of the noisy data produced by $\mech$. 
Then $\vw = \vq \mech^{-1}$ is an unbiased estimate for $\vtheta$, i.e. $\E[\vq] =\vtheta$.  
\end{lemma}
\begin{proof}
From $\vw = \vq \mech^{-1}$, it follows that $\E[\vw] = \E[\vq] \mech^{-1}$. 
We also have $\E[\vq] = \vtheta \, \mech$. Therefore, 
$\E[\vw] = \vtheta \, \mech \mech^{-1} = \vtheta$. 
\end{proof}

{\bf \autoref{prop:invKrrBound}}~{\rm (\inv{} with $k$-RR)}.~{\it
Consider any alphabet $\calx$ of size $k$, and let $\vtheta$ be any original distribution on $\calx$. 
Let $n$ data points be sampled from $\calx$ and then obfuscated by a $k$-RR mechanism with local privacy parameter $\leps$. 
Then the \inv{}-estimated vector $\vw$ satisfies  
$\E[ {\| \vw - \vtheta \|}_2^2 ] \le \frac{1}{n}\left( \frac{e^\leps + k -1}{e^\leps -1} \right)^2$. 
}

\begin{proof}
From the definition~\eqref{eq:krr} of the $k$-RR mechanism with parameter $\leps$, 
the inverse matrix $\mech^{-1}$ of the $k$-RR is   
\begin{equation}\label{eq:invkrr}
\mech^{-1}_{xz} = \frac{1}{e^\leps-1} 
\left\{ 
\begin{array}{l l}
e^\leps+k-2 & \mbox{if $z=x$}\\
-1                & \mbox{if $z\neq x$}. 
\end{array} 
\right.
\end{equation} 
From the inverse matrix~\eqref{eq:invkrr} we have
$w_x = \frac{e^\leps+k-1}{e^\leps -1} q_x - \frac{1}{e^\leps - 1}$. 
Let $m_x$ be the marginal probability that the mechanism produces $x$. Then 
$q_x$ is the number of observations (out of $n$) sampled with probability $m_x$ divided by $n$.  
Therefore $w_x$ can be expressed as  
$w_x = 1/n \sum_{i=1}^n T^i_x - \frac{1}{e^\leps - 1} $ where $T^i_x$ are i.i.d random variables which follow the distribution  
\begin{equation}
T^i_x =  
\left\{ 
\begin{array}{l l}
\frac{e^\leps+k-1}{e^\leps - 1} & \mbox{with probability $m_x$},\\
0                & \mbox{with probability $1- m_x$}. 
\end{array} 
\right.
\end{equation} 
Therefore 
\begin{align*}
\textit{Var}[w_x] &= \frac{1}{n} \sum_{i=1}^n \textit{Var}[T^i_x] 
    =\frac{1}{n} \left(\frac{e^\leps+k-1}{e^\leps - 1}\right)^2 m_x \, (1-m_x)\\
    &\leq \frac{1}{n} \left(\frac{e^\leps+k-1}{e^\leps - 1}\right)^2 m_x.
\end{align*}
where the last inequality follows from the fact $1-m_x\leq1$. 
By \autoref{lem:q_unbiased}, we have $\E[\vw]=\vtheta$. Therefore $\textit{Var}[w_x] = \E[(w_x - \theta_x)^2]$. 
Thus we obtain 
\[
\E[ {\| \vw - \vtheta \|}_2^2 ] = \sum_{x\in\calx} \textit{Var}[w_x] \leq \frac{1}{n} \left(\frac{e^\leps+k-1}{e^\leps - 1}\right)^2.
\]
\end{proof}

{\bf \autoref{prop:invTLinGeomBound}}~{\rm (\inv{} with truncated linear geometric)}.~{\it
Consider the alphabet $\calx=\{ 0,1,\cdots, k\}$ with spacing $s$, and any original distribution $\vtheta$ over $\calx$.
Let $n$ data be sampled from $\calx$ and then obfuscated by the truncated linear geometric mechanism $\mech:\calx\to\calx$ 
with parameter $0< \lambda \leq 0.94/s$. Let $\alpha = e^{-\lambda s}$ and $\beta = 1/(1- e^{-\lambda s})$.
Then the \inv{}-estimated vector $\vw$ satisfies  $\E[ {\| \vw - \vtheta \|}_2^2 ]\geq \frac{1}{n}(\beta^3 - 2 \alpha \beta^2 - 2)$. 
}
\begin{proof}
From the definition~\eqref{eq:TLinearGeometric} of the truncated linear geometric mechanism with parameters $\lambda , s$, 
the inverse matrix $\mech^{-1}$ of the mechanism is shown in Figure~\ref{fig:geometric_inverse}.  
%
%
%
By the definition of \inv{} we have $\vw = \vq \mech^{-1}$. We will express $w_x$ for all $x\in \calx$.  
Starting with $x=0$, we have from the expression of $\mech^{-1}$ that
$w_0 = \beta q_0 - \alpha \beta^2 q_1$. Let $m_0, m_1$ be the marginal probabilities that the mechanism produces noisy observations $0,1$ respectively. 
Then we can write 
$w_0 = \frac{1}{n} \sum_{i=1}^n T^i_0$ where $T^i_0$ is a sequence of i.i.d. random variables that follow the distribution
\[
T^i_0 =  
\left\{ 
\begin{array}{l l}
\beta                & \mbox{with probability $m_0$},\\
-\alpha\beta^2 & \mbox{with probability $m_1$},\\
0                      & \mbox{with probability $1- m_0 - m_1$}. 
\end{array} 
\right.
\] 
Note that 
\begin{equation}\label{eq:E_T0}
\E[T^i_0] =  m_0 \beta - m_1 \alpha\beta^2 = \theta_0
\end{equation} 
where the final equality holds because 
$m_0 \beta - m_1 \alpha\beta^2 = (\vtheta \mech) \mech^{-1}_{* 0} = \theta_0$ (where $\mech^{-1}_{* 0}$ denotes the $0$-th column of $\mech^{-1}$ ). 
Therefore, we can write the variance of $T^i_0$ as follows.  
\begin{align}
\textit{Var}[T^i_0] &= m_0 (\beta - \theta_0)^2 + m_1 (- \alpha\beta^2- \theta_0)^2 \nonumber \\
                            & \qquad  + (1-m_0-m_1)(0-\theta_0)^2 \nonumber\\
                            &= m_0 \beta^2 + m_1 \alpha^2\beta^4 - 2\theta_0(m_0 \beta - m_1\alpha\beta^2) \nonumber\\
                            &\qquad + (m_0+m_1)\theta_0^2 + (1-m_0-m_1)\theta_0^2 \nonumber\\
                            &= m_0 \beta^2 + m_1 \alpha^2\beta^4 - 2 \theta_0^2 + \theta_0^2 \label{eq:var_T0_2}\\
                            & \geq m_0 \beta^2 + m_1 \alpha^2\beta^4 - 1 \label{eq:var_T0_3} \\
                            & =      m_0 \beta^2 + \alpha\beta^2 (m_0 \beta - \theta_0) - 1 \label{eq:var_T0_4} \\
                            & \geq m_0 \beta^2 + \alpha\beta^2 (m_0 \beta - 1) - 1 \label{eq:var_T0_5} \\
                            & = m_0 (1 + \alpha\beta) \beta^2 - \alpha\beta^2 -1 \nonumber\\
                            & = m_0 \beta^3 - \alpha\beta^2 -1\label{eq:var_T0_6}
\end{align}
where \eqref{eq:var_T0_2} and \eqref{eq:var_T0_4} follow from \eqref{eq:E_T0}, the inequalities~\eqref{eq:var_T0_3} and \eqref{eq:var_T0_5} follow from the fact $\theta_0 \leq 1$,
and~\eqref{eq:var_T0_6} follows from the fact that $1 + \alpha\beta = \beta$. Since $w_0 = \frac{1}{n} \sum_{i=1}^n T^i_0$, we can write the 
variance of $w_0$. By the symmetry of $\mech^{-1}$, we have a similar form for the variance of $w_k$. Therefore
\begin{equation}\label{eq:var_w0}
\textit{Var}[w_x] \ge (m_x \beta^3 - \alpha\beta^2 -1)/n \quad \mbox{for $x \in \{0, k\}$}
\end{equation}
Similarly, we have for every $x\in \calx\setminus\{0,k\}$ that $w_x = \frac{1}{n}\sum_{i=1}^n T^i_x$ 
where the $T^i_x$ is a sequence of i.i.d. random variables taking their values from the entries 
of the $x$-th column of $\mech^{-1}$. Therefore, we can write for all $x\in \calx$ 
\begin{equation}\label{eq:E_Tx}
\E[ T^i_x ] = (\vtheta \mech) \mech^{-1}_{*x} = \theta_x 
\end{equation}
where $\mech^{-1}_{*x}$ is the $x$-th column of $\mech^{-1}$. 
Let $m_x$ be the probability that the mechanism reports $x$. 
Then from the form of $\mech^{-1}$ (Fig.~\ref{fig:geometric_inverse}),
we have for every $x\in \calx\setminus\{0,k\}$
\begin{align}
\textit{Var}[T^i_x]  & \geq m_x ((1+\alpha^2) \beta^2 - \theta_x)^2 \nonumber\\
                             & \geq m_x ( \beta^2 + \alpha^2 \beta^2 - 1)^2 \label{eq:var_Tx_1}\\
                             & \geq m_x \beta^3 \quad\quad{\text{for all}\quad \alpha \geq 0.3904} \label{eq:var_Tx_2}
 \end{align}
 where the inequality~\eqref{eq:var_Tx_1} follows from $\beta^2 > 1$ and $\theta_x \leq 1$, 
 and inequality~\eqref{eq:var_Tx_2} follows from 
 \begin{align*}
 ( \beta^2 + \alpha^2 \beta^2 - 1)^2 - \beta^3 
 &= \left(\frac{1+\alpha^2}{(1-\alpha)^2}-1\right)^2-\frac{1}{(1-\alpha)^3} \\ 
 &= \frac{4\alpha^2+\alpha-1}{(1-\alpha)^4} > 0 \quad\forall \alpha \geq 0.3904.
 \end{align*}
%
Since $\alpha=e^{-\lambda s}$, the condition $\alpha \geq 0.3904$ is satisfied by $\lambda \leq 0.94/s$. 
From the last inequality, we obtain  
\begin{equation}\label{eq:var_wx}
\textit{Var}[w_x] \ge (m_x \beta^3)/n \quad \mbox{for all $x\in\calx\setminus\{0,k\}$}
\end{equation}
Finally by~\eqref{eq:E_Tx} we have $\E[w_x] = 1/n \sum_{i=1}^n \E[ T^i_x] = \theta_x$ for all $x\in\calx$. Thus we obtain 
\begin{align}
\E[ {\| \vw - \vtheta \|}_2^2 ] &= \sum_{x\in\calx} \E [(w_x - \theta_x)^2] 
                                              = \sum_{x\in\calx} \textit{Var}[w_x] \label{eq:invTGeomVar}\\
                                              &\geq \frac{1}{n} \left( \beta^3 - 2\alpha\beta^2-2\right) \nonumber
\end{align}
where the last equality in~\eqref{eq:invTGeomVar} follows from that $\E[ w_x  ] = \theta_x$ for all $x\in\calx$, 
and the final inequality follows from summing \eqref{eq:var_w0} and \eqref{eq:var_wx}.  
\begin{figure*}
\begin{equation*} 
\begin{bmatrix}
\beta & -\alpha\beta &0 &\dots &\dots &\dots &0 &0\\
-\alpha\beta^2  &(1+\alpha^2)\beta^2 &-\alpha\beta^2 &0 &\dots&\dots&\dots&0\\ 
0 &-\alpha\beta^2 &(1+\alpha^2)\beta^2 &-\alpha \beta^2 &0 &\dots&\dots&\dots\\ 
\dots &0 &-\alpha\beta^2 &(1+\alpha^2)\beta^2 &-\alpha\beta^2 &0 &\dots&\dots\\ 
\dots &\dots &\dots &\dots &\dots &\dots &\dots &\dots\\    
0 &\dots&\dots&\dots&0 &-\alpha\beta^2 &(1+\alpha^2)\beta^2 &-\alpha\beta^2\\ 
0 &0 &\dots&\dots&\dots &0 &-\alpha\beta &\beta  
\end{bmatrix}
\end{equation*}
\caption{Inverse matrix of the truncated linear geometric mechanism 
$\mech:\calx\to\calx$, where $\calx$ is a finite range of integers, 
$\alpha = e^{-s \lambda}$, and $\beta = \nicefrac{1}{(1-\alpha)}$.}
\label{fig:geometric_inverse}
\end{figure*}
\end{proof}

{\bf \autoref{prop:invTTaxicabBound}}~{\rm (\inv{} with truncated taxicab mechanisms)}.~{\it
For two finite ranges $\calu, \calv$ of integers, consider the planar alphabet $\calx = \calu\times\calv$ with spacing $s$, 
and any original distribution $\vtheta$ over $\calx$. Let $n$ data points be sampled from $\calx$ and then obfuscated by the 
truncated taxicab mechanism $\mech: \calx\to\calx$ with parameter $0< \lambda \leq 1.09/s$. 
Let $\alpha = e^{-\lambda s}$ and $\beta = 1/(1- e^{-\lambda s})$. 
Then the \inv{}-estimated vector $\vw$ satisfies $\E[ {\| \vw - \vtheta \|}_2^2 ]\geq \frac{1}{n}(\beta^5 - 4 \alpha \beta^3 - 4)$. 
}
\begin{proof}
First, we recall that in this case, the elements of $\calx$ are in the form $(u,v)$, 
where $u\in\calu$ and $v\in\calv$. 
With no loss of generality we let $\calu = \{0,1,\dots, k_\calu\}$ and 
$\calv = \{0,1,\dots, k_\calv\}$. 
The mechanism $\mech$ is a composition of two truncated linear geometric mechanisms $U: \calu \to \calu$ 
and $V: \calv \to \calv$. Precisely, $\mech$ is the Kronecker product of $U$ and $V$. 
Therefore we have 
\begin{align}
\mech_{(u,v) (\mu,\nu)} &= U_{u \mu} \, V_{v \nu},  \label{eq:invBoundTaxicabEq1} \\
\mech^{-1}_{(\mu,\nu) (u,v)} &= U^{-1}_{\mu u} \, V^{-1}_{\nu v}  \label{eq:invBoundTaxicabEq2}.
\end{align}
By definition we have $\vw = \vq \mech^{-1}$.   
In the following we will express every $w_{(u,v)}$ using the inverse matrix $\mech^{-1}$~\eqref{eq:invBoundTaxicabEq2} which is expressed in terms 
of $U^{-1},V^{-1}$ shown by Figure~\ref{fig:geometric_inverse}. Starting with $(0,0)$ we have
\begin{align*}
w_{(0,0)} &= \beta^2 \,q_{(0,0)} - \beta (\alpha \beta^2) \,q_{(0,1)} - (\alpha \beta^2) \beta \,q_{(1,0)}\\
&\quad+ (\alpha \beta^2)^2 \,q_{(1,1)}.
\end{align*}
Let $m_{(0,0)}, m_{(0,1)}, m_{(1,0)}, m_{(1,1)}$ be the marginal probabilities that the mechanism produces the observations 
$(0,0),(0,1),(1,0),(1,1)$ respectively. Then we can write 
$w_{(0,0)} = \frac{1}{n} \sum_{i=1}^n T^i_{(0,0)}$ where $T^i_{(0,0)}$ is a sequence of i.i.d random variables that follow the distribution
\[
T^i_{(0,0)} =   
\left\{ 
\begin{array}{l l}
\beta^2 & \text{with prob. $m_{(0,0)}$},\\
-\alpha\beta^3 & \text{with prob. $m_{(0,1)}$},\\
-\alpha\beta^3 & \text{with prob. $m_{(1,0)}$},\\
\alpha^2 \beta^4  & \text{with prob. $m_{(1,1)}$},\\
0  & \text{with prob. $1-\sum_{j,k\in\{0,1\}}m_{(j,k)}$}. 
\end{array} 
\right.
\] 
Note that 
\begin{equation}
\begin{aligned}
\E[T^i_{(0,0)}]
&= m_{(0,0)}\beta^2 - \left(m_{(0,1)}+m_{(1,0)}\right)\alpha\beta^3\\
&\quad + m_{(1,1)}\alpha^2\beta^4 = \theta_{(0,0)},
\end{aligned}
\label{eq:invBoundTaxicabEq3}
\end{equation}
because the sum in the above equation is exactly $(\vtheta \mech) \mech^{-1}_{* (0,0)}$,   
where $\mech^{-1}_{* (0,0)}$ is the $(0,0)$-th column of $\mech^{-1}$. 
Therefore, we can write the variance of $T^i_{(0,0)}$ as follows.  
\begin{align}
&\textit{Var}[T^i_{(0,0)}] = m_{(0,0)} \left(\beta^2 - \theta_{(0,0)}\right)^2 \nonumber\\
&\qquad + \left(m_{(0,1)} + m_{(1,0)} \right) \left(- \alpha\beta^3 - \theta_{(0,0)} \right)^2 \nonumber\\
&\qquad + m_{(1,1)} (\alpha^2 \beta^4 - \theta_{(0,0)})^2  \nonumber \\
&\qquad + \left(1- m_{(0,0)} - m_{(0,1)} - m_{(1,0)} - m_{(1,1)}\right) \left(- \theta_{(0,0)} \right)^2 \nonumber\\
&= m_{(0,0)} \beta^4 +  \left(m_{(0,1)} + m_{(1,0)}\right) \alpha^2\beta^6 + m_{(1,1)} \alpha^4 \beta^8 \nonumber \\
&\qquad -2\theta_{(0,0)} \bigg( m_{(0,0)} \beta^2 - \left(m_{(0,1)}+m_{(1,0)} \right) \alpha\beta^3 \nonumber\\[-4mm]
&\hspace{6cm} + m_{(1,1)} \alpha^2 \beta^4 \bigg) \nonumber\\
&\qquad  + \left(\theta_{(0,0)}\right)^2 \left( m_{(0,0)} + m_{(0,1)} + m_{(1,0)} + m_{(1,1)} \right) \nonumber \\
&\qquad + \left(\theta_{(0,0)}\right)^2 \left(1- m_{(0,0)} - m_{(0,1)} - m_{(1,0)} - m_{(1,1)}\right). \nonumber \\
&
\begin{aligned}
&= m_{(0,0)} \beta^4 +  (m_{(0,1)} + m_{(1,0)}) \alpha^2\beta^6 + m_{(1,1)} \alpha^4 \beta^8\\
&\qquad - (\theta_{(0,0)})^2 
\end{aligned}\label{eq:invBoundTaxicabEq4}\\
&\geq m_{(0,0)} \beta^4 +  (m_{(0,1)} + m_{(1,0)}) \alpha^2\beta^6  - 1 \label{eq:invBoundTaxicabEq5} \\
&
\begin{aligned}
&=  m_{(0,0)} \beta^4 \\
&\qquad+  \left(m_{(0,0)} \beta^2 + m_{(1,1)} \alpha^2 \beta^4 - \theta_{(0,0)} \right) \alpha \beta^3  - 1
\end{aligned}\label{eq:invBoundTaxicabEq6}\\
&\geq  m_{(0,0)} \beta^4 +  (m_{(0,0)} \beta^2 - 1 ) \alpha \beta^3  - 1 \label{eq:invBoundTaxicabEq7}  \\
&= m_{(0,0)} \beta^4 ( 1 + \alpha\beta) - \alpha\beta^3 -1 \nonumber \\
&= m_{(0,0)} \beta^5 - \alpha\beta^3 -1 \label{eq:invBoundTaxicabEq8}
\end{align}
where~\eqref{eq:invBoundTaxicabEq4} and~\eqref{eq:invBoundTaxicabEq6} follow from~\eqref{eq:invBoundTaxicabEq3}.  
\eqref{eq:invBoundTaxicabEq5} follows from the fact $\theta_{(0,0)}\leq1$ and $m_{(1,1)} \alpha^4 \beta^8 \geq 0$. 
\eqref{eq:invBoundTaxicabEq7} follows from that $\theta_{(0,0)} \leq 1$ and $m_{(1,1)} \alpha^2 \beta^4 \geq 0$. 
Finally~\eqref{eq:invBoundTaxicabEq8} follows from the fact $1 + \alpha\beta = \beta$. 
Since $w_{(0,0)} = \frac{1}{n} \sum_{i=1}^n T^i_{(0,0)}$, we can write the variance of $w_{(0,0)}$. 
The element $(0,0)$ can be seen as one of corners of $\calx$, defined as 
\[
Q = \{  (0,0), (0, k_\calv),  (k_\calu, 0), (k_\calu, k_\calv) \}. 
\]
By the symmetry of $\mech^{-1}$ we have the same form of the variance for $T^i_{(u,v)}$ for all $(u,v) \in Q$. Thus  
\begin{equation}\label{eq:invBoundTaxicabEq9}
\textit{Var}[w_x] \geq (m_x \beta^5 - \alpha\beta^3 -1)/n \quad\mbox{for all $x \in Q$}
\end{equation}

Similarly, for other elements $(u,v)\not\in Q$, we have $w_{(u,v)} = \frac{1}{n} \sum_{i=1}^n T^i_{(u ,v )} $ where $T^i_{(u , v)}$ is a sequence 
of i.i.d random variables defined by the $(u,v)$-th column of $\mech^{-1}$ and the marginal probabilities $m_{(\mu,\nu)}$ (for all $(\mu,\nu)\in\calx$) as follows. 
For every $(\mu,\nu) \in\calx$, we have
\[
T^i_{(u,v)} = \mech^{-1}_{(\mu,\nu) (u,v)} = U^{-1}_{\mu u} \,V^{-1}_{\nu v} \quad\mbox{with probability $m_{(\mu,\nu)}$}, 
\]
where the second equality in the above equation follows from~\eqref{eq:invBoundTaxicabEq2}. 
We also have from the above distribution that 
\begin{equation}\label{eq:invBoundTaxicabEq10}
\E[T^i_{(u,v)}] = (\vtheta \mech) \mech^{-1}_{* (u,v)} = \theta_{(u,v)}  
\end{equation}
where $\mech^{-1}_{* (u,v)}$ in the above equation is the $(u,v)$-column of $\mech^{-1}$. Therefore we obtain 
\begin{align}
\textit{Var}[T^i_{(u,v)}] &\geq m_{(u,v)} ( (1+\alpha^2)\beta^3 - \theta_{(u,v)} )^2 \nonumber\\
                                    &\geq m_{(u,v)} ( \beta^3+ \alpha^2\beta^3 - 1)^2 \label{eq:invBoundTaxicabEq11} \\
                                    &\geq m_{(u,v)} \beta^5  \quad\quad{\text{for all}\quad \alpha \geq 0.334} \label{eq:invBoundTaxicabEq12} 
\end{align}
where~\eqref{eq:invBoundTaxicabEq11} follows from that $\beta^3>1$ and $\theta_{(u,v)}<1$, 
 and~\eqref{eq:invBoundTaxicabEq12} follows from 
 \begin{align*}
 &( \beta^3 + \alpha^2 \beta^3 - 1)^2 - \beta^5 = \left(\frac{1+\alpha^2}{(1-\alpha)^3}-1\right)^2-\frac{1}{(1-\alpha)^5} \\ 
 &\hspace{1cm} = \frac{\alpha^6 -4\alpha^5 +10\alpha^4 -12\alpha^3 +9\alpha^2 +\alpha -1}{(1-\alpha)^6} \\
 &\hspace{1cm}>0 \quad\text{for all $\alpha\geq 0.334$}.
 \end{align*}
Since $\alpha=e^{-\lambda s}$, the condition $\alpha \geq 0.334$ is satisfied by $\lambda \leq 1.09/s$. 
From \eqref{eq:invBoundTaxicabEq12}, we obtain  
\begin{equation}\label{eq:invBoundTaxicabEq13}
\textit{Var}[w_{(u,v)}] \geq ( m_{(u,v)} \beta^5 )/n \quad\mbox{for all $(u,v) \in \calx\setminus Q$}. 
\end{equation}
Finally by~\eqref{eq:invBoundTaxicabEq10} we have $\E[w_{(u,v)}] = 1/n \sum_{i=1}^n \E[ T^i_{(u,v)}] = \theta_{(u,v)}$ for all $(u,v)\in\calx$. Thus we obtain 
\begin{align*}
\E[ {\| \vw - \vtheta \|}_2^2 ] 
&= \sum_{(u,v)\in\calx} \E [(w_{(u,v)} - \theta_{(u,v)})^2] \\
&= \sum_{(u,v)\in\calx} \textit{Var}[w_{(u,v)}] \\
&\geq (\beta^5 - 4 \alpha\beta^3 - 4)/n
\end{align*}
where the final inequality follows from summing~\eqref{eq:invBoundTaxicabEq9} and~\eqref{eq:invBoundTaxicabEq13}.  
\end{proof}

\section{Proofs of Section \ref{sec:infinite_alphabet}}

{\bf \autoref{lem:unlikely}}.~{\it
Consider a mechanism that works on an alphabet $\calx$ and consider a set of observations. 
If $\hat\vtheta$ is an MLE on $\calx$ with respect to the mechanism and the observations, 
then it must hold that $\hat{\theta}_{x'}=0$ for every unlikely $x'\in\calx$. 
}

\begin{proof} 
The joint probability of the observed outputs as a function of any distribution $\vtheta$
can be written as  
$
P(\vtheta) = \prod_{i=1}^n (\sum_{u \in \calx} \theta_u\, \mg_{u i}). 
$
Let $\hat\vtheta$ be a likelihood maximizer on $\calx$ with $\hat{\theta}_x>0$ for some unlikely $x\in \calx$.  
Therefore by Definition \ref{def:unlikely}, there is $x'\in \calx$ such that $\mg_{x' i} \geq \mg_{x i}$ for all $i$ and 
the inequality is strict for some $i$. Define $\vtheta'$ such that $\theta'_{x} = 0$,  $\theta'_{x'} = \hat{\theta}_{x'} + \hat{\theta}_x$, and 
$\theta'_{x^{\prime\prime}}=\hat{\theta}_{x^{\prime\prime}}$ for all $x^{\prime\prime} \not\in \{x', x\}$.
Then for every $i$ it holds
$
\sum_{u \in \calx} \theta'_u\, \mg_{u i} - \sum_{u \in \calx} \hat{\theta}_u\, \mg_{u i} = 
(\hat{\theta}_{x'} + \hat{\theta}_x) \mg_{x' i} - (\hat{\theta}_{x'} \mg_{x' i} + \hat{\theta}_x \mg_{x i}) 
    = \hat{\theta}_x (\mg_{x' i} - \mg_{x i}) \geq 0, 
$
which means that $\sum_{u \in \calx} \theta'_u\, \mg_{u i} \geq \sum_{u \in \calx} \hat{\theta}_u\, \mg_{u i}$ for every $i$ and the inequality is
strict for $i$. Therefore $P(\vtheta') > P(\hat\vtheta)$ which contradicts that $\hat\vtheta$ is a likelihood maximizer on $\calx$. 
\end{proof}

{\bf \autoref{thm:likelysubset}}.~{\it
Consider a mechanism that works on an alphabet $\calx$ and consider a set of observations. 
Let $\hat\calx$ be a likely subset of $\calx$. 
Then with respect to the mechanism and observations, 
a distribution $\hat{\vtheta}$ is an MLE on $\calx$ if and only if 
$\hat{\theta}_x=0$ for all $x \not\in \hat\calx$, and the distribution $(\hat\theta_x : x\in \hat\calx)$
is an MLE on $\hat\calx$. 
}
\begin{proof} 
Let $\hat\vtheta$ be a maximizer on $\calx$. Then it must hold, by \autoref{lem:unlikely}, 
that $\hat{\theta}_x = 0$ for every $x \not\in \hat\calx$. Since also $\hat\vtheta$ is a maximizer on $\calx$, 
then $P(\hat\vtheta) = \prod_{i=1}^n (\sum_{u\in \hat\calx} \hat{\theta}_u \mg_{u i}) \geq P(\vtheta')$ for every 
distribution $\vtheta'$ on $\calx$ with $\theta'_x=0$ for all $x \not\in \hat\calx$. This means that 
$\prod_{i=1}^n (\sum_{u\in \hat\calx} \hat{\theta}_u \mg_{u i}) \geq \prod_{i=1}^n (\sum_{u\in \hat\calx} \theta'_u \mg_{u i})$. 
Since $\vtheta'$ is clearly an arbitrary distribution over $\hat\calx$, it follows from the last inequality that the distribution 
$\vtheta$ on $\hat\calx$ defined as $\theta_x=\hat{\theta}_x$ for all $x \in \hat\calx$ is a maximizer on $\hat\calx$.  

Now, for the 'if' implication, let $\vtheta$ be a maximizer on $\hat\calx$. 
Then we show that $\hat\vtheta$, defined as $\hat{\theta}_x = \theta_x$ for all $x \in \hat\calx$ and $\hat{\theta}_x = 0$ 
otherwise, is a maximizer on $\calx$. Suppose that $\hat\vtheta$ is not a maximizer on $\calx$. 
Then any maximizer $\vtheta'$ on $\calx$ satisfies $P(\vtheta') > P(\hat\vtheta)$. 
By \autoref{lem:unlikely} it holds that $\theta'_x=0$ for all $x \not\in \hat\calx$. 
Then $P(\vtheta') = \prod_{i=1}^n (\sum_{u \in \hat\calx} \theta'_u \mg_{u i}) > P(\hat\vtheta) = \prod_{i=1}^n (\sum_{u \in \hat\calx} \hat{\theta}_u \mg_{u i})$. 
Since $\hat{\theta}_x=\theta_x$ for all $x \in \hat\calx$, then the last inequality implies that 
$\prod_{i=1}^n (\sum_{u \in \hat\calx} \theta'_u \mg_{u i}) > \prod_{i=1}^n (\sum_{u \in \hat\calx} \theta_u \mg_{u i})$
which contradicts that $\vtheta$ is a maximizer on $\hat\calx$.
\end{proof}

{\bf \autoref{prop:likelylinear}}~{\rm (A linear likely subset)}.~{\it
Suppose that $\calx \subset \reals, \calz \subset \reals$. Consider the mechanism $\mech:\calx\to\calz$ 
that satisfies $\mech_{x z} > \mech_{x' z}$ whenever $| x - z | < | x' - z |$ for all $x,x'\in\calx$ and $z\in\calz$. 
Consider also the observations $\{z^i : i\in[n] \}$ produced by $\mech$. 
Then $[x_\textit{min},x_\textit{max}] \cap \calx$ is a likely subset of $\calx$. 
}
\begin{proof}
Consider any $x'\in\calx$ with $x'> x_\textit{max}$. 
Then for every observation in $\{z^i : i\in[n] \}$ we have by the assumption on $\mech$ that
$\mech_{x_\textit{max} z^i} > \mech_{x' z^i}$. Therefore by Definition \ref{def:unlikely}, $x'$ is unlikely. 
Consider also any $x'\in\calx$ with $x'< x_\textit{min}$. 
Then for every observation in $\{z^i : i\in[n] \}$ we have
$\mech_{x_\textit{min} z^i} > \mech_{x' z^i}$. Therefore  $x'$ is unlikely. 
Thus from Definition \ref{def:likely} we obtain that $[x_\textit{min},x_\textit{max}] \cap \calx$ is a likely subset of $\calx$
\end{proof}

\begin{restatable}{lemma}{lemmaprojection}
\label{lemma:projection-onto-hull}
Consider a convex hull $U$ and any point $y \notin U$. Let $x$ be the projection of $y$ onto $U$, i.e. the closest point to $y$ in $U$. 
Then for every $z\in U$ it holds $\cos(\widehat{y x z}) \leq 0$ and $|x z| < |y z|$. 
\end{restatable}
\begin{proof}
Let $\omega = \widehat{y x z}$ and let $v$ be an arbitrary point on $\overline{xz}$. 
Then it holds that
\[
2 |x y| |x v| \cos(\omega) = {|x y|}^2+{|x v|}^2 - {|y v|}^2.
\]
Since 
$x$ is the projection of $y$ onto $U$, we have ${|x y|}^2 \leq {|y v|}^2$, which implies that 
\[
2 |x y| |x v| \cos(\omega) \leq {|x v|}^2
\]
for all values of $|x v|$. Since $\cos(\omega)$ is fixed and 
$|x y|>0$, it must hold that $\cos(\omega) \leq 0$. 
Finally we have $|yz|^2 - |xz|^2 = |x y|^2 - 2 |xy| |xz| \cos(\omega)$. Since $|xy|>0$ and 
$\cos(\omega)\leq0$ we obtain $|yz|^2 - |xz|^2>0$ which completes the proof.
\end{proof}


{\bf \autoref{prop:likelyplanar}}~{\rm (A planar likely subset)}.~{\it
Suppose that $\calx \subset \reals^2$, $\calz \subset \reals^2$. Consider the mechanism
$\mech: \calx \to \calz$ that satisfies $\mech_{x z} > \mech_{x' z}$ whenever $d(x, z) < d(x',z)$ 
for every $x,x'\in \calx$ and $z\in\calz$. Let $\calb\subset\reals^2$ be the convex hull of 
the reported points $\{z^i: i\in[n]\}$.
Then $T_{\delta'}(\calb) \cap \calx$ is a likely subset of $\calx$.  
}

\begin{proof}
For convenience let $\calb'= T_{\delta'}(\calb)$. Let $z$ be any reported point in $\reals^2$ and $x'$ be any point in $\calx\setminus\calb'$. 
We show that there is $x''\in \calb'$ such that $|x''z|<|x'z|$. To this end, let $x$ be the projection of $x'$ onto $\calb$ and then choose 
$x''$ to be the closest point to $x$ in $\calx$. Note that $x''\in \calb'$ because $|x x''|\leq \delta < \delta'$ (by the definitions of $\delta, \delta'$). 
Now we proceed to show that $|x''z|<|x'z|$. From the triangle inequality and the fact $|x''z| \leq \delta$ we obtain
\begin{equation}\label{eq:likelyplanarproof1}
|x'' z| \leq |x z| + \delta.
\end{equation}
We have ${|x'z|}^2 = {|x z|}^2 + {|x x'|}^2 - 2 {|x z|}{|x x'|} \cos(\omega)$. 
Since $\cos(\omega) \leq 0$ (by \autoref{lemma:projection-onto-hull}) and $|x x'|>\delta'$ (because $x'$ is external to $\calb'$) it follows
${|x'z|}^2 > {|x z|}^2 + {\delta'}^2$ which implies by the definition~\eqref{eq:extension} of $\delta'$ that ${|x'z|}^2 > (|x z| + \delta)^2$. From 
this inequality and~\eqref{eq:likelyplanarproof1} we get $|x''z|<|x'z|$. 
\end{proof}


{\bf \autoref{prop:likelykrr}}~{\rm (likely subset under $k$-RR)}.~{\it
Consider any finite alphabet $\calx$, and suppose that a $k$-RR mechanism is used 
to obfuscate the users' data from $\calx$ to produce the observations $\calx' = \{z^i : i\in[n]\}$. 
Then $\calx'$ is a likely subset of $\calx$. 
}

\begin{proof}
Consider an $x' \not\in \calx'$, i.e. $x'$ is not observed. We show that $x'$ is unlikely.
Consider any $z^i \in \calx'$. Then it holds that  $\Pr[z^j | z^i] \geq \Pr[z^j | x']$ for every $j\in[n]$,
and in particular $\Pr[z^i | z^i] > \Pr[z^i | x']$. Thus, by \autoref{def:unlikely}, $x'$ is unlikely. 
\end{proof}


\section{planar geometric mechanisms}\label{sec:plan_geom}

%
%
Suppose that the planar space is discretized by an infinite grid of square cells, where the side length
of each cell is $s$. Then the center points of these cells can be indexed by $\integers^2$. That is, each 
point is denoted by $(u,v)$ where $u,v$ are integers describing its horizontal 
and vertical coordinates. Let the alphabet $\calx$ be a subset of these points, that is $\calx\subseteq\integers^2$. 
Then a \emph{planar geometric mechanism}, with a privacy parameter $\geps>0$, 
maps every $(u,v) \in \calx$ to another point $(\mu,\nu) \in \integers^2$ according to the probability 
\begin{align}
\Pr[ (\mu,\nu) | (u,v) ] &= a \, e^{-\geps\, s\, \sqrt{(\mu - u)^2 + (\nu - v)^2 } } \label{eq:pg} \\
	        \text{where }
	        a &= 1 / \sum_{(u,v) \in \mathbb{Z}^2} e^ {- \geps\, s \, \sqrt{u^2+v^2}}.
	                \nonumber
\end{align}
Note that the obfuscation probability~\eqref{eq:pg} depends on the planar (Euclidean) distance between the original and reported points. 
If $\calx$ is a finite subset of $\integers^2$, we can define for the above mechanism a \emph{truncated} version which works as follows. 
Given the real location of the user $(u,v) \in \calx$, another location $(\mu,\nu)$ is 
sampled from the infinite grid $\integers^2$ according to the distribution~\eqref{eq:pg}; then $(\mu,\nu)$ 
is remapped to the nearest point $(\mu',\nu')$ in $\calx$ (according to the Euclidean distance), which is finally reported.


\section{Measuring the estimation error by total variation}\label{sec:total_variation}
\begin{figure}
\centering 
\subfigure[obfuscation using truncated \newline linear geometric mechanism]{
      \label{fig:box_ages_vlgeom_tv}
      \includegraphics[width=0.23\textwidth]{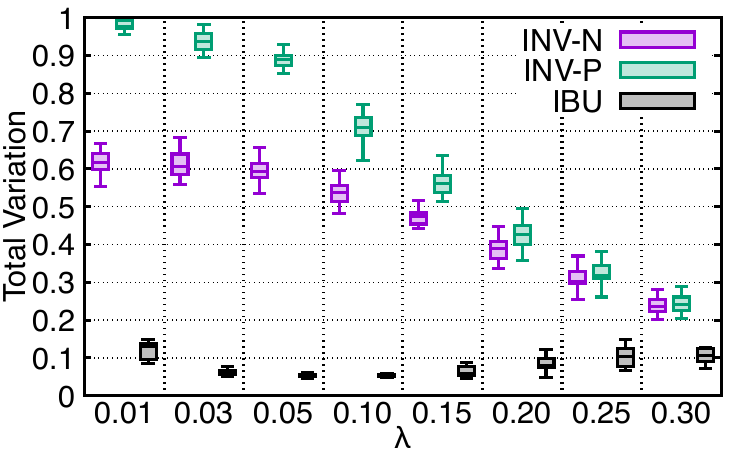}
      }
\hspace{-10pt}
\subfigure[obfuscation using truncated \newline linear Laplace mechanism]{
      \label{fig:box_ages_vllap_tv}
      \includegraphics[width=0.23\textwidth]{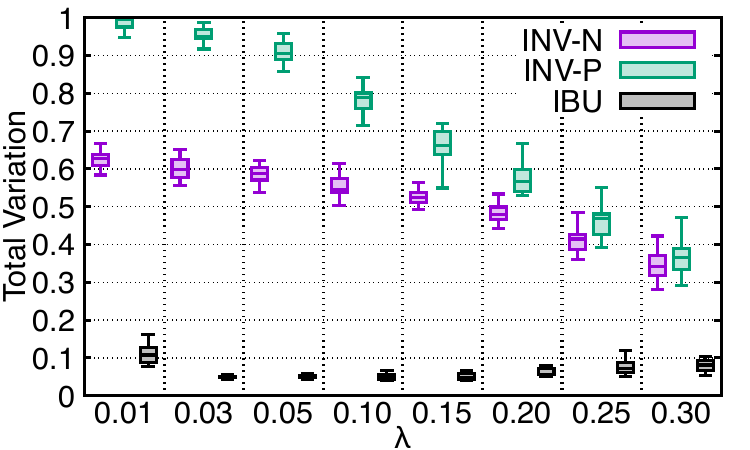}
      }
\hspace{-10pt}
\subfigure[obfuscation using \newline exponential mechanism]{
      \label{fig:box_ages_exp_tv}
      \includegraphics[width=0.23\textwidth]{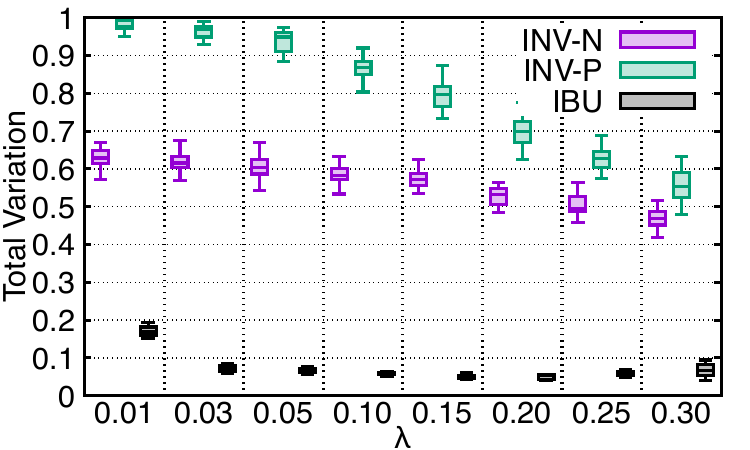}
      }
\caption{The Total Variation between the original and estimated distributions 
using \ibu{}, \invn{}, and \invp{}. Noisy data are produced using different mechanisms applied to the age data in the UCI adult dataset.}
\label{fig:box_estimators_dprivacy_tv}
\end{figure}

\begin{figure}
\centering 
\subfigure[obfuscation using truncated \newline taxicab mechanism]{
      \label{fig:box_taxicab_tv}
      \includegraphics[width=0.23\textwidth]{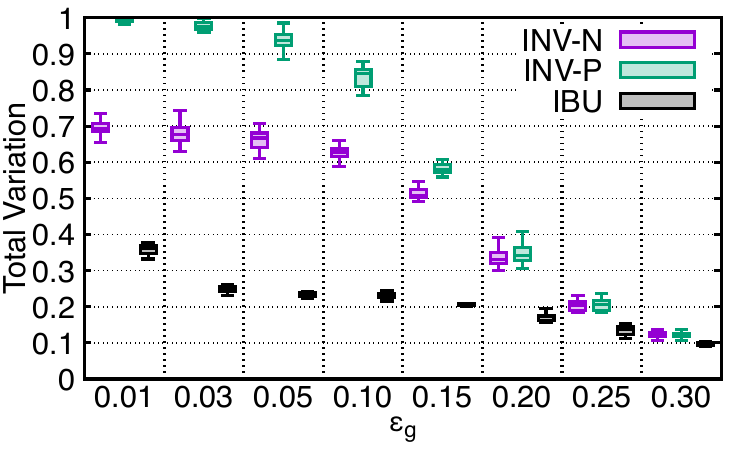}
      }
\hspace{-10pt}
\subfigure[obfuscation using truncated \newline planar geometric mechanism]{
      \label{fig:box_planar_pgeom_tv}
      \includegraphics[width=0.23\textwidth]{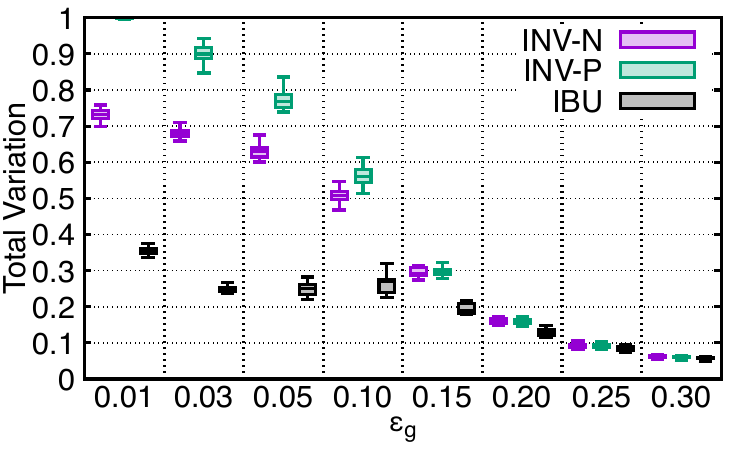}
      }
\hspace{-10pt}
\subfigure[obfuscation using truncated\newline planar Laplace mechanism]{
      \label{fig:box_planar_plap_tv}
      \includegraphics[width=0.23\textwidth]{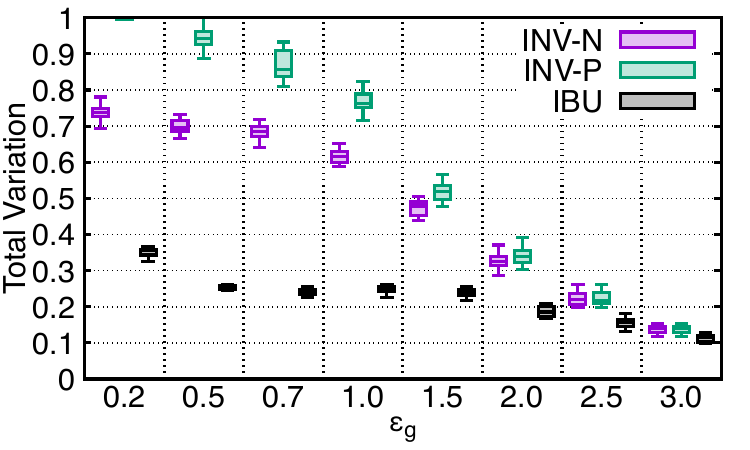}
      }
\hspace{-10pt}
\subfigure[obfuscation using \newline exponential mechanism]{
      \label{fig:box_planar_exp_tv}
      \includegraphics[width=0.23\textwidth]{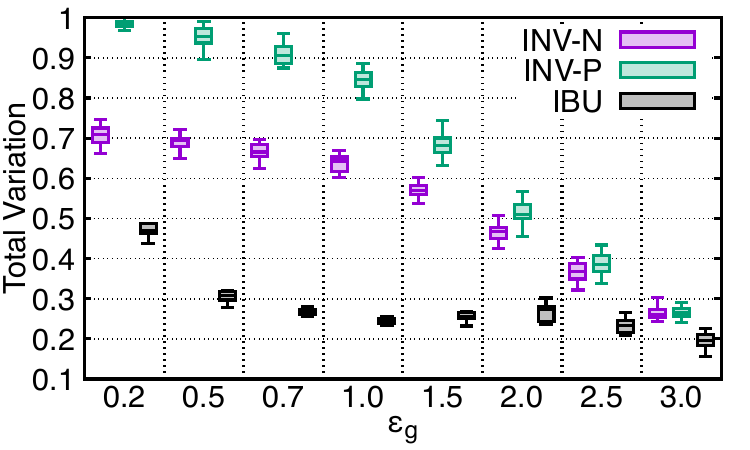}
      }
\caption{The Total Variation between the original and estimated distributions using \ibu{}, \invn{}, and \invp{}. Noisy data are produced using 
different mechanisms applied to location data constructed from the Gowalla check-ins within Manhattan.}
\label{fig:box_estimators_geind_tv}
\end{figure}

\begin{figure}
\centering 
\subfigure[age data sanitized by $k$-RR]{
      \label{fig:box_ages_krr_tv}
      \includegraphics[width=0.23\textwidth]{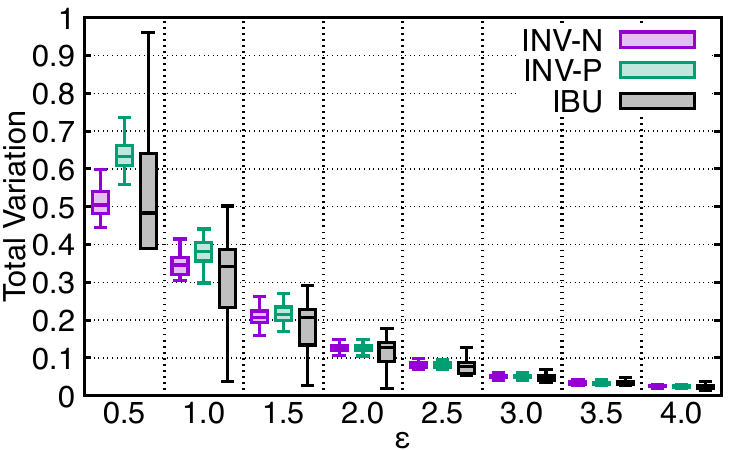}
      }
\subfigure[location data sanitized by $k$-RR]{
      \label{fig:box_planar_krr_tv}
      \includegraphics[width=0.23\textwidth]{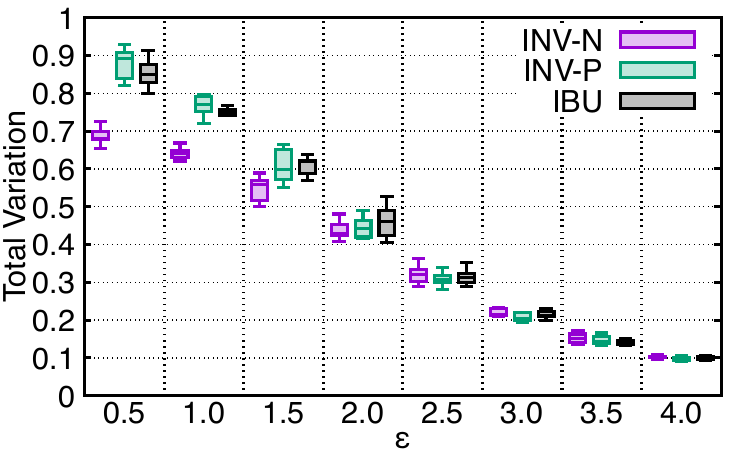}
      }
\caption{The Total Variation between the original and estimated distributions using the indicated methods. 
Noisy data are produced using a $k$-RR mechanism with the shown values of $\leps$.}
\label{fig:box_estimators_krr_tv}
\end{figure}

\begin{figure}
\centering 
\subfigure[original data sampled from\newline binomial distribution]{
      \label{fig:box_rappor_binomial_tv}
      \includegraphics[width=0.23\textwidth]{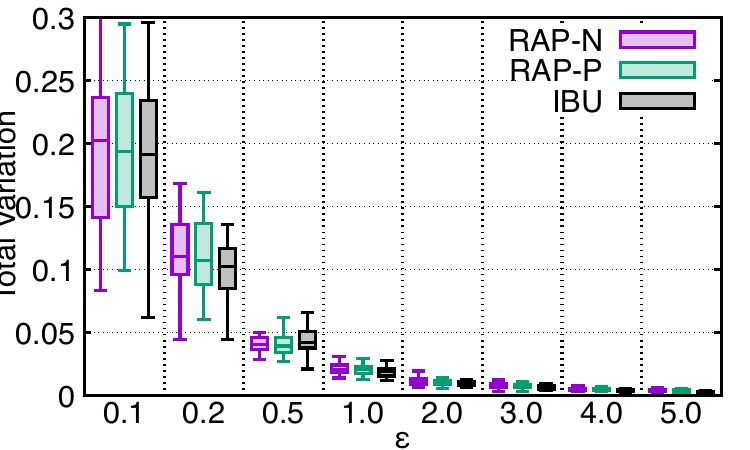}
      }
\subfigure[original data sampled from\newline uniform distribution]{
      \label{fig:box_rappor_uniform_tv}
      \includegraphics[width=0.23\textwidth]{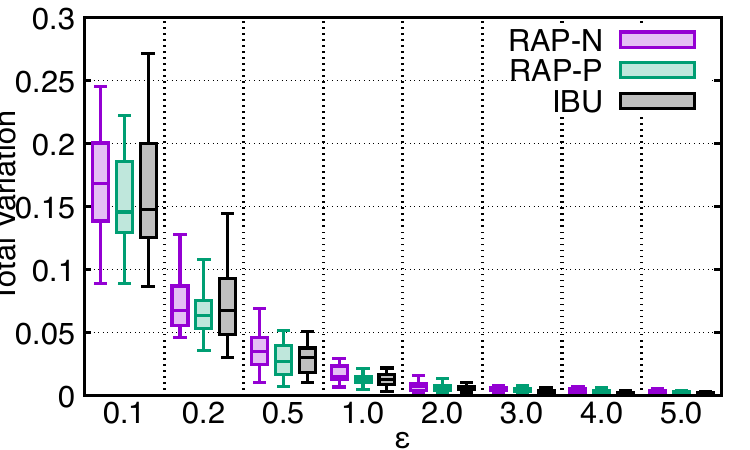}
      }
\caption{The Total Variation between the original and estimated distributions using 
\ibu{}, \rapn{}, and \rapp{},  
after applying \textsc{Rappor} mechanism for obfuscation. Original data of users are sampled from a Binomial distribution (a), 
and from uniform distribution on $\{3,4,5,6\}$ (b).}
\label{fig:box_estimators_rappor_tv}
\end{figure}

\begin{figure}
\centering 
\subfigure[$\lambda = 0.05$]{
      \label{fig:box_var_data_lgeom_0.05_tv}
      \includegraphics[width=0.23\textwidth]{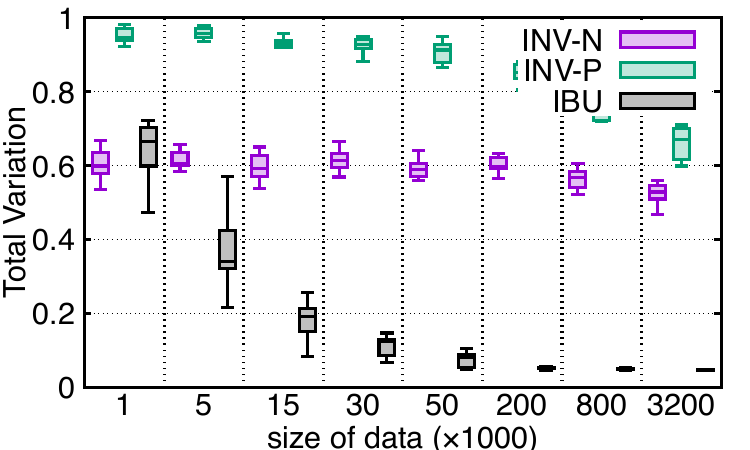}
      }
\subfigure[$\lambda = 0.1$]{
      \label{fig:box_var_data_lgeom_0.1_tv}
      \includegraphics[width=0.23\textwidth]{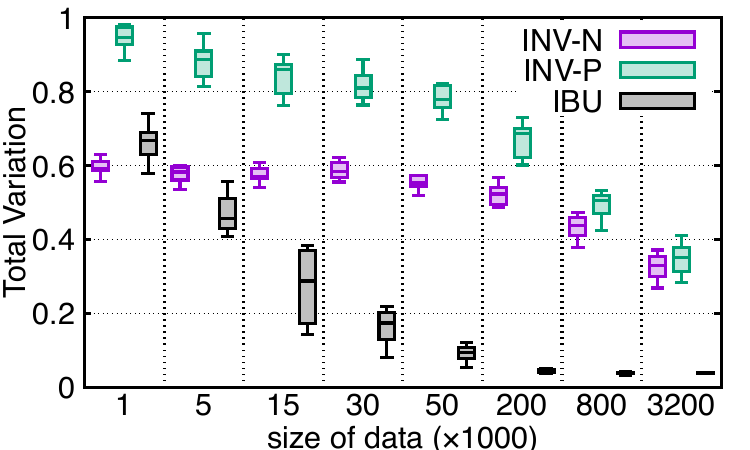}
      }
\caption{The total variation of \ibu{}, \invn{}, and \invp{} 
applied to noisy age data of various sizes. 
The noisy data are produced by the truncated linear 
geometric mechanism that satisfies $\lambda \dist$-privacy.}
\label{fig:box_var_data_lgeom_tv}
\end{figure}

\begin{figure}
\centering 
\subfigure[$\geps=0.5$]{
      \label{fig:box_var_data_taxicab_0.5_tv}
      \includegraphics[width=0.23\textwidth]{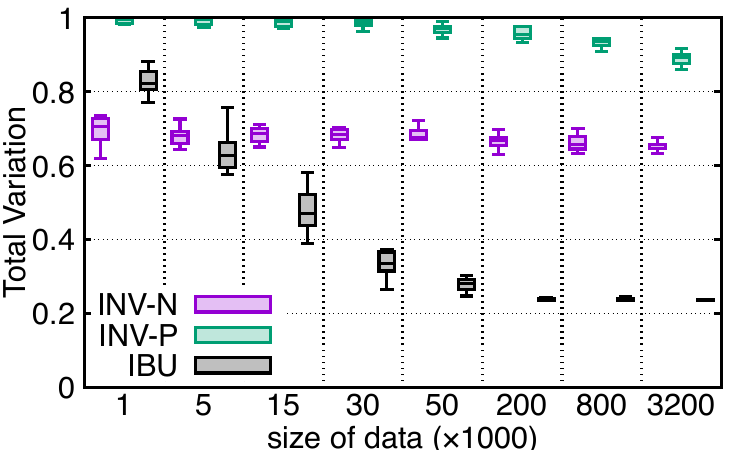}
      }
\subfigure[$\geps = 1.0$]{
      \label{fig:box_var_data_taxicab_1.0_tv}
      \includegraphics[width=0.23\textwidth]{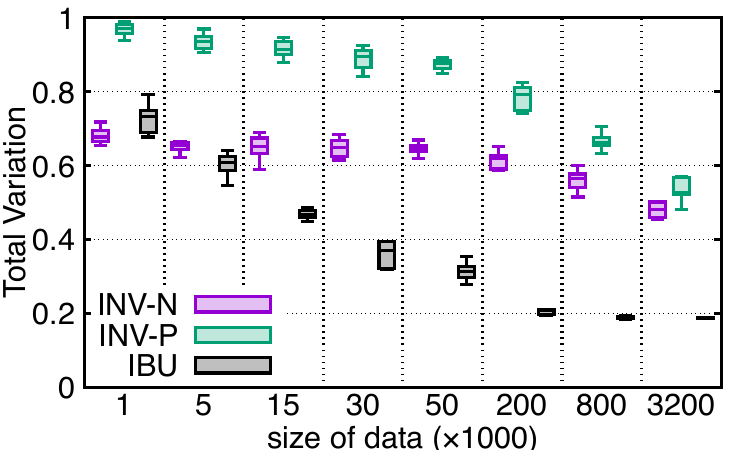}
      }
\caption{
The total variation of \ibu{}, \invn{}, and \invp{}, applied to noisy location data of various sizes. 
The noisy data are produced by a taxicab mechanism, with parameter $\geps$, applied to locations within Manhattan.}
\label{fig:box_var_data_taxicab_tv}
\end{figure}

\begin{figure}
\centering 
\subfigure[Ages]{
      \label{fig:box_var_data_krr_ages_tv}
      \includegraphics[width=0.23\textwidth]{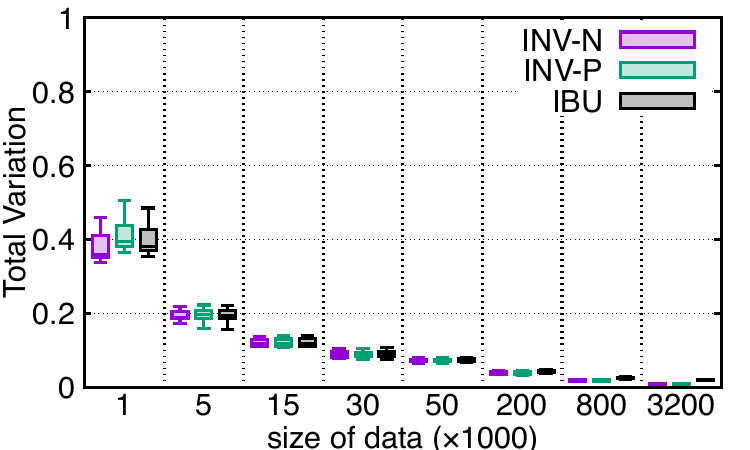}
      }
\subfigure[Locations (in Manhattan)]{
      \label{fig:box_var_data_krr_locations_tv}
      \includegraphics[width=0.23\textwidth]{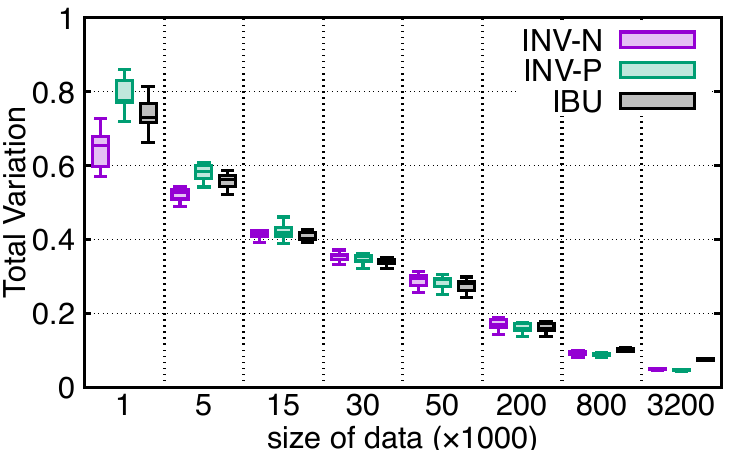}
      }
\caption{The total variation of \ibu{}, \invn{}, and \invp{} 
applied to various data sizes of ages (a), and locations (b). 
These data are produced using a $k$-RR mechanism that satisfies 
$3.0$-LDP.}
\label{fig:box_var_data_krr_tv}
\end{figure}

The total variation (TV) distance is a standard measure of the discrepancy between two probability distributions. For distributions $\vtheta$ and $\hat{\vtheta}$ over the same alphabet $\calx$, it is defined as
\[
d_{\textit{TV}}(\vtheta,\hat{\vtheta})
=\frac{1}{2}\sum_{x\in\calx}\left|\theta_x-\hat{\theta}_x\right|.
\]
The TV distance ranges from $0$ to $1$, where $0$ indicates that the two distributions are identical and $1$ indicates that they have disjoint support. Intuitively, it represents the largest possible difference between the probabilities that the two distributions assign to the same event. In this section, we use this measure to show the estimation error resulting from our experiments.   
Figures \ref{fig:box_estimators_dprivacy_tv},  
\ref{fig:box_estimators_geind_tv}, 
\ref{fig:box_estimators_krr_tv},
\ref{fig:box_estimators_rappor_tv}
show the experimental results of Sections \ref{sec:metric_mechanisms}, \ref{sec:var_priv-krr}, and \ref{sec:exp_rappor} respectively. 
The results of Section \ref{sec:var_datasize} measured in TV are given by Figures
\ref{fig:box_var_data_lgeom_tv}, 
\ref{fig:box_var_data_taxicab_tv}
\ref{fig:box_var_data_krr_tv}.

\section{Experiments on Lab test counts from the UCI Diabetes 130-US Hospitals dataset}\label{sec:labTests}
The Diabetes 130-US Hospitals 
for Years 1999--2008 dataset
is a publicly available collection of electronic health records from $130$ hospitals in the United States spanning the years 1999 to 2008 
\footnote{\url{https://archive.ics.uci.edu/dataset/296/diabetes+130-us+hospitals+for+years+1999-2008}}. 
The dataset contains more than $100,000$ patient records and includes demographic information, diagnoses, medications, and various clinical measurements. 
We consider the attribute ``num-lab-procedures'', 
which records the number of laboratory tests performed during a patient's hospital encounter. This attribute is a non-negative integer-valued personal measurement ranging from $1$ to $140$. 

\begin{figure}
\centering 
\subfigure[obfuscation using truncated \newline linear geometric mechanism]{
      \label{fig:box_labTest_vlgeom_emd}
      \includegraphics[width=0.23\textwidth]{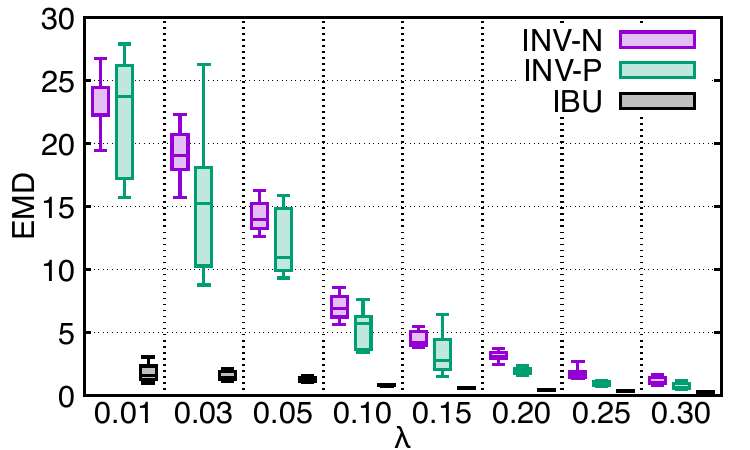}
      }
\hspace{-10pt}
\subfigure[obfuscation using truncated \newline linear Laplace mechanism]{
      \label{fig:box_labTest_vllap_emd}
      \includegraphics[width=0.23\textwidth]{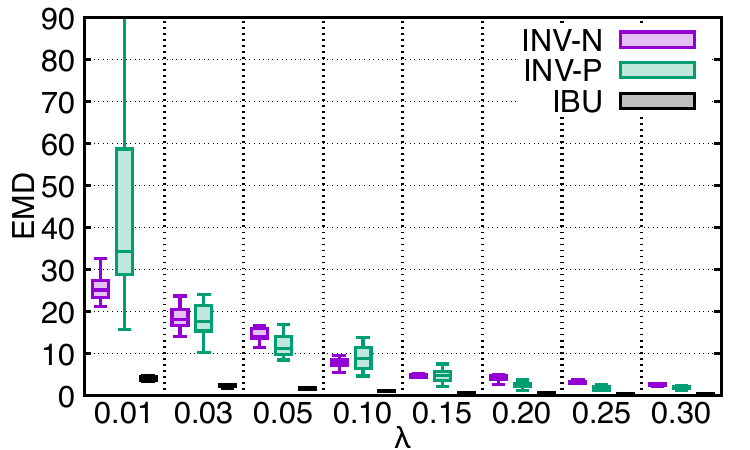}
      }
\hspace{-10pt}
\subfigure[obfuscation using \newline exponential mechanism]{
      \label{fig:box_labTest_exp_emd}
      \includegraphics[width=0.23\textwidth]{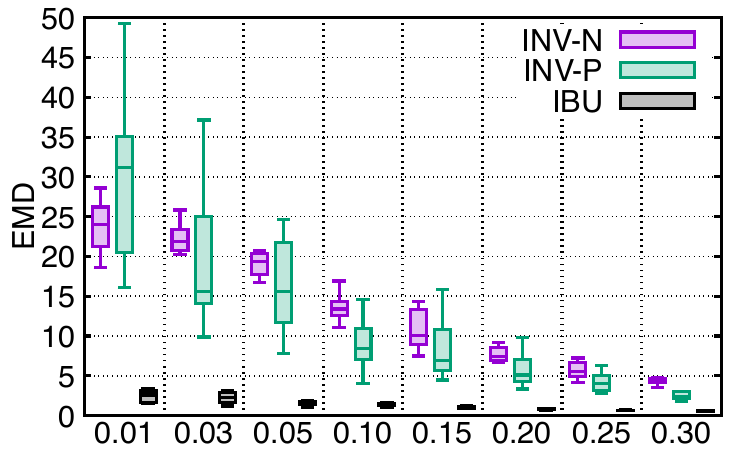}
      }
\hspace{-10pt}
\subfigure[obfuscation using $k$-RR]{
      \label{fig:box_labTest_krr_emd}
      \includegraphics[width=0.23\textwidth]{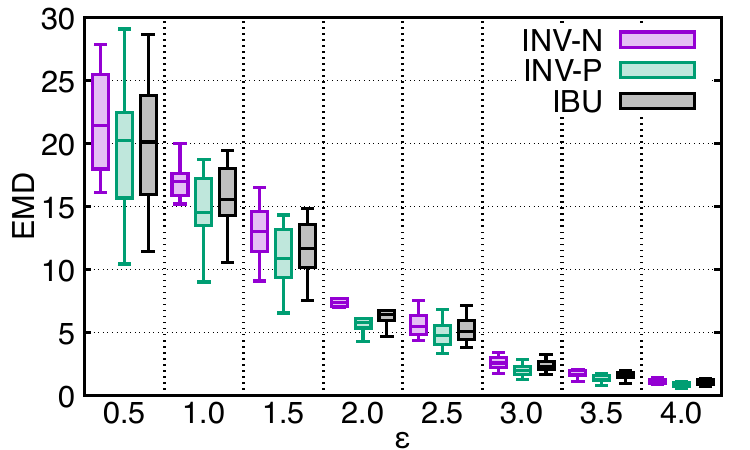}
      }

\caption{The EMD between the original and estimated distributions using \ibu{}, \invn{}, and \invp{}. The original data is from the ``num-lab-procedures'' attribute of UCI Diabetes 130-US Hospitals dataset. The noisy data points are produced by the shown metric privacy mechanisms and the $k$-RR.}
\label{fig:box_estimators_labTests_emd}
\end{figure}

We perform the experiments described in Sections~\ref{sec:metric_mechanisms}, 
and \ref{sec:var_priv-krr} on this dataset and the num-lab-procedures attribute instead of the UCI Adults and the "age" attribute, and we show the results via  
Figure \ref{fig:box_estimators_labTests_emd}.

\section{Experiments on the Gowalla check-ins in Paris}\label{sec:paris}
\begin{figure}
\centering 
\subfigure[obfuscation using truncated \newline taxicab mechanism]{
      \label{fig:box_taxicab_paris_emd}
      \includegraphics[width=0.23\textwidth]{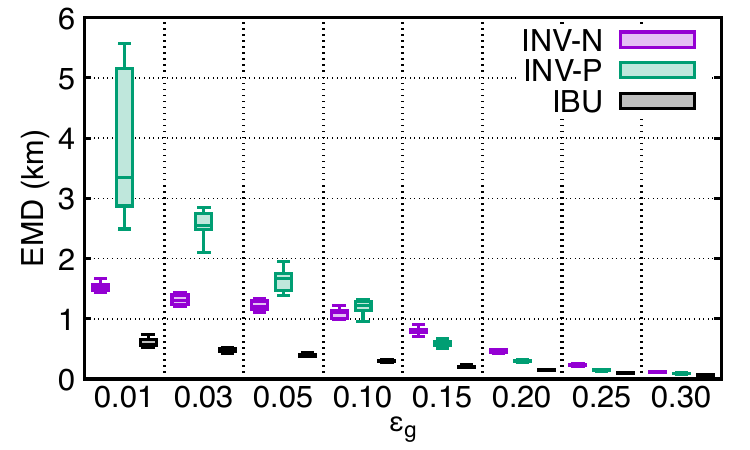}
      }
\hspace{-10pt}
\subfigure[obfuscation using truncated \newline planar geometric mechanism]{
      \label{fig:box_planar_pgeom_paris_emd}
      \includegraphics[width=0.23\textwidth]{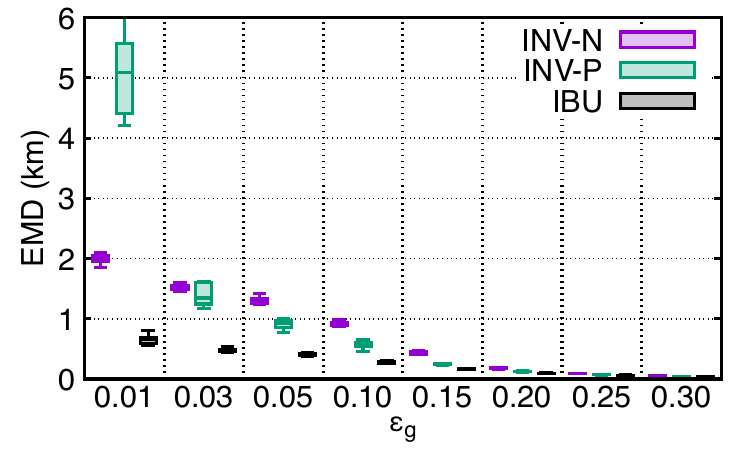}
      }
\hspace{-10pt}
\subfigure[obfuscation using truncated\newline planar Laplace mechanism]{
      \label{fig:box_planar_plap_paris_emd}
      \includegraphics[width=0.23\textwidth]{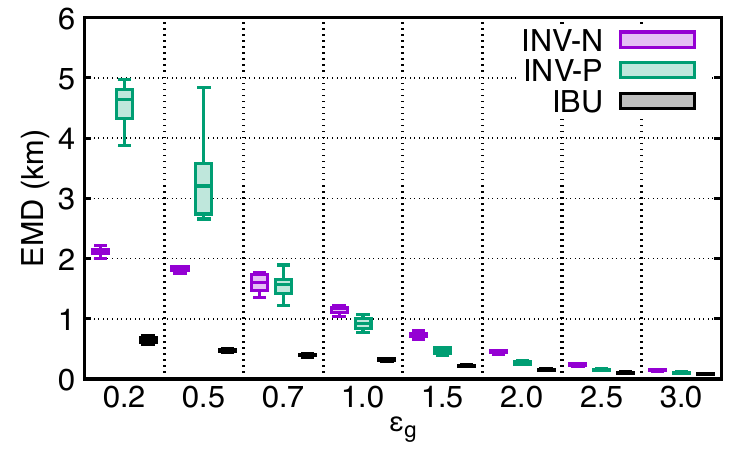}
      }
\hspace{-10pt}
\subfigure[obfuscation using \newline exponential mechanism]{
      \label{fig:box_planar_exp_paris_emd}
      \includegraphics[width=0.23\textwidth]{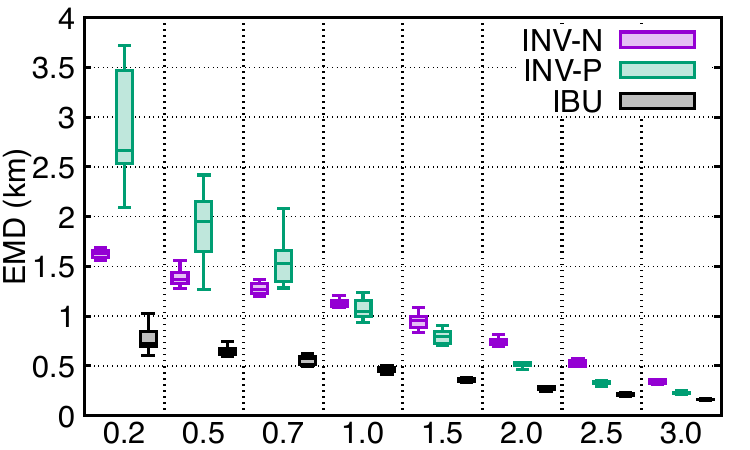}
      }
\subfigure[obfuscation using $k$-RR]{
      \label{fig:box_planar_krr_paris_emd}
      \includegraphics[width=0.23\textwidth]{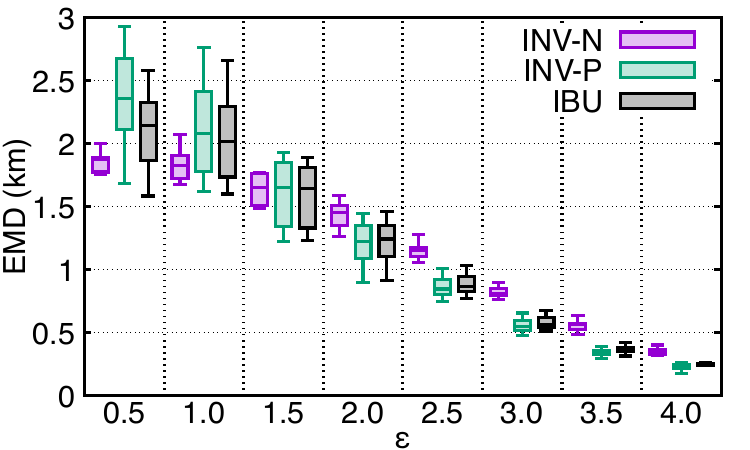}
      }
\caption{The EMD between the original and estimated distributions using \ibu{}, \invn{}, and \invp{}. Noisy data are produced using 
different mechanisms applied to location data constructed from the Gowalla check-ins within Paris.}
\label{fig:box_estimators_paris_emd}
\end{figure}

We repeat the experiments described in Sections~\ref{sec:metric_mechanisms}, 
and \ref{sec:var_priv-krr} on a planar alphabet and an original distribution defined by the Gowalla dataset. However, we now focus on the check-ins in the Paris region, specifically the area bounded by the latitudes $48.8176$ and $48.9074$ and the longitudes $2.2509$ and $2.4149$, covering an area of $10$ km $\times$ $12$ km. We define the alphabet $\calx$ by discretizing 
this region into a grid of $20\times 24$ square cells of side length $0.5$km.
Figure~\ref{fig:box_estimators_paris_emd} shows the estimation quality of \ibu{}, \invp{}, and \invn{} for the privacy mechanisms considered in the paper.

\end{document}